\pdfoutput=1

\documentclass[11pt,twoside,a4paper,cmspaper,final,collab]{cms-tdr}
\def\svnVersion{5cc5a36-D}\def\svnDate{2019/10/19}\def\cmsCernNoTag{CERN-EP-2019-152}\def\cmsCernDate{\today}\def\cmsMessage{"Published in the Journal of High Energy Physics as \href{http://dx.doi.org/10.1007/JHEP10(2019)244}{\doi{10.1007/JHEP10(2019)244}.}"}

\begin{document}\cmsNoteHeader{SUS-19-006}

\hyphenation{had-ron-i-za-tion}
\hyphenation{cal-or-i-me-ter}
\hyphenation{de-vices}
\newlength\cmsFigWidth
\ifthenelse{\boolean{cms@external}}{\setlength\cmsFigWidth{0.49\textwidth}}{\setlength\cmsFigWidth{0.9\textwidth}}
\ifthenelse{\boolean{cms@external}}{\providecommand{\cmsLeft}{top\xspace}}{\providecommand{\cmsLeft}{left\xspace}}
\ifthenelse{\boolean{cms@external}}{\providecommand{\cmsRight}{bottom\xspace}}{\providecommand{\cmsRight}{right\xspace}}
\ifthenelse{\boolean{cms@external}}{\providecommand{\cmsAppendix}{}}{\providecommand{\cmsAppendix}{Appendix~}}
\providecommand{\cmsTable}[1]{\resizebox{\textwidth}{!}{#1}}
\newlength\cmsTabSkip\setlength{\cmsTabSkip}{1ex}
\newcommand{\sieie}{\ensuremath{\sigma_{\eta\eta}}\xspace}
\newcommand{\njets}{\ensuremath{N_{\text{jet}}}\xspace}
\newcommand{\nbjets}{\ensuremath{N_{{\cPqb}\text{-jet}}}\xspace}
\newcommand{\mlsp}{\ensuremath{m_{\PSGczDo}}\xspace}
\newcommand{\mgluino}{\ensuremath{m_{\PSg}}\xspace}
\newcommand{\msquark}{\ensuremath{m_{\PSQ}}\xspace}
\newcommand{\mstop}{\ensuremath{m_{\sTop}}\xspace}
\newcommand{\zll}{\ensuremath{\cPZ\to \ell^{+}\ell^{-}}\xspace}
\newcommand{\gjets}{{{\cPgg}\!+jets}\xspace}
\newcommand{\ngdata}{\ensuremath{N_{\gamma}^\text{data}}\xspace}
\newcommand{\ngsim}{\ensuremath{N_{\gamma}^\text{sim}}\xspace}
\newcommand{\nlldata}{\ensuremath{N_{\zll}^\text{data}}\xspace}
\newcommand{\nllsim}{\ensuremath{N_{\zll}^\text{sim}}\xspace}
\newcommand{\nznn}{\ensuremath{N_{\znn}^\text{pred}}\xspace}
\newcommand{\rznnsim}{\ensuremath{\mathcal{R}_{\znn/\gamma}^{\text{sim}}}\xspace}
\newcommand{\zjets}{{{\cPZ}\!+jets}\xspace}
\newcommand{\wjets}{{{\PW}\!+jets}\xspace}
\newcommand{\znn}{\ensuremath{\cPZ\to\cPgn\cPagn}\xspace}
\newcommand{\zlljets}{{\ensuremath{\cPZ(\to\ell^{+}\ell^{-})}\!+jets}\xspace}
\newcommand{\znnjets}{{\ensuremath{\cPZ(\to\cPgn\cPagn)}\!+jets}\xspace}
\newcommand{\mtop}{\ensuremath{m_{\PQt}}\xspace}
\newcommand{\dmass}{\ensuremath{\Delta m}\xspace}
\newcommand{\dR}{\ensuremath{\Delta R}\xspace}
\newcommand{\imini}{\ensuremath{I}\xspace}
\newcommand{\dphimht}{\ensuremath{\Delta\phi_{\mht,\mathrm{j_i}}}\xspace}
\newcommand{\dpmht}[1]{\ensuremath{\Delta\phi_{\mht,\mathrm{j}_{#1}}}\xspace}
\newcommand{\dphib}{\ensuremath{\Delta{\phi}_{\mathrm{j}_{1(\PQb)}}}\xspace}
\newcommand{\dphibtrue}{\ensuremath{\Delta{\phi}_{\mathrm{j}_{1(\PQb)},\text{true}}}\xspace}
\newcommand{\SFdmcg}{\ensuremath{\mathcal{C}_{\text{data/sim}}^\gamma}\xspace}
\newcommand{\SFdmcll}{\ensuremath{\mathcal{C}_{\text{data/sim}}^{\ell\ell}}\xspace}
\newcommand{\Fdir}{\ensuremath{\mathcal{F}_{\text{dir}}}\xspace}
\newcommand{\DblR}{\ensuremath{\rho}\xspace}
\newcommand{\avgDblR}{\ensuremath{\langle\rho\rangle}\xspace}
\newcommand{\njetsisr}{\ensuremath{N_{\text{jet}}^{\text{ISR}}}\xspace}
\newcommand{\dphi}{\ensuremath{\Delta \phi}\xspace}
\newcommand{\rscale}{\ensuremath{\mu_{\text{R}}}\xspace}
\newcommand{\fscale}{\ensuremath{\mu_{\text{F}}}\xspace}
\newcommand{\mll}{\ensuremath{m_{\ell\ell}}\xspace}
\newcommand{\fextrap}{\ensuremath{\mathcal{F}_{\mathrm{j},\PQb}^{\text{data}}}\xspace}
\newcommand{\Zpurity}{\ensuremath{\beta^{\text{data}}_{\ell\ell}}\xspace}
\newcommand{\htmisstrue}{\ensuremath{\vec{H}^{\text{miss}}_{\mathrm{T},\text{true}}}\xspace}

\newcolumntype{R}{>{$}r<{$}}
\newcolumntype{L}{>{$}l<{$}}
\newcolumntype{M}{R@{$\;$}L}
\newcolumntype{S}{r@{$\,\pm\,$}r}

\cmsNoteHeader{SUS-19-006}
\title{Search for supersymmetry in proton-proton collisions at 13\TeV in final states with jets and missing transverse momentum}

\date{\today}

\abstract{
Results are reported from a search for supersymmetric particles
in the final state with multiple jets
and large missing transverse momentum.
The search uses a sample of proton-proton collisions at
$\sqrt{s}=13\TeV$ collected with the CMS detector in 2016--2018,
corresponding to an integrated luminosity of 137\fbinv,
representing essentially the full LHC Run 2 data sample.
The analysis is performed in a four-dimensional search region defined
in terms of the number of jets, the number of tagged bottom quark jets,
the scalar sum of jet transverse momenta, and the magnitude of the
vector sum of jet transverse momenta.
No significant excess in the event yield is observed relative to the
expected background contributions from standard model processes.
Limits on the pair production of gluinos and squarks are obtained in the framework
of simplified models for supersymmetric particle production and decay processes.
Assuming the lightest supersymmetric particle to be a neutralino,
lower limits on the gluino mass as large as
2000 to 2310\GeV are obtained at $95\%$ confidence level,
while lower limits on the squark mass as large as 1190 to 1630\GeV are obtained,
depending on the production scenario.
}

\hypersetup{%
pdfauthor={CMS Collaboration},%
pdftitle={Search for supersymmetry in proton-proton collisions at 13 TeV in final states with jets and missing transverse momentum},%
pdfsubject={CMS},%
pdfkeywords={CMS, physics, SUSY, LHC}}

\maketitle

\section{Introduction}
\label{sec:introduction}

The search for particles and interactions beyond the standard model (SM)
is a major goal of experiments at the CERN LHC.
The search described here focuses on experimental signatures in which
a proton-proton ($\Pp\Pp$) collision
produces at least two jets (collimated sprays of particles),
in conjunction with large unbalanced (``missing'') momentum in the direction
transverse to the beam axis.
The jets result from the production and hadronization of energetic quarks or gluons
that could be generated in the decay chains of new heavy particles.
The jets are classified according to whether
their properties are consistent with a jet
initiated by the production of a bottom quark ({\PQb} jet),
a key experimental signature in many models of new-particle production.
The large missing transverse momentum is
typically associated with the production of a stable, weakly interacting particle
that is not detected by the apparatus.
In this analysis, this quantity is inferred from
the total momentum of the observed jets in the transverse plane,
which should sum to approximately zero if there are no unobserved particles.
Signatures of this type have been studied extensively by both the ATLAS and CMS
Collaborations~\cite{Aad:2016eki,Aaboud:2017vwy,Aad:2019pfy,
Khachatryan:2016kdk,Khachatryan:2016xvy,
Khachatryan:2016epu,Khachatryan:2016dvc,Sirunyan:2017cwe}.
This signature arises frequently in theoretical models based on
supersymmetry (SUSY)~\cite{Ramond:1971gb,Golfand:1971iw,Neveu:1971rx,Volkov:1972jx,Wess:1973kz,Wess:1974tw,Fayet:1974pd,Fayet:1976cr,Nilles:1983ge,Martin:1997ns}
as well as in a broad range of other
theories~\cite{Gripaios:2009dq,Dimopoulos:1979sp,Perelstein:2005ka,Grojean:2011vu,
LopezHonorez:2012kv,No:2015xqa}
extending the SM.

The analysis uses a sample of $\Pp\Pp$ collision events at $\sqrt{s}=13\TeV$
recorded with the CMS detector in 2016--2018,
corresponding to an integrated luminosity of 137\fbinv.
This represents essentially the complete CMS Run 2 data sample
and is about four times larger than the 2016 data sample alone,
which was used in the previous analysis
based on this methodology~\cite{Sirunyan:2017cwe}.

The motivation for searches for new physics in the final state with jets
and large missing transverse momentum arises from several considerations.
Astrophysical observations provide compelling
evidence for the existence of dark matter,
known empirically to be the dominant component of matter in the universe.
A weakly interacting massive particle (WIMP) is one class of candidates
for dark matter.
However,
the SM does not contain such a particle.
Within the SM,
the Higgs boson presents special theoretical challenges.
Assuming that the Higgs boson is a fundamental particle,
its spin-0 nature implies
that the physical mass of the Higgs boson, as a quantity
in the SM, is unstable against corrections from quantum-loop processes.
In the absence of extreme fine
tuning~\cite{Barbieri:1987fn,Dimopoulos:1995mi,Barbieri:2009ev,Papucci:2011wy}
that would precisely cancel these effects,
the Higgs boson mass is generically driven to the
cutoff scale of validity of the theory,
which could be as high as the Planck scale of quantum gravity.
The instability of the Higgs boson mass,
and with it, that of the entire electroweak scale
(including the $\PW$ and $\PZ$ boson masses),
is known as the gauge hierarchy problem.
This problem has been a major challenge confronting theoretical
particle physics for several decades.
The discovery by ATLAS and CMS of a
Higgs boson with a mass around 125\GeV has strongly
highlighted this puzzle.
The concept of ``naturalness,''~\cite{Dimopoulos:1995mi,Barbieri:2009ev,Papucci:2011wy},
which refers to the degree of fine tuning of parameters,
has been discussed extensively as an important, yet difficult to quantify,
consideration in assessing theoretical scenarios.

Theories postulating physics beyond the SM, such as SUSY,
can potentially address these problems.
Supersymmetry relates each SM
bosonic field degree of freedom to a corresponding fermionic superpartner field,
and vice versa.
Each spin $J=1/2$ particle in the SM (the quarks and leptons)
therefore has a spin $J=0$ superpartner,
so the SUSY spectrum contains a large number of
scalar quarks (squarks, $\PSQ$) and scalar leptons (sleptons, $\PSl$).
The SUSY partners
of the SM gauge bosons ($J=1$) are referred to as gauginos ($J=1/2$).
For example, the superpartner of the gluon is a gluino ($\PSg$).
The minimal supersymmetric SM (MSSM)~\cite{Fayet:1976cr,Nilles:1983ge,Martin:1997ns}
contains five Higgs bosons ($J=0$)
plus the usual four electroweak gauge bosons ($J=1$) of the SM.
In the MSSM,
the partners of the Higgs and gauge bosons map onto a set of four $J=1/2$
higgsinos and four electroweak gauginos.
Because of possible mixing among these particles,
these superpartners are generically referred to as electroweakinos,
four of which are electrically neutral
(neutralinos, $\PSGc_i^0$, $i=1,\ldots,4$) and four of which are
charged (charginos, $\PSGc_j^\pm$, $j=1, 2$).
Supersymmetry provides a dark matter candidate if the
lightest supersymmetric particle (LSP) is stable and has no electric or color charge.
Stability of the LSP is guaranteed if the model
conserves $R$ parity~\cite{Fayet:1974pd,bib-rparity},
which also implies that SUSY particles are produced in pairs.
In this scenario,
which is assumed in this paper,
the lightest neutralino \PSGczDo is the LSP and
could be a WIMP dark matter candidate.

Because gluinos and squarks carry color charges,
like their SM partners,
they can be produced via the strong interaction:
they therefore have the highest production cross sections
among SUSY particles for a given mass.
The absence of signals for these particles has so far led to lower limits on their
masses of roughly $m_{\PSg}\approx 2\TeV$ for gluinos
and $m_{\sQua}\approx 1\TeV$ for light-flavored
squarks~\cite{Aad:2016eki,Aaboud:2016zdn,Khachatryan:2016kdk,Khachatryan:2016xvy,
Khachatryan:2016epu,Khachatryan:2016dvc,Sirunyan:2017cwe},
although these results are model dependent.
The present search focuses on processes involving the
production of colored SUSY particles, either gluinos or squarks.
Once the SUSY particles are produced,
they typically decay via a sequence of processes that generates jets,
leptons, and large missing transverse momentum (\ptmiss),
where \ptmiss is the vector \pt sum of the particles in an event.
Large \ptmiss is a feature of models in which the masses involved
in the decay chains allow the LSP to carry substantial transverse momentum (\pt).
So that this study is orthogonal to ones explicitly requiring leptons,
and to help enable a well-structured and independent set of SUSY searches in CMS,
the present search vetoes events in which leptons
(electrons or muons) are detected above a certain threshold in~\pt.

\section{Analysis methodology}
\label{sec-methodology}

The basic approach of the analysis involves defining search regions
in a four-dimensional space specified by key event variables
that characterize the topology and kinematics of the events:
the total number of jets (\njets),
the number of tagged {\PQb} jets (\nbjets),
the scalar sum of jet \pt (\HT),
and the magnitude of the vector \pt sum of the jets (\mht).
The \mht variable is used
to estimate the missing transverse momentum in the event.
For all-hadronic events, \mht is similar to \ptmiss,
but \mht is less susceptible to uncertainties in the modeling of soft energy deposits.

In total, there are 174 exclusive analysis bins in the four-dimensional search region,
which together provide sensitivity to a wide range of SUSY scenarios.
In each of the 174 analysis bins, the background from SM processes
is evaluated using event yields measured in
corresponding control samples in the data,
in conjunction with correction factors
obtained from Monte Carlo (MC) simulated event samples.
The principal sources of background arise from several SM processes:
production of a top quark,
either through top quark-antiquark (\ttbar) pair production or,
less often, a single top quark;
production of an on- or off-mass-shell $\PW$ or $\PZ$ boson
(\wjets and \zjets events, respectively);
and production of multijet events through quantum chromodynamics (QCD) processes.
Both top quark and \wjets events can exhibit
significant \mht and thus contribute to the background
if a {\PW} boson decays to a neutrino and an
undetected or out-of-acceptance charged lepton, including a $\tau$
lepton with either a leptonic or hadronic decay. These backgrounds
are determined using a single-lepton control sample.
Similarly, \zjets events can exhibit significant \mht if
the {\cPZ} boson decays to two neutrinos.
This background is determined using a control sample of \gjets events,
in conjunction with a control sample
in which a \PZ boson decays into an $\EE$ or $\MM$ pair.
Significant \mht in QCD multijet events can arise if the \pt of a jet is mismeasured,
if a jet falls outside the acceptance of the jet selection,
or from \PQb jets that produce one or more neutrinos.
The QCD background contribution is evaluated using
specially defined control samples together with the
``rebalance and smear''
technique~\cite{Collaboration:2011ida,Chatrchyan:2014lfa,Sirunyan:2017cwe}.

The search is performed using methodologies similar to those presented
in Ref.~\cite{Sirunyan:2017cwe}.
The search regions, however, have been optimized for the larger amount of data,
and refinements to the background estimation
procedures have been implemented.
The main difference with respect to Ref.~\cite{Sirunyan:2017cwe}
is that for the evaluation of
background from top quark and \wjets events,
we now implement a transfer factor method
rather than construct event-by-event background predictions separately
for events with a hadronic tau lepton decay
and for events with an electron or a muon.
Also, the larger data set of the current analysis allows
us to evaluate the background from \znnjets events,
in the cases with $\nbjets>0$,
using extrapolation factors based entirely on data,
rather than relying on simulation for these extrapolations
when $\njets\geq9$.

\begin{figure*}[tb]
\centering
\includegraphics[width=0.32\textwidth]{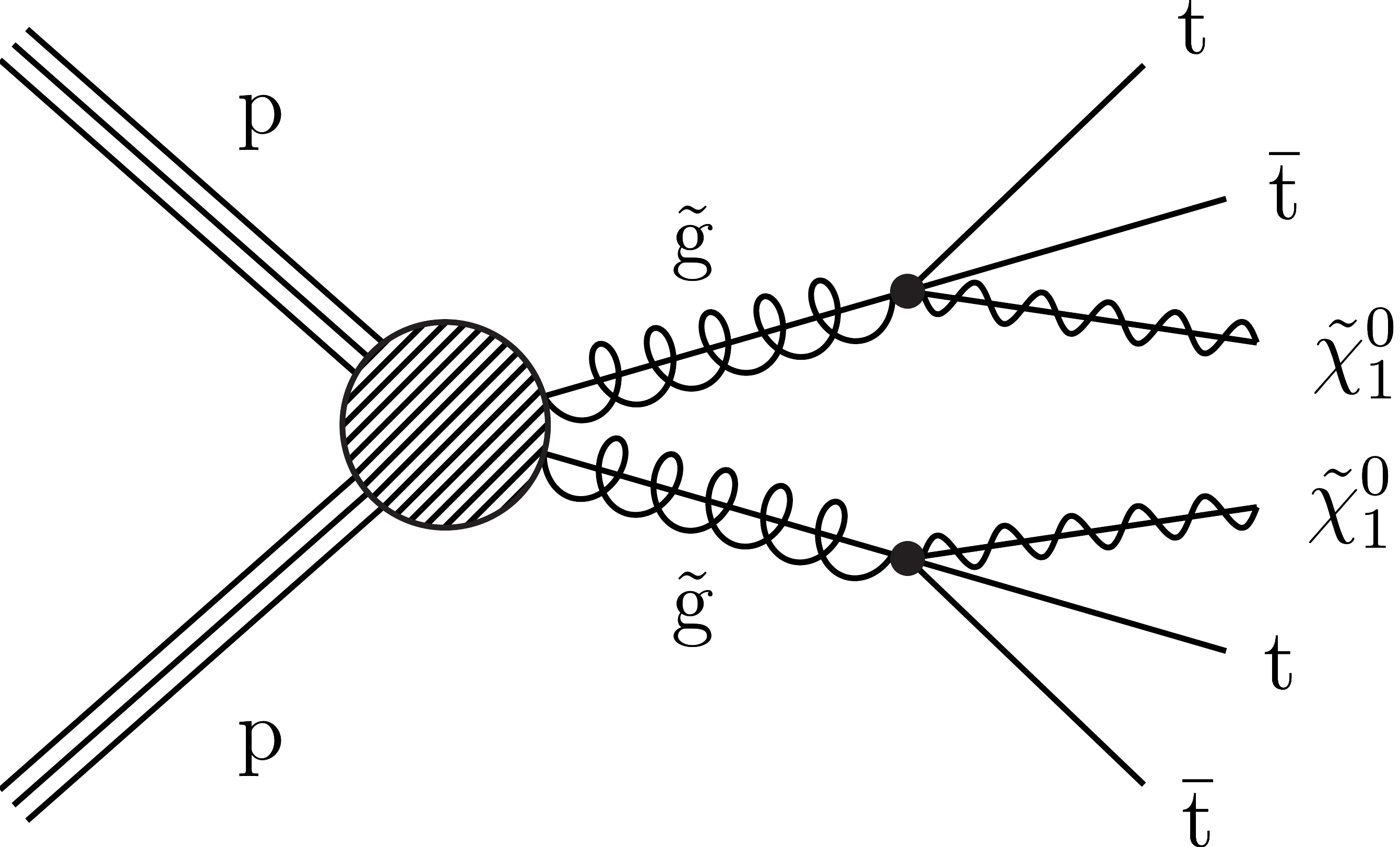}
\includegraphics[width=0.32\textwidth]{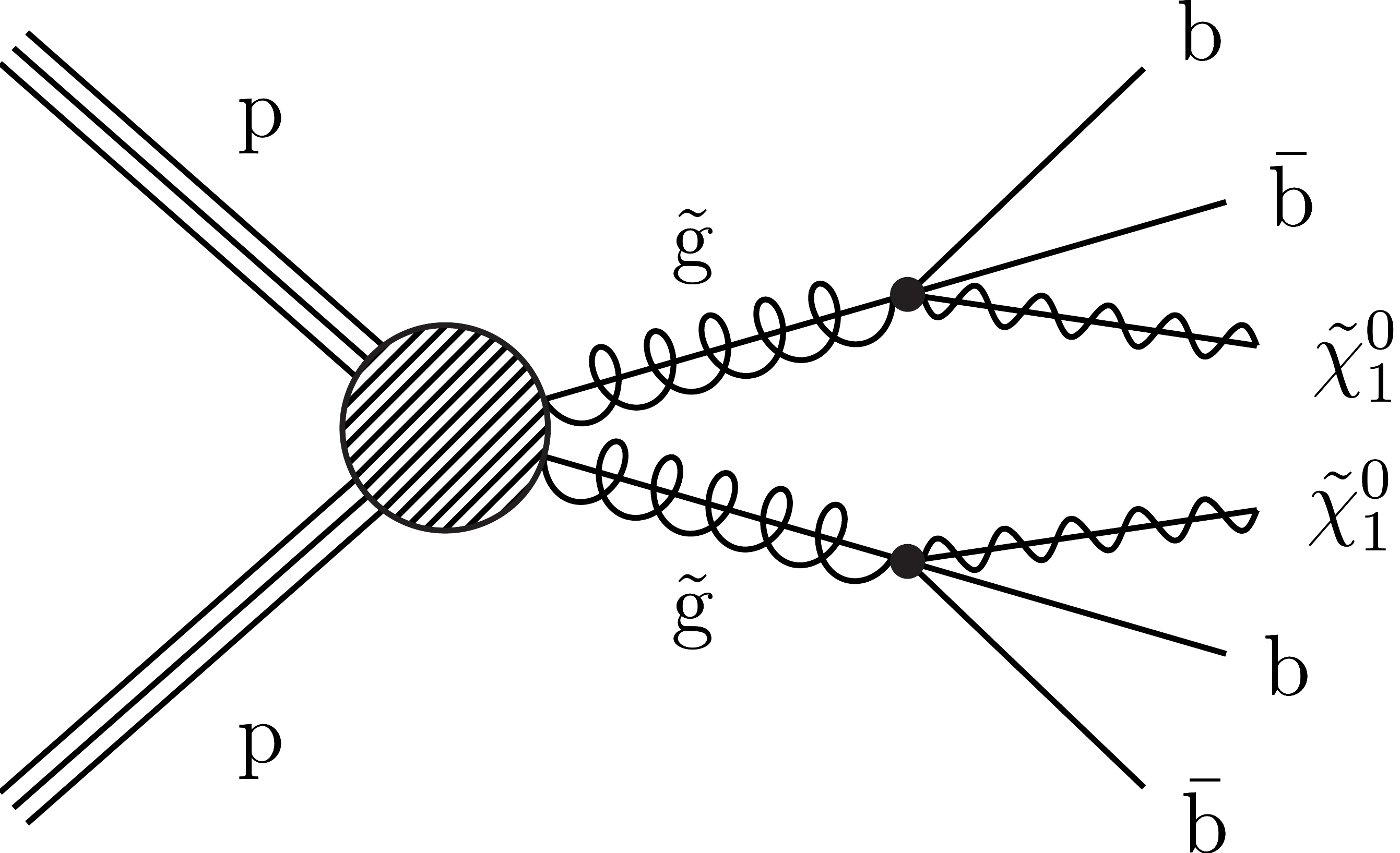}\\[2mm]
\includegraphics[width=0.32\textwidth]{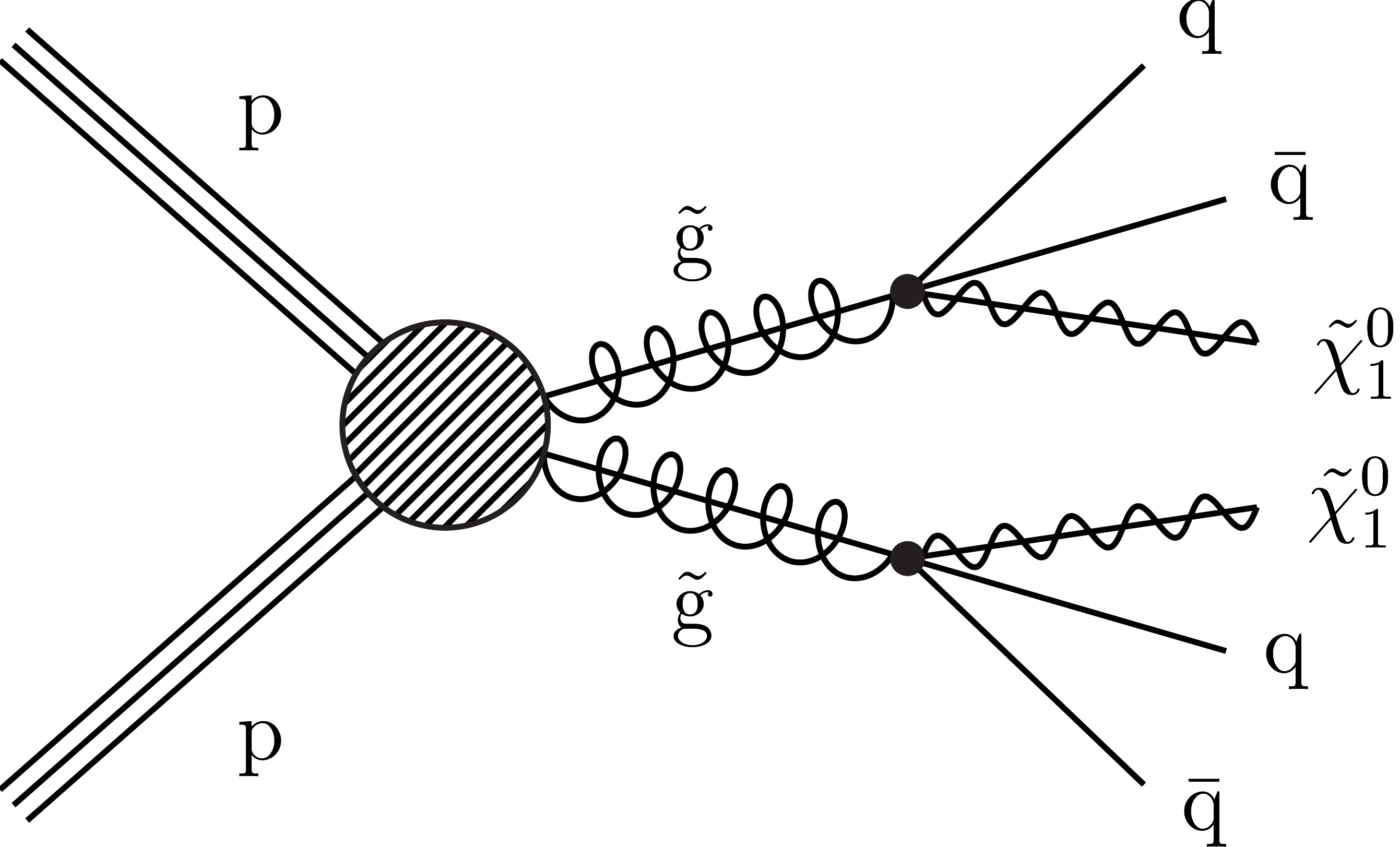}
\includegraphics[width=0.32\textwidth]{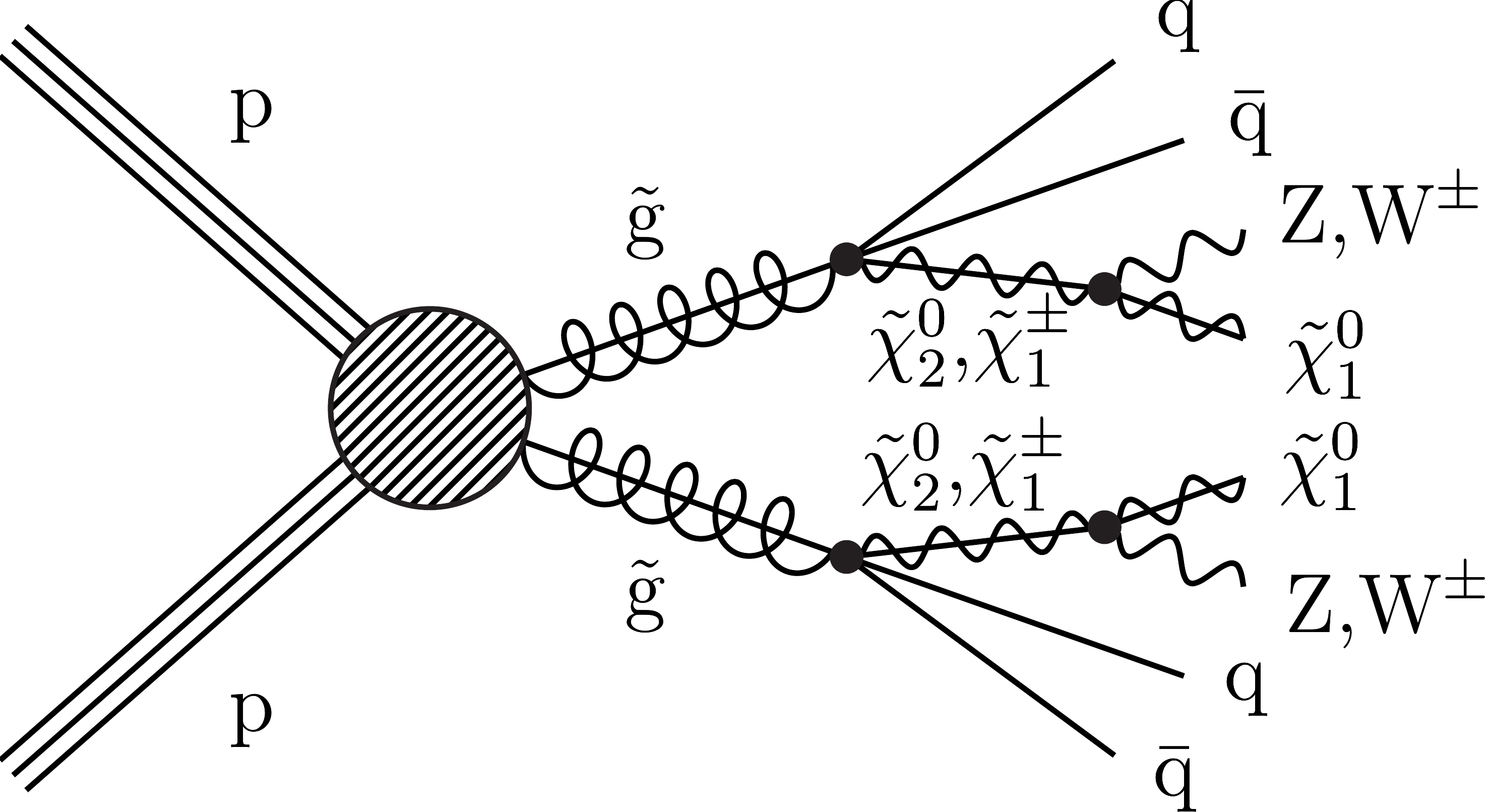}
\caption{
  Diagrams for the simplified models with direct gluino pair production
  considered in this study:
  (upper left) T1tttt, (upper right) T1bbbb,
  (lower left) T1qqqq, and (lower right) T5qqqqVV.
}
\label{fig:gg-event-diagrams}
\end{figure*}

The interpretation of the results
is performed using a set of representative SUSY models,
each of which is characterized by a small
number of mass parameters. For this purpose, we use so-called simplified
models~\cite{bib-sms-1,bib-sms-3,bib-sms-2,bib-sms-4}.
For gluino pair production, the T1tttt, T1bbbb, T1qqqq,
and T5qqqqVV~\cite{Chatrchyan:2013sza}
simplified models are considered (Fig.~\ref{fig:gg-event-diagrams}).
In the T1tttt model, each gluino undergoes a three-body decay
$\PSg\to\ttbar\PSGczDo$, where $\PSGczDo$ is the LSP.
The T1bbbb and T1qqqq models are the same as the T1tttt model,
except the \ttbar system is replaced by
bottom quark-antiquark ($\bbbar$) or
light-flavored (\cPqu, \cPqd, \cPqs, \cPqc) quark-antiquark ($\qqbar$) pairs,
respectively.
In the T5qqqqVV scenario,
each gluino decays to a light-flavored \qqbar pair
and either to the next-to-lightest neutralino \PSGczDt
or to the lightest chargino $\PSGc_1^{\pm}$.
The probability for the decay to proceed via the \PSGczDt,
$\PSGcpDo$, or $\PSGc_1^-$ is 1/3 for each channel.
The \PSGczDt ($\PSGc_1^{\pm}$) subsequently decays to the \PSGczDo
and to an on- or off-mass-shell {\cPZ} ({$\PW^\pm$}) boson.
In this model, we assign
\mbox{$m_{\PSGc_1^{\pm}} =m_{\PSGczDt} = 0.5(m_{\PSGczDo}+m_{\PSg})$.}

\begin{figure*}[tb]
\centering
\includegraphics[width=0.32\textwidth]{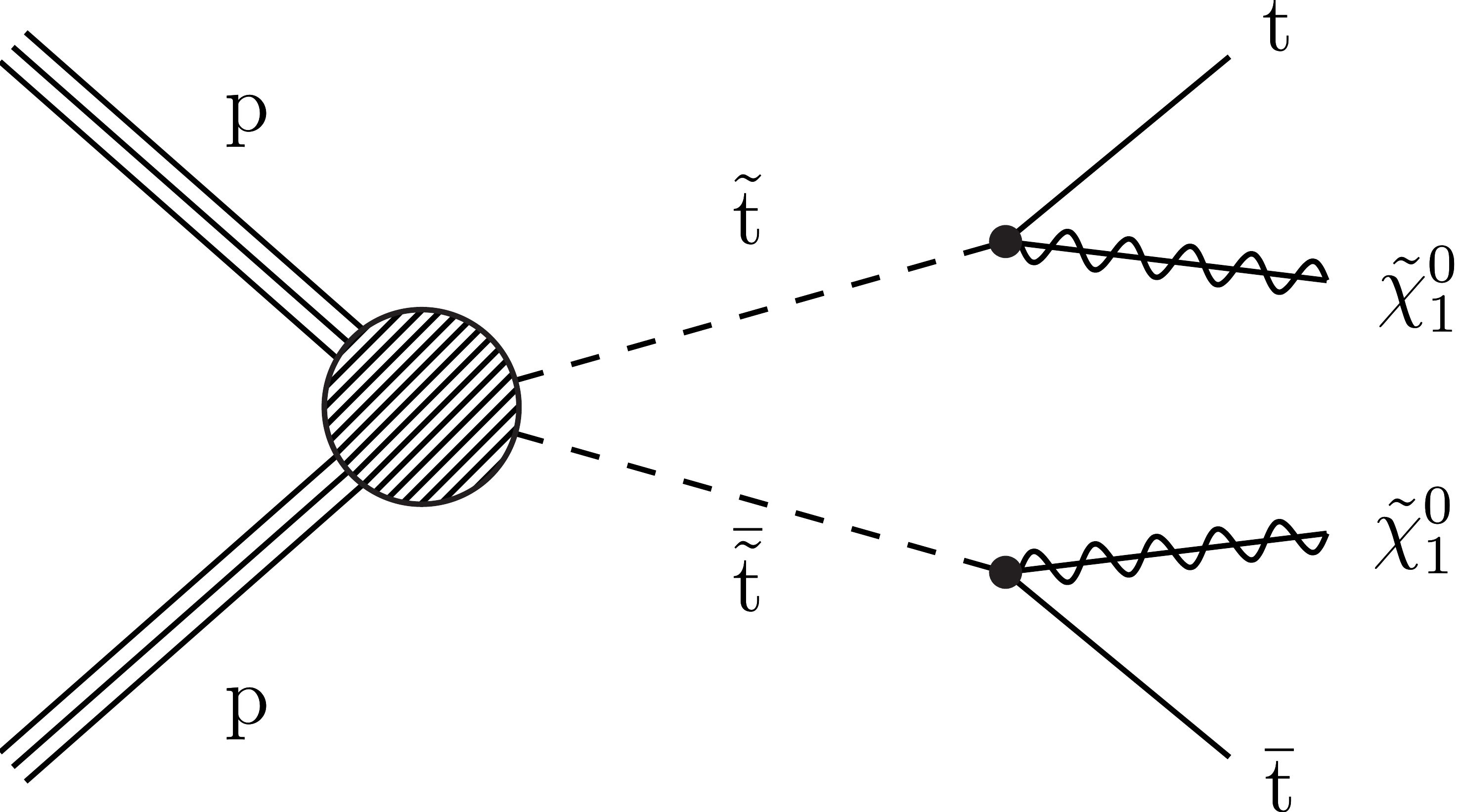}
\includegraphics[width=0.32\textwidth]{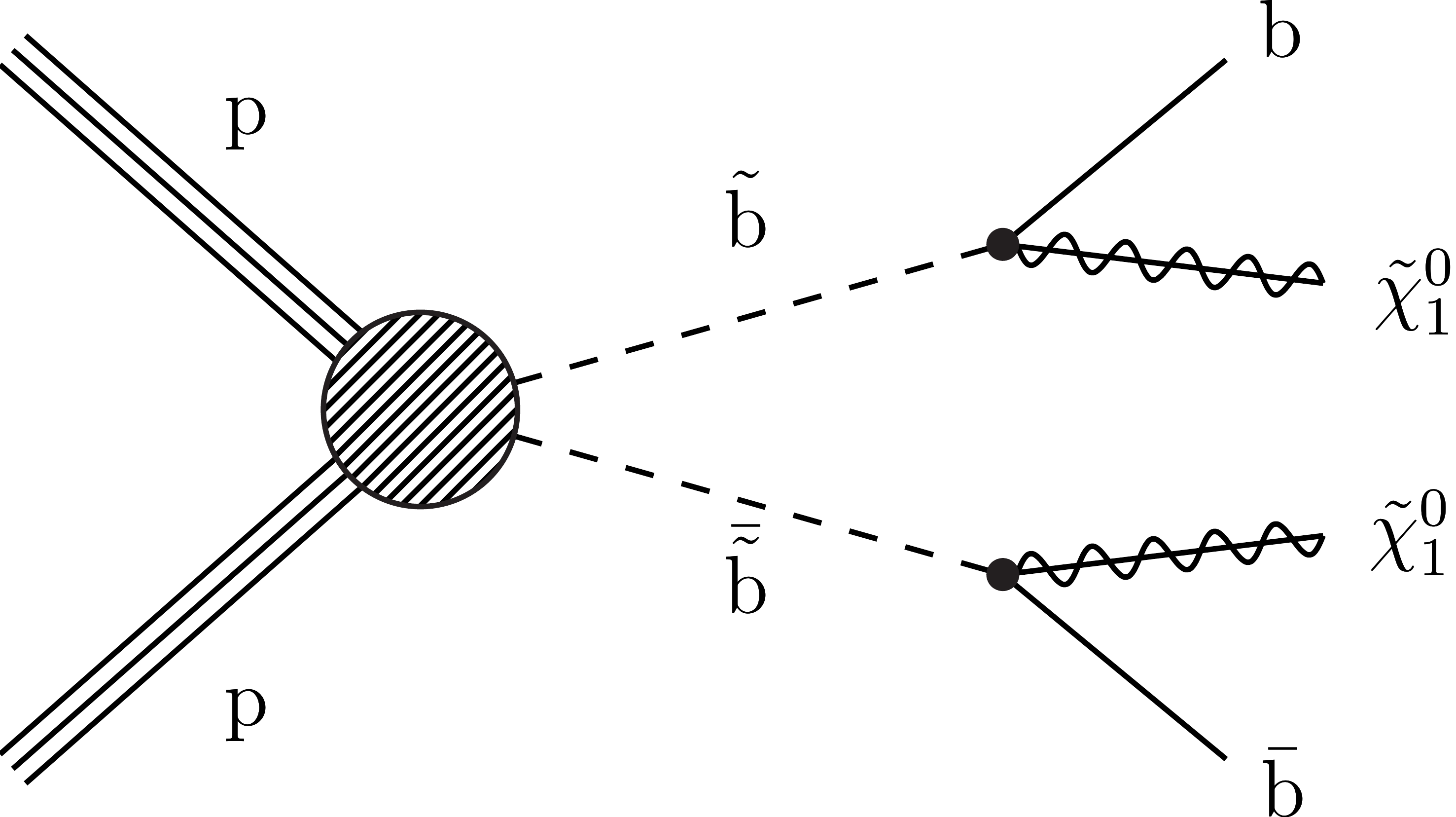}
\includegraphics[width=0.32\textwidth]{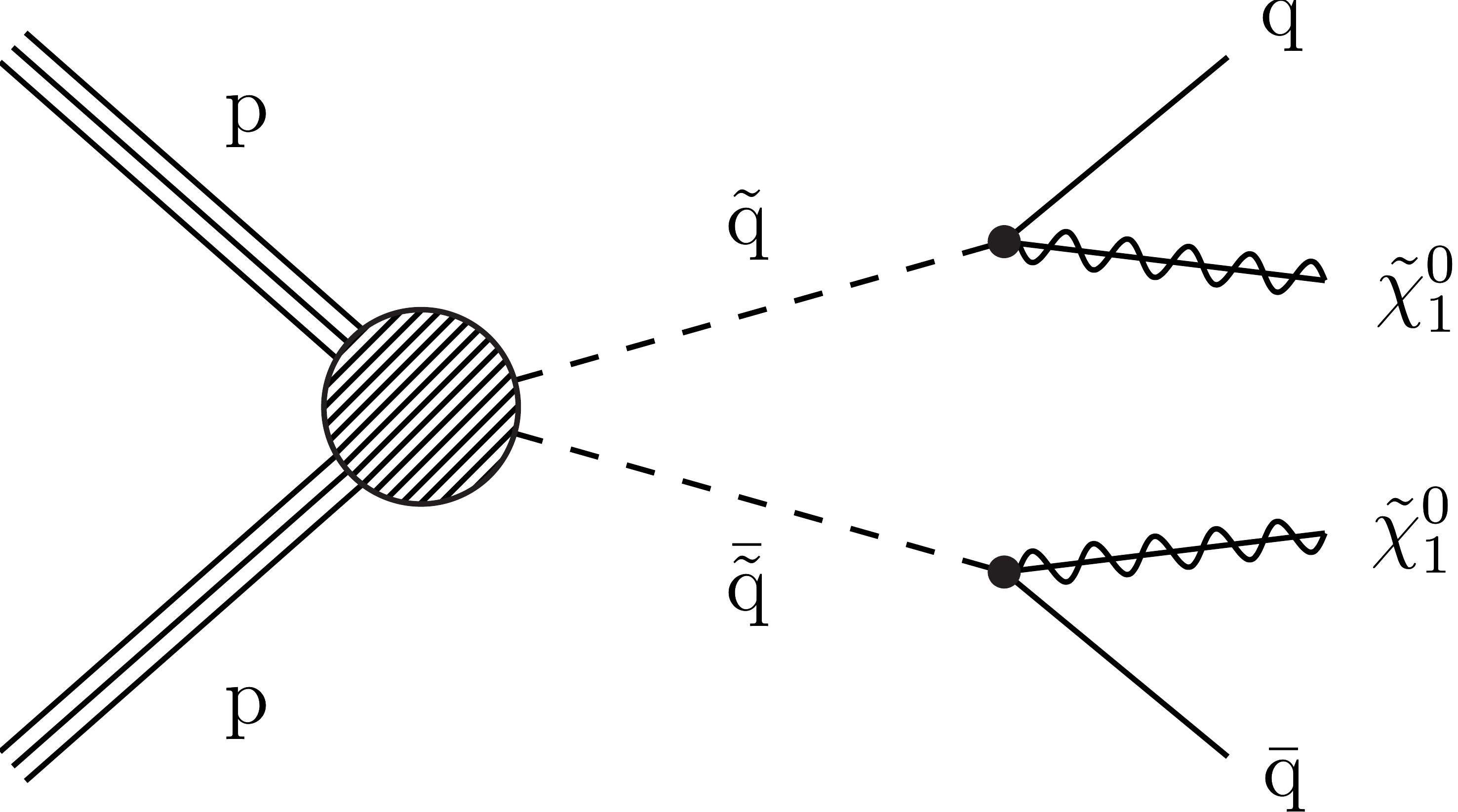}
\caption{
  Diagrams for the simplified models with direct squark pair production
  considered in this study:
  (left) T2tt, (middle) T2bb, and (right) T2qq.}
\label{fig:qq-event-diagrams}
\end{figure*}

For squark-antisquark production,
three simplified models are considered,
denoted T2tt, T2bb, and T2qq (Fig.~\ref{fig:qq-event-diagrams}).
In the T2tt model,
top squark-antisquark production is followed by the decay of
the (anti)squark to a top (anti)quark and the \PSGczDo.
The T2bb and T2qq models are the same as T2tt
except with bottom squarks and quarks,
or light-flavored squarks and quarks,
respectively, in place of the top squarks and quarks.

\section{Detector and trigger}
\label{sec:detector}

A detailed description of the CMS detector,
along with a definition of the coordinate system and
pertinent kinematic variables,
is given in Ref.~\cite{Chatrchyan:2008aa}.
Briefly,
a cylindrical superconducting solenoid with an inner diameter of 6\unit{m}
provides a 3.8\unit{T} axial magnetic field.
Within the cylindrical volume
are a silicon pixel and strip tracker,
a lead tungstate crystal electromagnetic calorimeter (ECAL),
and a brass and scintillator hadron calorimeter (HCAL).
The tracking detectors cover the range $\abs{\eta}<2.5$,
where $\eta$ is the pseudorapidity.
The ECAL and HCAL,
each composed of a barrel and two endcap sections,
cover $\abs{\eta}<3.0$.
Forward calorimeters extend the coverage to $3.0<\abs{\eta}<5.2$.
Muons are measured within $\abs{\eta}<2.4$ by gas-ionization detectors
embedded in the steel flux-return yoke outside the solenoid.
The detector is nearly hermetic,
permitting accurate measurements of~\mht.

The CMS trigger is described in Ref.~\cite{Khachatryan:2016bia}.
For this analysis, signal event candidates were recorded
by requiring \mht at the trigger level
to exceed a threshold that varied between 100 and 120\GeV,
depending on the LHC instantaneous luminosity.
The efficiency of this trigger is measured in data and is found to
exceed 97\% for events satisfying the event selection
criteria described below.
Additional triggers requiring the presence of charged leptons,
photons, or minimum values of \HT are used to select control samples for
the evaluation of backgrounds, as described below.

\section{Event reconstruction}
\label{sec:reconstruction}

Individual particles are reconstructed
with the CMS particle-flow (PF)
algorithm~\cite{CMS-PRF-14-001},
which identifies them as photons,
charged hadrons, neutral hadrons, electrons, or muons.
To improve the quality of the photon and electron reconstruction,
additional criteria are imposed on the
\sieie variable~\cite{Khachatryan:2015iwa},
which is a measure of the width of the ECAL
shower shape with respect to the $\eta$ coordinate,
and on the ratio of energies associated with the photon or electron candidate
in the HCAL and ECAL~\cite{Khachatryan:2015iwa,Khachatryan:2015hwa}.
For muon candidates~\cite{Sirunyan:2018fpa},
more stringent requirements are imposed on the matching between
silicon tracker and muon detector track segments.
Photon and electron candidates are restricted to $\abs{\eta}<2.5$
and muon candidates to $\abs{\eta}<2.4$.

The reconstructed vertex with the largest value of summed physics-object
$\pt^2$ is taken to be the primary $\Pp\Pp$ interaction vertex,
where the physics objects are the jets,
clustered using the jet finding algorithm~\cite{Cacciari:2008gp,Cacciari:2011ma}
with the charged particle tracks assigned to the vertex as inputs,
and the associated missing transverse momentum,
taken as the negative vector sum of the \pt of those jets.
Charged particle tracks associated with vertices other than
the primary vertex are removed from further consideration.
The primary vertex is required to lie within 24\unit{cm} of the
center of the detector in the direction along the beam axis
and within 2\unit{cm} in the plane transverse to that axis.

To suppress jets erroneously identified
as leptons and genuine leptons from hadron decays,
electron and muon candidates are subjected to an
isolation requirement.
The isolation criterion is based on the variable~$\imini$,
which is the scalar \pt sum of charged hadron,
neutral hadron, and photon PF candidates within a cone
of radius $\dR=\sqrt{\smash[b]{(\Delta\phi)^2+(\Delta\eta)^2}}$
around the lepton direction,
divided by the lepton~\pt,
where $\phi$ is the azimuthal angle.
The expected contributions of neutral particles from
extraneous $\Pp\Pp$ interactions (pileup)
are subtracted~\cite{Cacciari:2007fd}.
The radius of the cone is 0.2 for lepton $\pt<50\GeV$,
$10\GeV/\pt$ for $50\leq\pt\leq 200\GeV$,
and 0.05 for $\pt>200\GeV$.
The decrease in cone size with increasing lepton \pt
accounts for the increased collimation of the
decay products from the lepton's parent particle
as the Lorentz boost of the parent particle
increases~\cite{Rehermann:2010vq}.
The isolation requirement is $\imini<0.1$ (0.2) for electrons (muons).

To further suppress leptons from hadron decays
and also single-prong hadronic $\tau$ lepton decays,
charged particle tracks not identified as an isolated electron or muon,
including PF electrons and muons,
are subjected to a track isolation requirement.
(Note that PF electrons and muons that do not satisfy the
isolation requirements of the previous paragraph
are not considered to be electron and muon candidates in this analysis.)
To be identified as an isolated track,
the scalar \pt sum of all other charged particle tracks
within a cone of radius 0.3 around the track direction,
divided by the track~\pt,
must be less than 0.2 if the track is identified
as a PF electron or muon
and less than 0.1 otherwise.
Isolated tracks are required to satisfy $\abs{\eta}<2.4$.

Similarly,
we require photon candidates to be isolated.
The photon isolation requirement is based on
the individual sums of energy from
charged hadrons,
neutral hadrons,
and electromagnetic particles,
excluding the photon candidate itself,
within a cone of radius
$\dR=0.3$ around the photon candidate's direction,
corrected for pileup~\cite{Khachatryan:2015iwa}.
Each of the three individual sums is required to lie
below a (different) threshold that depends on whether the
photon appears in the barrel or endcap calorimeter.

Jets are defined by clustering PF candidates
using the anti-\kt jet algorithm~\cite{Cacciari:2008gp,Cacciari:2011ma}
with a distance parameter of~0.4.
Jet quality criteria~\cite{cms-pas-jme-10-003,CMS-PAS-JME-16-003}
are imposed to eliminate jets from
spurious sources such as electronics noise.
The jet energies are corrected for the nonlinear response of the
detector~\cite{Khachatryan:2016kdb}
and to account for the expected contributions of neutral
particles from pileup~\cite{Cacciari:2007fd}.
Jets are required to have $\pt>30\GeV$.

The identification of {\PQb} jets ({\PQb} jet tagging)
is performed by applying,
to the selected jet sample,
a version of the combined secondary vertex algorithm based on
deep neural networks (DeepCSV)~\cite{Sirunyan:2017ezt}.
The medium working point of this algorithm is used.
The tagging efficiency for {\PQb} jets with $\pt\approx30\GeV$ is 65\%.
The corresponding misidentification probability for gluon and
up, down, and strange quark jets is 1.6\%
while that for charm quark jets is 13\%.

\section{Event selection and search regions}
\label{sec:event-selection}

Events considered as signal candidates are required to satisfy:
\begin{itemize}
\item $\njets\geq 2$, where jets must appear within $\abs{\eta}<2.4$;
\item $\HT>300\GeV$, where \HT is the scalar \pt sum of jets with $\abs{\eta}<2.4$;
\item $\mht>300\GeV$, where \mht is the magnitude of \htvecmiss,
  the negative of the vector \pt sum of jets with $\abs{\eta}<5$;
  an extended $\eta$ range is used to calculate \mht
  so that it better represents the total missing momentum in an event;
\item $\mht<\HT$, because events with $\mht>\HT$ are likely to arise from mismeasurement;
\item no identified isolated electron or muon candidate with $\pt>10\GeV$;
\item no isolated track with $\mT<100\GeV$
  and $\pt>10\GeV$
  ($\pt>5\GeV$ if the track is identified as a PF electron or muon),
  where \mT is the transverse mass~\cite{Arnison:1983rp}
  formed from \ptvecmiss and the isolated-track \pt vector,
  with \ptvecmiss the
  negative of the vector \pt sum of all PF objects with appropriate
  calibration applied as explained in Ref.~\cite{Sirunyan:2019kia};
  the \mT requirement restricts the veto to situations
  consistent with a {\PW} boson decay;
\item no identified, isolated photon candidate with $\pt>100\GeV$;
  this requirement has a minimal impact on signal efficiency and is
  implemented to make the analysis orthogonal to
  SUSY searches based on events with
  photons and missing transverse energy,
  which typically require photon $\pt\gtrsim100\GeV$
  (e.g., Ref.~\cite{Sirunyan:2019hzr});
\item
  $\dphimht>0.5$ for the two highest \pt jets j$_1$ and j$_2$,
  with \dphimht the azimuthal angle between \htvecmiss
  and the \pt vector of jet j$_{\mathrm{i}}$;
  if $\njets\geq3$, then, in addition, $\dpmht3>0.3$
  for the third-highest \pt jet j$_3$;
  if $\njets\geq4$, then, yet in addition, $\dpmht4>0.3$
  for the fourth-highest \pt jet j$_4$;
  all considered jets must have $\abs{\eta}<2.4$;
  these requirements suppress background from QCD events,
  for which \htvecmiss is usually aligned along a jet direction.
\end{itemize}
In addition,
anomalous events with reconstruction failures or
that arise from noise or beam halo interactions
are removed~\cite{Sirunyan:2019kia}.

\begin{figure*}[tb]
\centering
    \includegraphics[width=0.95\textwidth]{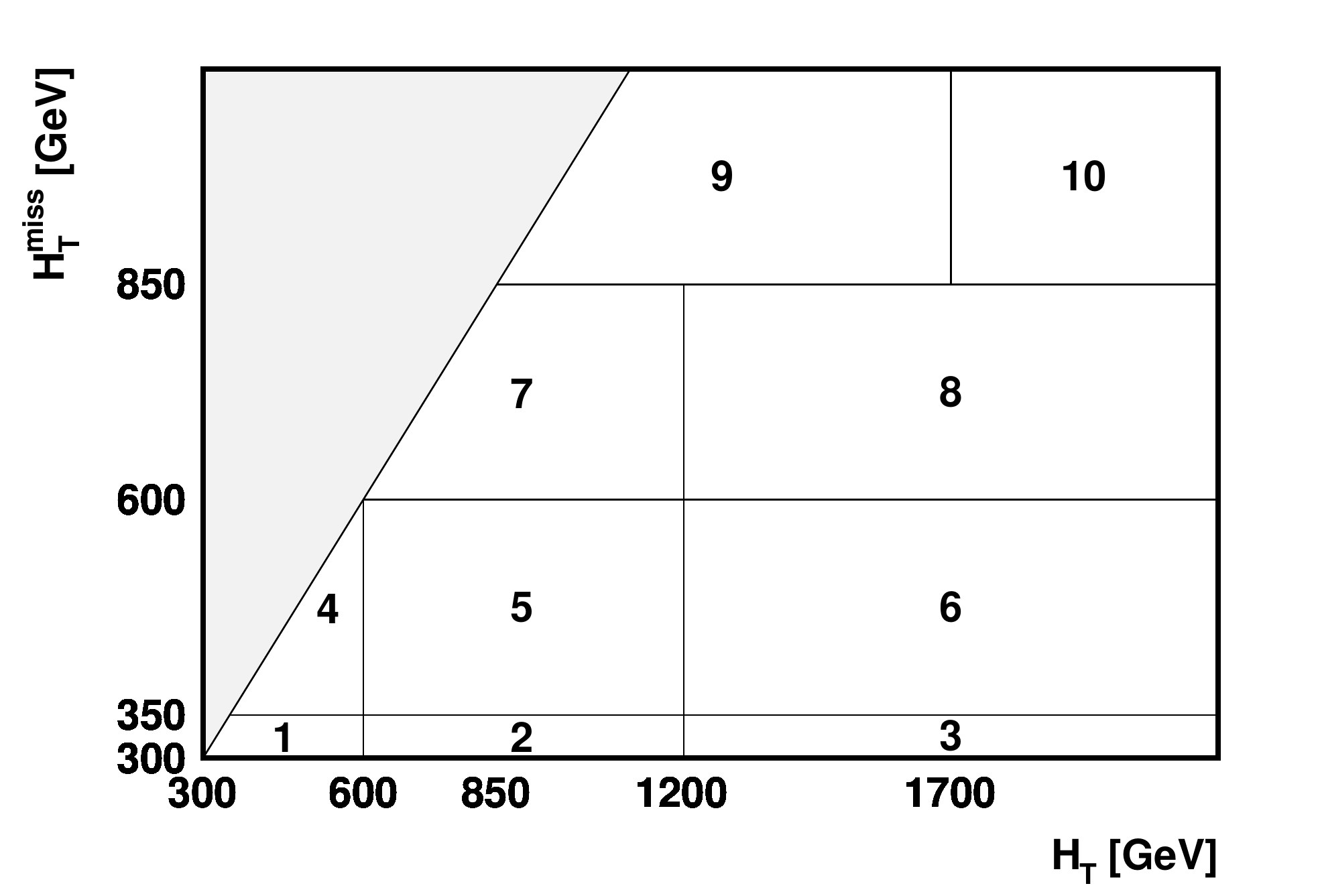}
    \caption{
      Schematic illustration of the 10 kinematic search intervals
      in the \mht versus \HT plane.
      The diagonal line delineating the leftmost edge of regions
      1, 4, 7, and 9 corresponds to the restriction $\mht<\HT$.
      Regions 1 and 4 are excluded for $\njets\geq8$.
      The rightmost and topmost bins are unbounded,
      extending to $\HT=\infty$ and $\mht=\infty$, respectively.
    }
    \label{fig:HT-MHT}
\end{figure*}

The search is performed in a four-dimensional region defined by exclusive
intervals in \njets, \nbjets, \HT, and \mht.
The search intervals in \njets and \nbjets are:
\begin{itemize}
\item \njets: 2--3, 4--5, 6--7, 8--9, $\geq$10;
\item \nbjets: either 0, 1, 2, $\geq$3 (for intervals with $\njets \geq 4$), or
  0, 1, $\geq$2 (for the $\njets=2$--3 interval).
\end{itemize}
For \HT and \mht,
10 kinematic intervals are defined,
as indicated in Table~\ref{tab:kine-bins} and Fig.~\ref{fig:HT-MHT}.
For $\njets\geq8$,
the kinematic intervals labeled 1 and 4 are discarded
because of the small numbers of events.
The total number of search bins in the four-dimensional space is~174.

\begin{table}[tb]
\centering
\topcaption{
Definition of the search intervals in
the \mht and~\HT variables.
Intervals 1 and 4 are discarded for $\njets\geq8$.
In addition,
regions with $\mht>\HT$ are excluded as illustrated in Fig.~\ref{fig:HT-MHT}.
}
\label{tab:kine-bins}
\begin{tabular}{ccc}
Interval & \mht [\GeVns{}] & \HT [\GeVns{}]  \\ \hline
1 & 300--350 & 300--600 \\
2 & 300--350 & 600--1200 \\
3 & 300--350 & $>$1200 \\
4 & 350--600 & 350--600  \\
5 & 350--600 & 600--1200  \\
6 & 350--600 & $>$1200 \\
7 & 600--850 & 600--1200 \\
8 & 600--850 & $>$1200 \\
9 & $>$850 & 850--1700 \\
10 & $>$850 & $>$1700 \\
\end{tabular}
\end{table}

\section{Simulated event samples}
\label{sec:mc}

The evaluation of background (Section~\ref{sec:background})
is primarily based on data control regions.
Samples of MC simulated SM events are used to evaluate
multiplicative transfer factors that account for
kinematic or other selection criteria differences
between the data control and signal regions and
to validate the analysis procedures.

The SM production of \ttbar, \wjets, \zjets, \gjets,
and QCD events is simulated using the
{\MGvATNLO}\,2.2.2~\cite{Alwall:2014hca,Alwall:2007fs}
event generator with leading order (LO) precision.
The \ttbar events are generated with
up to three additional partons in the matrix element calculations.
The \wjets, \zjets, and \gjets events are generated
with up to four additional partons.
Single top quark events produced through the $s$ channel,
diboson events such as those originating from
$\PW\PW$, $\cPZ\cPZ$, or $\cPZ\PH$ production
(with $\PH$ a Higgs boson),
and rare events such as those from $\ttbar\PW$,
$\ttbar\cPZ$, and $\PW\PW\cPZ$ production,
are generated with {\MGvATNLO}\,2.2.2
at next-to-leading order (NLO)~\cite{Frederix:2012ps},
except that $\PW\PW$ events in which both {\PW} bosons decay leptonically
are generated using the
{\POWHEG}\,v2.0~\cite{Nason:2004rx,Frixione:2007vw,Alioli:2010xd,Alioli:2009je,Re:2010bp}
program at NLO.
This same \POWHEG generator is used to describe
single top quark events produced through the $t$ and $\cPqt\PW$ channels.
The detector response is modeled with
the \GEANTfour~\cite{Agostinelli:2002hh} suite of programs.
Normalization of the simulated background samples is performed using
the most accurate cross section calculations
available~\cite{Alioli:2009je,Re:2010bp,Alwall:2014hca,Melia:2011tj,Beneke:2011mq,
Cacciari:2011hy,Baernreuther:2012ws,Czakon:2012zr,Czakon:2012pz,Czakon:2013goa,
Gavin:2012sy,Gavin:2010az},
which generally correspond to NLO or next-to-NLO (NNLO) precision.

Samples of simulated signal events are generated at LO using {\MGvATNLO}\,2.2.2,
with up to two additional partons included in the matrix element calculations.
The production cross sections are
determined with approximate NNLO plus next-to-next-to-leading logarithmic (NNLL)
accuracy~\cite{bib-nlo-nll-01,bib-nlo-nll-02,bib-nlo-nll-03,bib-nlo-nll-04,bib-nlo-nll-05,
Beenakker:2016lwe,Beenakker:2011sf,Beenakker:2013mva,Beenakker:2014sma,
Beenakker:1997ut,Beenakker:2010nq,Beenakker:2016gmf}.
Events with gluino (squark) pair production are generated
for a range of gluino \mgluino (squark \msquark) and LSP \mlsp mass values,
with $\mlsp<\mgluino$ ($\mlsp<\msquark$).
The ranges of mass considered vary according to the model,
but are generally from around 600--2500\GeV for \mgluino,
200--1700\GeV for \msquark,
and 0--1500\GeV for~\mlsp
(see Section~\ref{sec:results}).
For the T5qqqqVV model,
the masses of the intermediate \PSGczDt and $\PSGc_1^{\pm}$
are given by the mean of \mlsp and $\mgluino$,
as was already stated in the introduction.
The gluinos and squarks decay according to
the phase space model~\cite{Sjostrand:2014zea}.
To render the computational requirements manageable,
the detector response is described using the CMS fast
simulation program~\cite{Abdullin:2011zz,Giammanco:2014bza},
which yields results that are generally
consistent with the {\GEANTfour}-based simulation.
To improve the consistency of the fast simulation description
with respect to that based on \GEANTfour,
we apply a correction of 1\% to account for differences in the efficiency
of the jet quality requirements~\cite{cms-pas-jme-10-003,CMS-PAS-JME-16-003},
corrections of 5--12\% to account for differences
in the {\cPqb} jet tagging efficiency,
and corrections of 0--14\% to account for differences
in the modeling of \HT and \mht.

All simulated samples make use of the
{\PYTHIA}\,8.205~\cite{Sjostrand:2014zea} program
to describe parton showering and hadronization.
The CUETP8M1~\cite{ Khachatryan:2015pea} (CP5~\cite{Sirunyan:2019dfx})
{\PYTHIA}\,8.205 tune was used to produce the SM background samples
for the analysis of the 2016 (2017 and 2018) data,
with signal samples based on the CUETP8M1 tune for 2016
and on the CP2 tune~\cite{Sirunyan:2019dfx} for 2017 and 2018.
Simulated samples generated at LO (NLO) with the CUETP8M1 tune use the
\textsc{nnpdf2.3lo} (\textsc{nnpdf2.3nlo})~\cite{Ball:2013hta}
parton distribution function (PDF),
while those using the CP2 or CP5 tune use the \textsc{nnpdf3.1lo}
(\textsc{nnpdf3.1nnlo})~\cite{Ball:2017nwa} PDF.
The simulated events are generated with a distribution
of $\Pp\Pp$ interactions per bunch crossing that is adjusted to match
the corresponding pileup distribution measured in data.

To improve the description of initial-state radiation (ISR),
the \MGvATNLO prediction is compared to data in a control
region enriched in \ttbar events:
two leptons ($\Pe\Pe$, $\Pgm\Pgm$, or $\Pe\Pgm$)
and two tagged {\cPqb} jets are required.
The number of all remaining jets in the event is denoted \njetsisr.
A correction factor is applied to simulated \ttbar and signal events
so that the \njetsisr distribution agrees with that in data.
The correction is found to be unnecessary for \ttbar samples that are
generated with the CP5 tune,
so it is not applied to those samples.
The central value of the correction ranges from 0.92 for $\njetsisr = 1$
to 0.51 for $\njetsisr \geq 6$.
From studies with a single-lepton data control sample,
dominated by \ttbar events,
the associated systematic uncertainty is taken to be 20\%
of the correction for \ttbar events
and 50\% of the correction for signal events,
where the larger uncertainty in the latter case accounts
for possible differences between signal and \ttbar event production.

\section{Background evaluation}
\label{sec:background}

The evaluation of the SM backgrounds is primarily based on data control regions (CRs).
Signal events,
if present,
could populate the CRs,
an effect known as signal contamination.
The impact of signal contamination is accounted for in the
interpretation of the results (Section~\ref{sec:results}).
Signal contamination is negligible for all CRs except for
the single-lepton CR described in Section~\ref{sec:lost_lepton}.
Similarly, it is negligible for all signal models except those
that can produce an isolated track or lepton.
With respect to the models examined here,
signal contamination is relevant only for the
T1tttt, T5qqqqVV, and T2tt models.

\subsection{Background from top quark and \texorpdfstring{\wjets}{W+jets} events:
``lost leptons''}
\label{sec:lost_lepton}

The background from the SM production of
\ttbar, single top quark, and \wjets events
originates from {\PW} bosons that decay leptonically
to yield a neutrino and a charged lepton.
The charged lepton can be an electron, a muon, or a \PGt lepton.
The \PGt lepton can decay leptonically to produce an electron or a muon
or it can decay hadronically,
in each case yielding at least one additional neutrino.
For {\PW} boson decays that produce electrons or muons,
top quark and \wjets events can enter as background to the signal region
if there is large  \mht from the neutrino(s) and
if the electron or muon lies outside the analysis acceptance,
is not reconstructed,
or is not isolated.
For {\PW} boson decays that produce a hadronically decaying \PGt lepton,
top quark and \wjets events can enter as background
if there is large \mht from the neutrinos.
Collectively, the background from events with top quark and \wjets
production is referred to in this paper as the ``lost-lepton'' background.

To evaluate the lost-lepton background,
a single-lepton (\Pe, \PGm) CR is selected using the same trigger
and event selection criteria used for signal events,
except the electron and muon vetoes are inverted
and the isolated-track veto is not applied.
Exactly one isolated electron or muon is required to be present.
The single-electron and single-muon samples are
combined to form a single CR.
The transverse mass \mT formed from \ptvecmiss and
the lepton \pt vector is required to satisfy $\mT<100\GeV$.
This requirement has a high efficiency for SM events
while reducing potential contamination from signal events with large \ptmiss.

The signal contamination in the resulting CR is generally small,
with a typical value of 7, 3, and 1\% for the T1tttt, T5qqqqVV, and T2tt model,
respectively.
The contamination tends to be larger in search regions with
large values of \njets, \nbjets, \HT, and/or \mht,
while it is usually negligible in search regions with small \HT and \mht.
For certain values of \mgluino or \mstop and~\mlsp,
the contamination can be as large as 30--50, 4--12, and 20--50\%
for the respective model.
In a narrow diagonal range in the \mlsp versus \mstop plane,
for which $\mstop-\mlsp\approx\mtop$,
the signal contamination for the T2tt model
can even be as large as around 90\% at small \mlsp.
Because of this large contamination,
this diagonal region is excluded from the analysis
as explained in Section~\ref{sec:results}
(see Fig.~\ref{fig:limits-squarks} (upper left)).

\begin{figure*}[ht]
  \centering
  \includegraphics[width=0.95\textwidth]{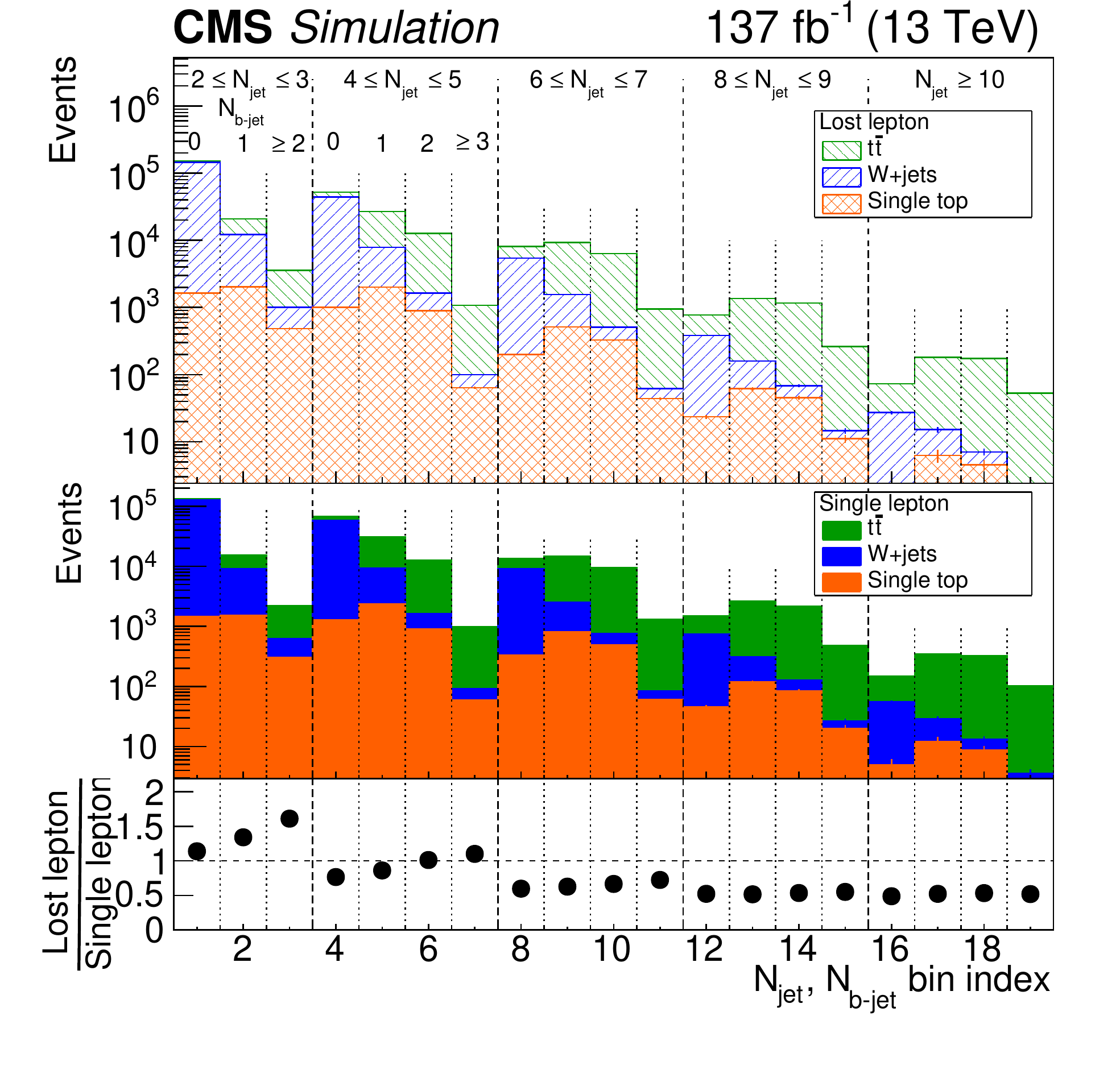}
  \caption{
  (upper) The number of lost-lepton events in simulation,
  integrated over \HT and \mht,
  as a function of \njets and \nbjets.
  (middle) Corresponding results from simulation for the number of events
  in the single-lepton control region.
  (lower) The ratio of the simulated lost-lepton to the single-lepton results,
  with statistical uncertainties (too small to be visible).
  These ratios are equivalent to the transfer factors used in the
  evaluation of the lost-lepton background,
  except integrated over \HT and \mht.
  }
  \label{fig:TFVariation}
\end{figure*}

The lost-lepton background is evaluated by applying an MC-derived multiplicative
transfer factor to the observed single-lepton CR yields,
with a separate transfer factor determined for each of the 174 search bins.
The transfer factor is defined by the ratio, in simulation,
of the number of lost-lepton events in a search bin
to the number of events in the corresponding bin of the single-lepton CR,
following normalization to the same integrated luminosity.
The simulated events are corrected to account for
differences with respect to data
in the lepton, isolated track, and \PQb jet tagging efficiencies.

The upper panel of Fig.~\ref{fig:TFVariation} shows the simulated results,
as a function of \njets and \nbjets,
for the number of lost-lepton events.
The corresponding results from simulation for the number of events in
the single-lepton CR are shown in the middle panel of Fig.~\ref{fig:TFVariation}.
The ratio of the results in the upper to the middle panels,
equivalent to the transfer factor integrated over \HT and \mht,
is shown in the lower panel.
At lower values of \njets,
the distributions are enhanced in \wjets events,
for which a larger fraction of leptons lie outside
the kinematic acceptance of the analysis compared to \ttbar events.
This reduces the event acceptance in the single-lepton CR,
increasing the value of the integrated transfer factors above unity
as seen in the lower panel of Fig.~\ref{fig:TFVariation}
for $2\leq\njets\leq3$.
As \nbjets increases,
the probability for a lepton to fail the isolation requirement increases,
leading to a larger rate of lost-lepton events and to an increase in the
integrated transfer factors.
This latter effect is especially visible for $2\leq\njets\leq3$
in Fig.~\ref{fig:TFVariation} (lower).

The dominant uncertainty in the lost-lepton background prediction is statistical,
arising from the limited number of events in the CR.
Other uncertainties are evaluated to account for the lepton
and {\PQb} jet tagging scale factors,
the \mT selection requirement,
the PDFs,
the renormalization and factorization scales~\cite{Kalogeropoulos:2018cke},
and the jet energy corrections.
These uncertainties are summed in quadrature
to obtain the total uncertainty in the lost-lepton background prediction.

\subsection{Background from \texorpdfstring{\znnjets}{Z(->nu nu)+jets} events}
\label{sec:zinv}

The background from \zjets events with \znn decay
is evaluated using a CR with a single photon (\gjets CR),
in conjunction with a \zlljets CR in which the \cPZ boson
decays to an $\EE$ or $\MM$ pair.
The method relies on the kinematic similarity between
the production of \cPZ bosons and photons.
The \znnjets background in search bins with
$\nbjets=0$ is determined by applying multiplicative
transfer factors from simulation to the observed rate of \gjets events,
analogous to the method described in Section~\ref{sec:lost_lepton}
for the evaluation of the lost-lepton background.
A correction is made to the normalization based on
the observed rate of \zlljets CR events.
An extrapolation to the search bins with $\nbjets\geq1$
is then made based on factors constructed from the \zlljets data.
We follow this procedure in order to take advantage of both the
higher statistical precision of the \gjets CR
and the more direct transfer factors of the \zll CR,
while preserving the \nbjets--\njets correlation observed in the latter.

\subsubsection{The \texorpdfstring{\gjets}{g+jets} events}
\label{sec:gamma_jets}

Events in the \gjets CR were collected using a single-photon trigger,
with an online threshold that varied between 180 and 190\GeV,
depending on the data collection period.
In the offline analysis,
events are required to contain exactly one photon with $\pt>200\GeV$.
In each CR event, the photon serves as a proxy for a \PZ boson and
is removed to emulate the undetected \cPZ boson in \znn decays.
To ensure that the kinematics of the \gjets events match those expected
for \znnjets events,
jets are reclustered after removing the photon
and all event-level variables are recomputed.
The same event selection criteria used to select signal events are then applied except,
in addition,
we require $\nbjets=0$.

The \gjets CR contains nonnegligible contributions from photons
produced in neutral meson decays.
These photons are referred to as ``nonprompt.''
The contamination of the CR from nonprompt photons,
and thus the purity of signal photons in the sample,
is evaluated using a binned maximum likelihood fit to the distribution of
the photon candidate's charged hadron isolation variable.
The fit is based on templates for nonprompt and signal photons.
For signal photons,
the template is taken from simulation,
using the nominal photon selection criteria.
For nonprompt photons,
three different versions of the template are made:
i)~from simulation using the nominal criteria;
ii)~from simulation in a high-\sieie sideband
(defined by inverting the \sieie selection criterion),
where nonprompt photon production is expected to dominate;
and iii)~from data in this same high-\sieie sideband.
The arithmetic mean of the three nonprompt templates is used in the fit,
with the variation in the results obtained using the three templates individually
defining  a systematic uncertainty.
The purity is determined as a function of \mht and typically exceeds 90\%.

In the generation of simulated \gjets events,
photons that are approximately collinear with a parton ($\dR<0.4$)
are removed to improve the fraction of events with well-isolated photons
and thus the statistical precision of the sample.
A correction denoted \Fdir
is evaluated to account for a bias from this requirement,
using simulated events with a looser restriction
on the angular separation between the generator-level photons and partons.
The corrections are typically less than 10\%.
A systematic uncertainty in the correction
given by $0.30 (1 - \Fdir)$
is determined by evaluating
the level of agreement between simulation and data in the distribution
of the angular separation between a photon and the nearest jet,
and the effect of changing the definition of collinear photons in the simulation.

\subsubsection{The \zlljets events}
\label{sec:zll_jets}

The \zlljets CR,
collected using single-lepton triggers,
is selected by requiring two oppositely charged electrons or muons
with a dilepton invariant mass \mll within 15\GeV of the {\PZ} boson mass.
The selection requirements for electrons and muons are the same as those described
in Section~\ref{sec:reconstruction},
including the isolation requirements.
To suppress \ttbar events,
the \pt of the dilepton system is required to exceed 200\GeV.
Similar to the \gjets CR,
the lepton pair in each \zlljets event is removed to emulate
the undetected {\PZ} boson in \znnjets events,
following which jets are reclustered and
the event-level quantities recalculated.

Top quark pair production typically constitutes $<$5\%
of the observed dilepton event yield,
except for events with $\nbjets\geq 2$
where it can comprise up to $\approx$15\% of the sample.
Using fits to the observed \mll distribution,
the purity \Zpurity of the \zlljets sample
is evaluated for each individual \njets and \nbjets region.

\subsubsection{The \znnjets background prediction}
\label{sec:zjets-prediction}

\begin{figure}[!htb]
\begin{center}
\includegraphics[width=0.95\textwidth]{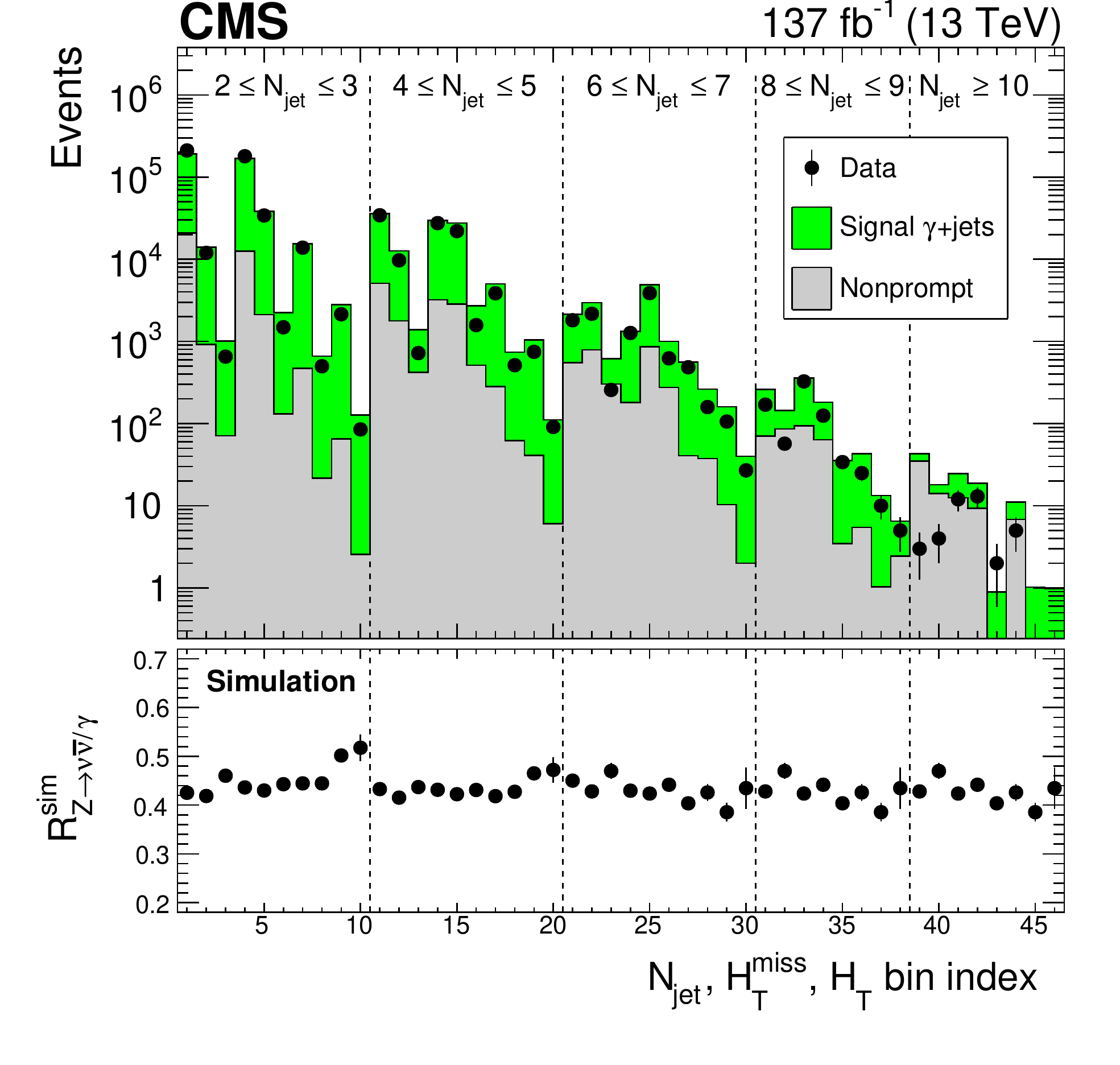}
\caption{
  (upper) The number of events in the \gjets control region for data and simulation.
  (lower) The transfer factors \rznnsim from simulation.
  The respective results are shown for the 46 search bins with $\nbjets=0$.
  The 10 results (8 for $\njets\geq8$) within each region delineated
  by vertical dashed lines correspond sequentially to the 10
  (8) kinematic intervals in \HT and \mht
  listed in Table~\ref{tab:kine-bins} and Fig.~\ref{fig:HT-MHT}.
  The uncertainties are statistical only.
  For the upper plot, the simulated results show the stacked event rates for
  the \gjets and nonprompt MC event samples,
  where ``nonprompt'' refers to SM MC events other than \gjets.
  The simulated nonprompt results are dominated by events from the QCD sample.
  Because of limited statistical precision in the simulated event samples
  at large \njets,
  the transfer factors determined for
  the $8\leq\njets\leq 9$ region are also used
  for the $\njets>10$ region.
}
\label{fig:Ngammayields-RzG}
\end{center}
\end{figure}

For each of the 46 search bins with $\nbjets=0$,
the \znnjets background is evaluated according to:
\begin{linenomath}
\begin{equation}
  \left.\nznn\right|_{\nbjets=0}
     = \avgDblR {\rznnsim} \Fdir \beta_\gamma \ngdata \,/\, \SFdmcg,
\label{eq:gjet1}
\end{equation}
\end{linenomath}
where \ngdata is the number of observed events in the \gjets CR,
\rznnsim is the transfer factor,
\SFdmcg accounts for the trigger efficiency and
for differences between data and simulation in the
photon reconstruction efficiency~\cite{Khachatryan:2015iwa},
and $\beta_\gamma$ is the photon purity.
The transfer factors,
which account for known differences between photon and \PZ boson production,
are given by the ratio from simulation of the rates of
\znnjets events to \gjets events in the 46 bins.
For the photon selection criteria used in this analysis,
the transfer factor has a value of around 0.5,
with a relatively mild dependence on the signal region kinematics.
The distribution of \ngdata,
along with the simulated results for signal and nonprompt \gjets events,
is shown in Fig.~\ref{fig:Ngammayields-RzG} (upper).
Figure~\ref{fig:Ngammayields-RzG} (lower)
shows the transfer factors \rznnsim.

The term denoted \avgDblR in Eq.~(\ref{eq:gjet1})
accounts for possible residual mismodeling of \rznnsim.
The value of \avgDblR is expected to be close to unity,
with possible deviations due to differences in missing higher-order terms
between the \gjets and \zjets simulation.
It is the average over all search bins with $\nbjets=0$ of the double ratio
\begin{linenomath}
\begin{equation}
\DblR = \frac{\mathcal{R}_{\zll/\gamma}^{\text{data}}}{\mathcal{R}_{\zll/\gamma}^{\text{sim}}}
 = \frac{\nlldata}{\nllsim}
   \frac{\ngsim}{\ngdata}
\frac{\Zpurity}{\SFdmcll} \frac{\SFdmcg}{\Fdir\beta_\gamma},
\label{eq:doubleratio}
\end{equation}
\end{linenomath}
where \nlldata and \nllsim represent the number of events in
the observed and simulated \zll CR,
respectively,
\ngsim is the number of events in the simulated \gjets CR,
and \SFdmcll accounts for the trigger efficiency and
for differences between data and simulation in the lepton reconstruction
efficiencies in \zlljets events~\cite{Khachatryan:2015hwa,Sirunyan:2018fpa}.
The event yields in the \zlljets CR are too small to allow a meaningful
determination of \DblR in all search bins
and thus we calculate the average \avgDblR and apply it to all bins.

From studies of the variation of \DblR with \HT, \mht, and \njets,
we observe a mild trend in \DblR with respect to \HT.
This trend is parameterized as
$\DblR(\HT) = 0.86+(2.0\times10^{-4})\text{min}(\HT,900\GeV)$.
Using this parameterization,
an event-by-event weight is applied to each simulated \gjets CR event
before it enters Eq.~(\ref{eq:gjet1}).
Prior to this event weighting,
we find $\avgDblR=0.95$.
Following the event weighting, $\avgDblR=1.00$.
It is this latter value of \avgDblR,
along with its uncertainty,
that enters Eq.~(\ref{eq:gjet1}).
In each bin,
the residual deviation of \DblR from unity as a function of \HT
is added in quadrature with the associated statistical uncertainty,
and analogously but separately for \mht and \njets,
and the largest of the three resulting terms is taken
as the corresponding systematic uncertainty
in the background prediction.
The values of this bin-dependent uncertainty range from 1 to 13\%.

\begin{figure}[!htb]
\begin{center}
\includegraphics[width=0.95\textwidth]{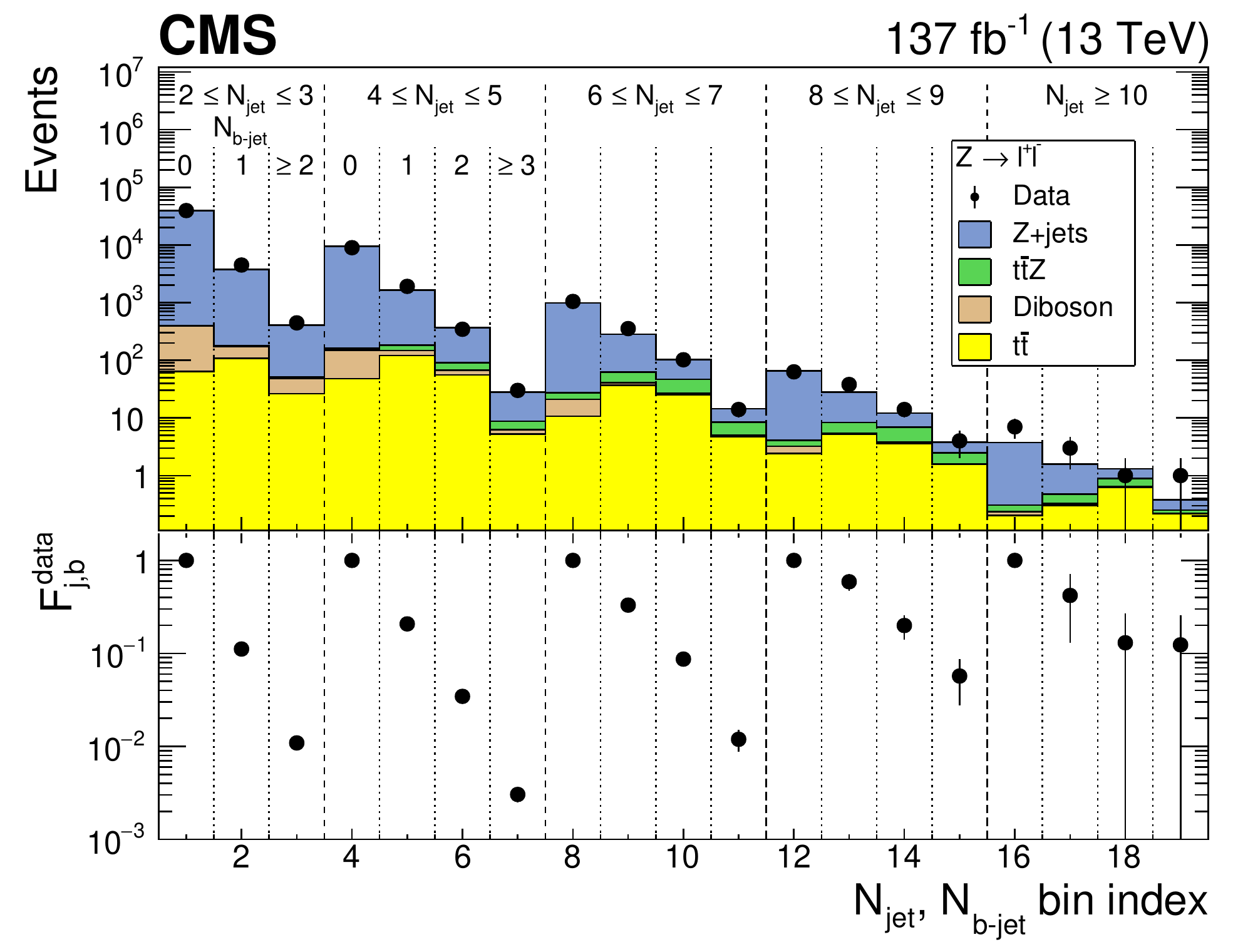}
\caption{
  (upper) The observed event yield in the \zlljets control region,
  integrated over \HT and \mht,
  as a function of \njets and \nbjets.
  The uncertainties are statistical only.
  The stacked histograms show the corresponding results from simulation.
  (lower) The extrapolation factors \fextrap
  with their statistical uncertainties.
}
\label{fig:Zllyields-Fvsjb}
\end{center}
\end{figure}

To evaluate the \znnjets background for search bins with $\nbjets\geq 1$,
we assume that the relative population of \znnjets events
in the \HT--\mht plane is independent of \nbjets for fixed \njets.
A systematic uncertainty deduced from a closure test (described below)
is assigned to account for this assumption,
where ``closure test'' refers to a check of the ability of the method,
applied to simulated event samples,
to correctly predict the genuine number of background events in simulation.
We extend the result from
Eq.~(\ref{eq:gjet1}) using extrapolation factors \fextrap from \zlljets data,
as follows:
\begin{linenomath}
\begin{equation}
\left(N^{\text{pred}}_{\znn}\right)_{j,b,k}
= \left(N^{\text{pred}}_{\znn}\right)_{j,0,k}\fextrap
\equiv \left(N^{\text{pred}}_{\znn}\right)_{j,0,k}
  \frac{ \left(\nlldata\Zpurity\right)_{j,b}}
  {\left(\nlldata\Zpurity\right)_{j,0}},
\label{eq:nbextrp}
\end{equation}
\end{linenomath}
where $j$ corresponds to the five \njets regions,
$b$ to the four \nbjets regions (three for $\njets\leq 3$),
and $k$ to the 10 kinematic regions of \HT and \mht (Table~\ref{tab:kine-bins}).
The data used and the resulting extrapolation factors \fextrap
are shown in Fig.~\ref{fig:Zllyields-Fvsjb}.

The rare process $\ttbar\cPZ$ and the even more rare processes $\PZ\PZ$, $\PW\PW\PZ$,
$\PW\PZ\PZ$, and $\PZ\PZ\PZ$ can also contribute to the background.
Those processes with a counterpart when the \PZ boson is replaced with a photon
are already accounted for in \ngdata
and thus are automatically included in the background estimate.
We assume that the ratio of the rate of the
rare process to its counterpart with a photon,
\eg, the ratio of $\ttbar\PZ$ (with $\PZ\to\cPgn\cPagn)$
to $\ttbar\gamma$ events,
equals \rznnsim.

\begin{figure}[htb]
\begin{center}
\includegraphics[width=0.95\textwidth]{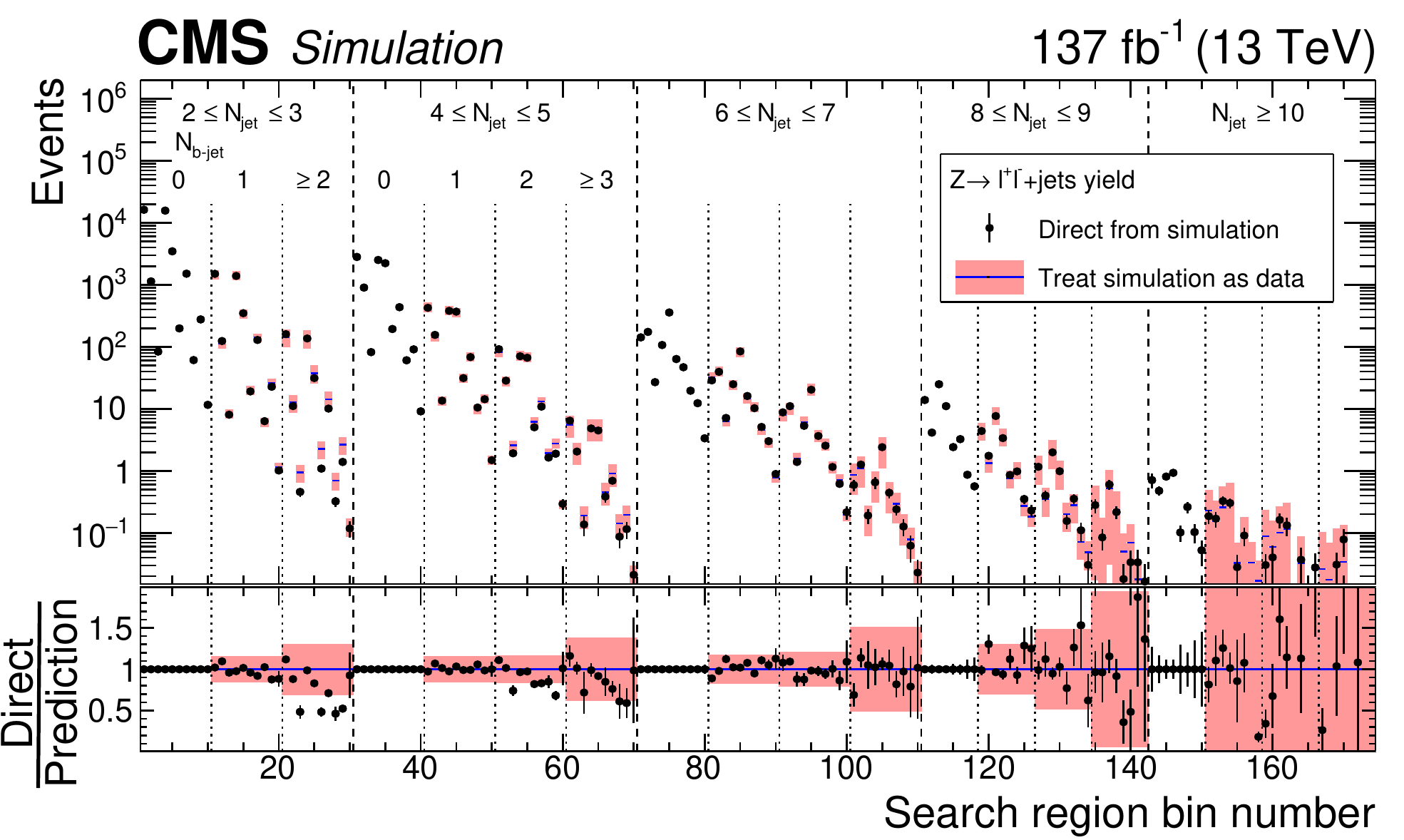} \\
\caption{
  Prediction from simulation for the \zlljets event yields
  in the 174 search bins
  as determined by computing the \fextrap factors (Eq.~(\ref{eq:nbextrp}))
  and the $\nbjets=0$ event yields in the same manner as for data,
  in comparison to the corresponding direct \zlljets prediction from simulation.
  The 10 results (8 for $\njets\geq8$) within each region delineated
  by vertical dashed lines correspond sequentially to the 10
  (8) kinematic intervals in \HT and \mht
  listed in Table~\ref{tab:kine-bins} and Fig.~\ref{fig:HT-MHT}.
  For bins with $\njets\geq10$,
  some points do not appear in the upper panel because they lie
  below the minimum of the displayed range.
  In the case that the direct expected yield is zero,
  there is no result in the lower, ratio panel.
  The pink bands show the statistical uncertainties in the prediction,
  scaled to correspond to the integrated luminosity of the data,
  combined with the systematic uncertainty attributable
  to the kinematic (\HT and \mht) dependence.
  The black error bars show the statistical uncertainties
  in the simulation.
  For bins corresponding to $\nbjets = 0$,
  the agreement is exact by construction.
}
\label{fig:extrapMCtest}
\end{center}
\end{figure}

A closure test of the procedure is performed
by treating event yields from the \zlljets simulation as data,
as shown in Fig.~\ref{fig:extrapMCtest}.
Based on this study,
the following systematic uncertainties are assigned.
For $\njets=2$--3,
a systematic uncertainty of 15 and 30\% is assigned to the
$\nbjets=1$ and $\geq$2 regions, respectively.
For $\njets\geq4$,
a systematic uncertainty of 15 and 30\% is assigned to the
$\nbjets=1$--2 and $\geq$3 regions.
These uncertainties account for correlations between \nbjets and
the \HT and/or \mht variables
in the shape of the \znn prediction.

\subsection{Background from QCD events}
\label{sec:qcd}

The QCD background comprises only a small fraction ($<$5\%)
of the total background but,
because it typically arises from the mismeasurement of jet \pt,
is difficult to evaluate with simulation.
We use data to model this background,
exploiting knowledge of the jet energy resolution.
Briefly, the method employs a set of CR events collected using triggers
requiring \HT to exceed various thresholds between 200 and 1050\GeV,
with no condition on~\mht.
Corresponding prescale factors ranging from around 10\,000 to 1 are applied,
where a prescale factor
reduces the recorded event rate relative to the raw trigger rate
in order to maintain a manageable data flow.
The jet momenta in each CR event are adjusted so that the
event has well-balanced jet~\pt,
consistent with the kinematics of a generator-level
(\ie, ideally measured) QCD event.
This step is called rebalancing.
The rebalancing step removes the intrinsic \ptmiss from the event,
thus effectively eliminating the contributions of events like \wjets
and \zjets events that can have genuine \ptmiss~\cite{Collaboration:2011ida}.
The jet momenta are then smeared according to the known detector
jet \pt resolution in order to determine the probability that a given event
will populate a given search bin.
This latter step is the smear stage.
The so-called rebalance and smear (R\&S) method
was introduced in Refs.~\cite{Collaboration:2011ida,Chatrchyan:2014lfa}
and was further developed in Ref.~\cite{Khachatryan:2016kdk}.

To rebalance an event,
a Bayesian inference procedure is used,
in which the \pt of each jet in a CR event
is varied within its uncertainty to maximize the probability:
\begin{linenomath}
\begin{equation}
\mathcal{P}( \vec{\mathcal{J}}_{\text{true}} | \vec{\mathcal{J}}_{\text{meas}})\sim
\mathcal{P}( \vec{\mathcal{J}}_{\text{meas}} | \vec{\mathcal{J}}_{\text{true}})
   \,\pi(\htmisstrue,\dphibtrue),
\label{eq:rs_posterior}
\end{equation}
\end{linenomath}
where $\mathcal{P}(\vec{\mathcal{J}}_{\text{true}} | \vec{\mathcal{J}}_{\text{meas}})$
is the posterior probability density for a given configuration of jets
with true (or ideal)
momentum assignments $\vec{\mathcal{J}}_{\text{true}}$,
given a configuration of measured jet momenta $\vec{\mathcal{J}}_{\text{meas}}$.
The $\mathcal{P}( \vec{\mathcal{J}}_{\text{meas}} | \vec{\mathcal{J}}_{\text{true}})$ term,
taken to be the product of the individual jet response functions
of all jets in an event,
is the likelihood to observe a configuration of measured jet momenta
given a configuration of jets with a particular set of true momenta.
The jet response functions are constructed from the distributions in simulation
of the ratio of the reconstructed jet \pt to the \pt of well-matched
generator-level jets.
The response functions are derived as a function of
jet \pt and $\eta$ and are corrected to account for differences in the
jet response shape between data and simulation.
The $\pi(\htmisstrue,\dphibtrue)$ term is the prior distribution,
determined as a function of the true (i.e., generator level) \htvecmiss and \dphib,
where \dphib is the azimuthal angle
between \htvecmiss and either the highest \pt jet in the event (for $\nbjets=0$),
or the highest \pt tagged \PQb jet (for $\nbjets\geq 1$).
This prior represents the distribution of the magnitude and direction of the
genuine \mht expected in QCD events.

After the transverse momenta of the individual jets
have been adjusted according to the posterior
probability density in Eq.~(\ref{eq:rs_posterior}),
the jet \pt values are smeared by rescaling them using factors sampled randomly
from the jet response functions.
This sampling is performed numerous times for each rebalanced event.
Each event is then weighted by the inverse of
the number of times it is smeared.
Events are smeared in up to 1000 independent trials,
with a final target event weight of~0.05,
equal to the prescale value of the trigger that collected the seed event
divided by the number of times the event was reused in the smearing step.

The R\&S procedure produces a sample of events
that closely resembles the original sample of CR events,
except with the contributions of the
electroweak backgrounds effectively removed.
The resulting events are subjected to the signal event selection criteria
of Section~\ref{sec:event-selection}
to obtain the QCD background prediction in each search bin.
The overall normalization is adjusted based on a scaling factor derived
from a QCD-dominated CR selected by inverting the \dphi selection criteria
and requiring $\nbjets=0$ and $250<\mht<300\GeV$.
The \dphi selection criteria are inverted by requiring
at least one of the two (for $\njets = 2$),
three (for $\njets = 3$), or four (for $\njets\geq 4$)
highest \pt jets in an event to fail at least one of the corresponding \dphimht
requirements given in Section~\ref{sec:event-selection}.
The normalization scale factors typically have values around 1.4.

\begin{figure}[tp]
\centering
\includegraphics[width=0.45\textwidth]{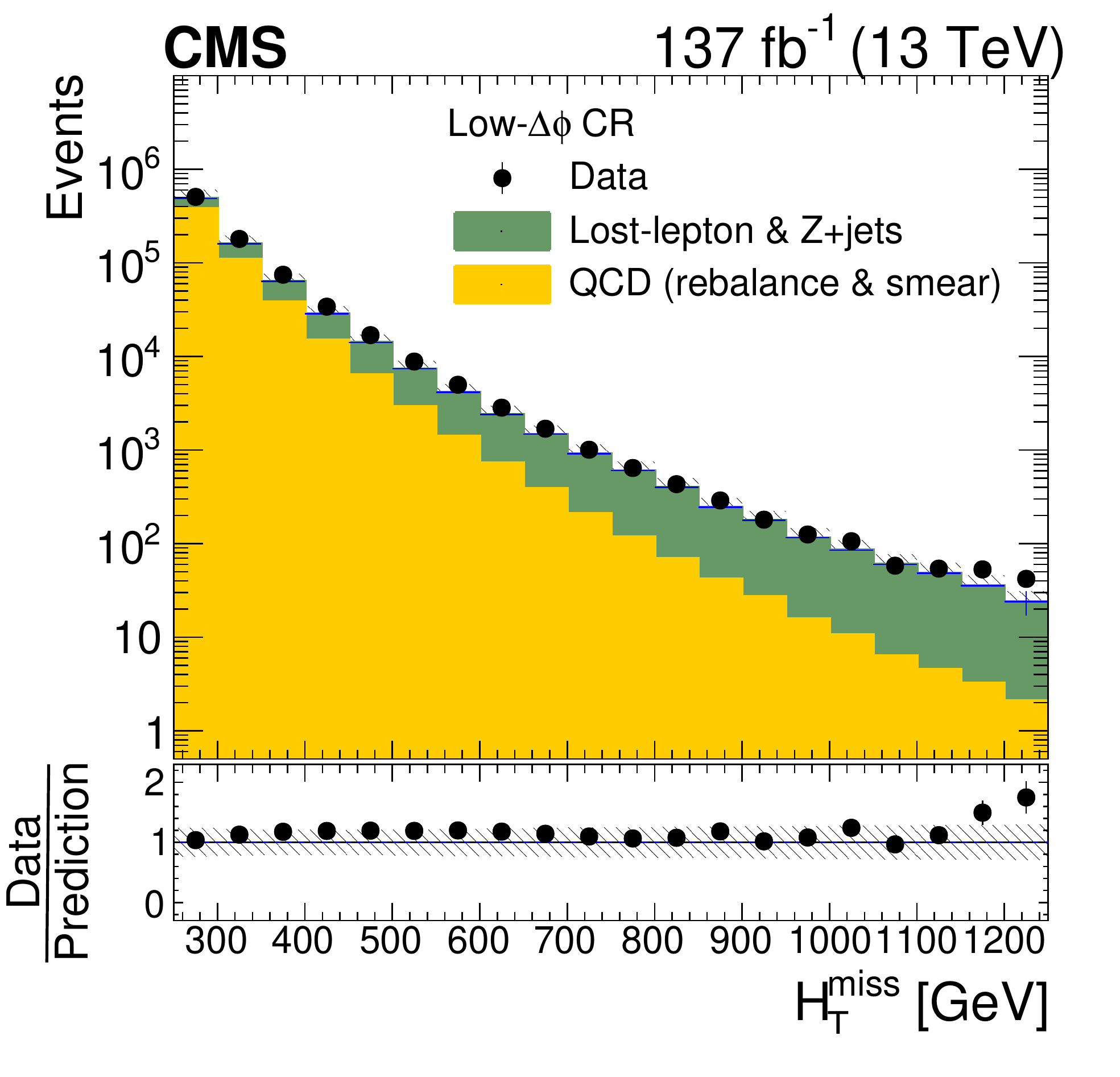}
\includegraphics[width=0.45\textwidth]{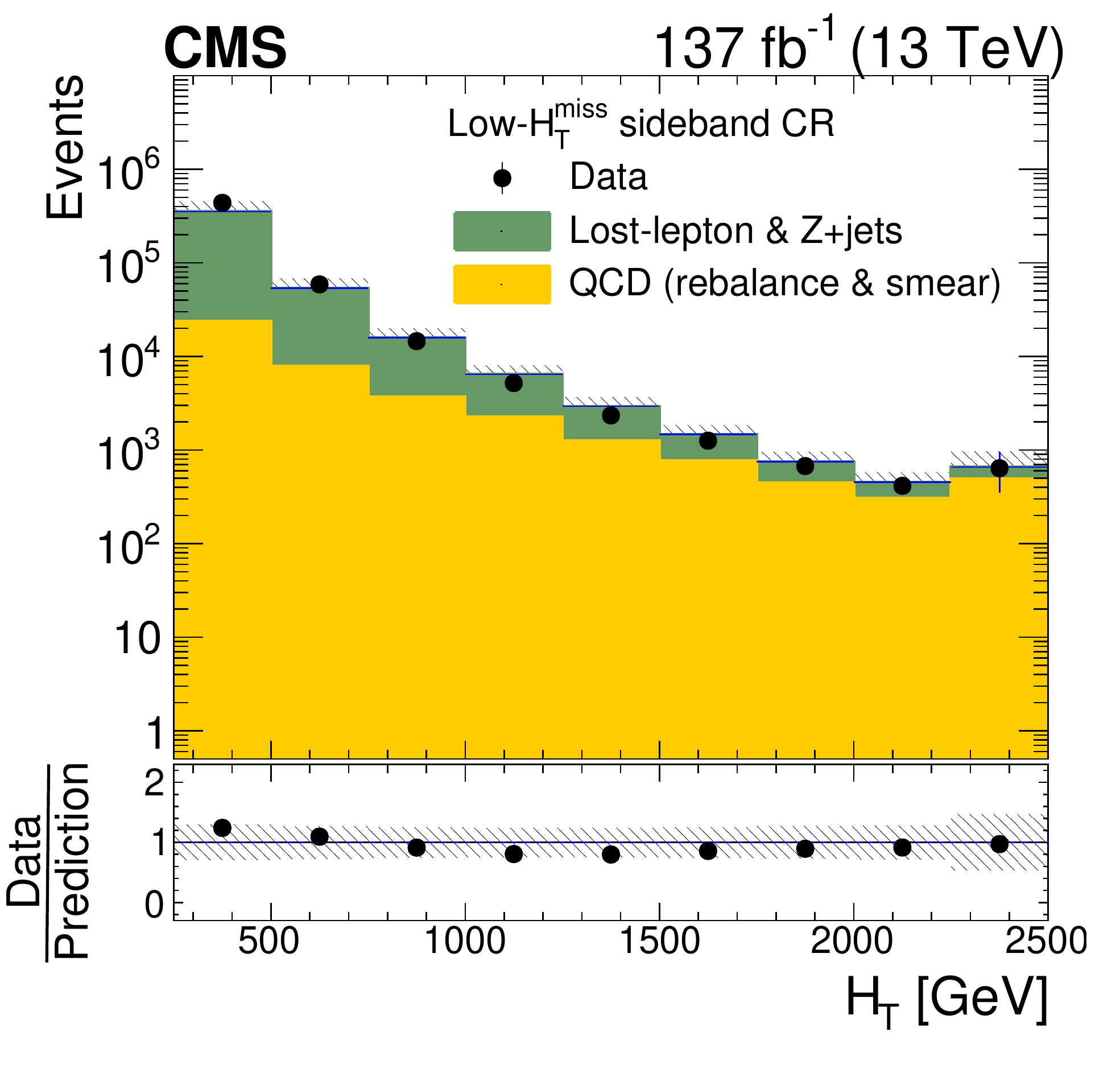}
\caption{
The observed and predicted distributions of (left) \mht
in the inverted-\dphi control region
and (right) \HT in the low-\mht sideband.
The uncertainties are statistical only.
The lower panels show the ratios of the observed to the predicted distributions,
with their statistical uncertainties.
The hatched regions indicate the total uncertainties in the predictions,
with statistical and systematic uncertainties combined in quadrature.
}
\label{fig:qcd_validation}
\end{figure}

Comparisons between the predicted QCD background yields and observations
are examined as a function of \mht, \HT, \njets, and \nbjets,
both in a CR defined by inverting the \dphi requirements
and in a low-\mht sideband defined by $250<\mht<300\GeV$.
As examples,
Fig.~\ref{fig:qcd_validation} shows the distribution of \mht
in the inverted-\dphi CR
and the distribution of \HT in the low-\mht sideband.

\begin{figure}[tbh]
\centering
\includegraphics[width=0.95\textwidth]{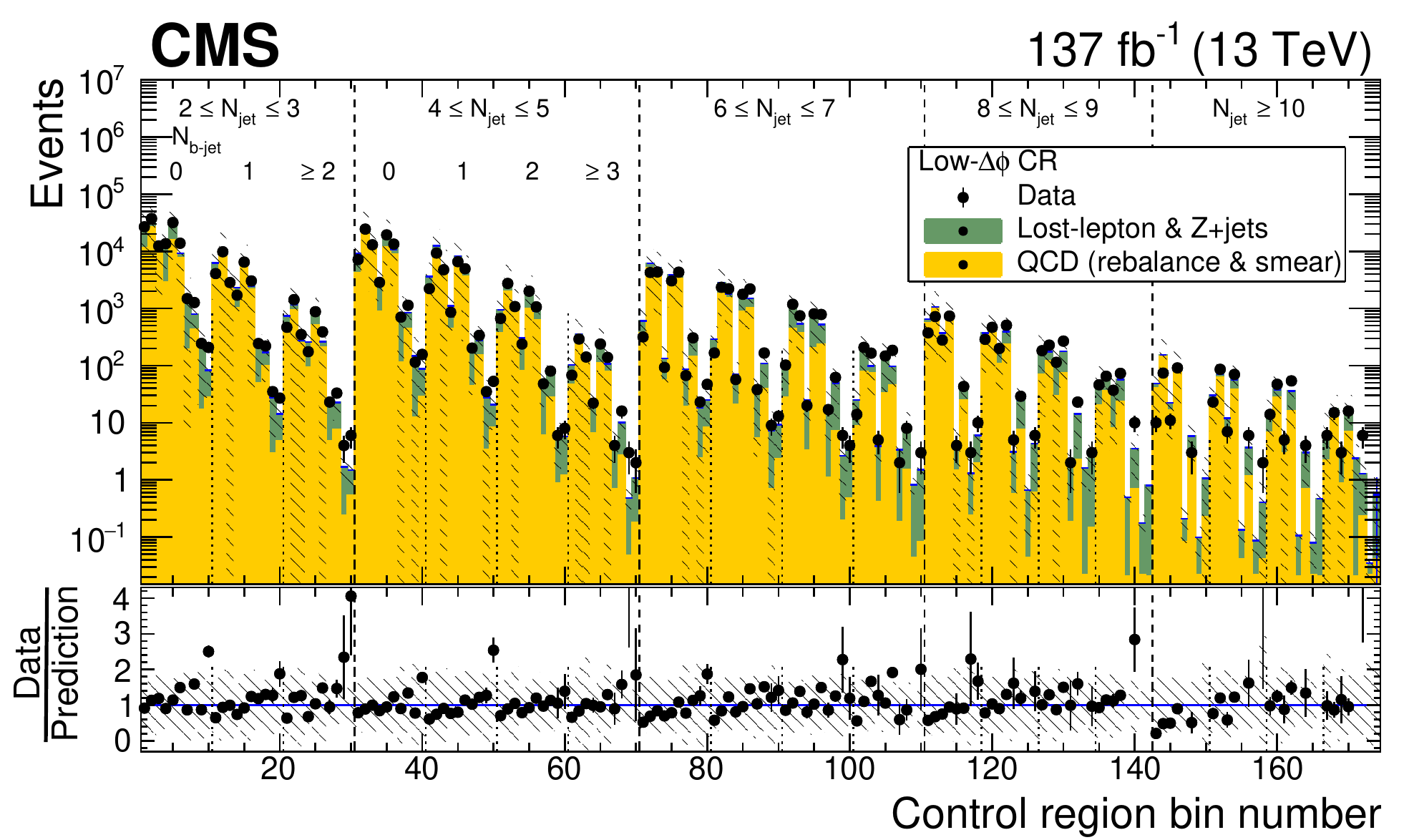}
\caption{
  Distribution of observed and predicted event yields in the
  inverted-\dphi control region analysis bins.
  The uncertainties are statistical only.
  The labeling of the bin numbers is the same as in
  Fig.~\ref{fig:extrapMCtest}.
  The lower panel shows the ratio of the observed to the predicted event yields,
  with their statistical uncertainties.
  The hatched region indicates the total uncertainty in the prediction,
  with statistical and systematic uncertainties combined in quadrature.
}
\label{fig:qcd_validation2}
\end{figure}

Figure~\ref{fig:qcd_validation2} shows the observed and predicted
event yields in 174 analysis control bins defined using the same criteria
as for the search bins except with the inverted-\dphi requirement.
For all these validation tests,
contributions from QCD events are evaluated using the R\&S method,
contributions from top quarks and \wjets events are evaluated
using the lost-lepton method described in Section~\ref{sec:lost_lepton},
and contributions from \znnjets events are taken from simulation.

The principal uncertainty in the R\&S QCD background prediction is systematic,
associated with the uncertainty in the shape of the jet response functions.
This uncertainty is evaluated by varying the jet energy resolution scale factors
within their uncertainties,
resulting in uncertainties in the prediction that range from 30--70\%,
depending on the search bin.
Smaller uncertainties related to the trigger and the
finite size of the seed sample are evaluated,
as well as a nonclosure uncertainty that accounts for inaccuracies
identified from simulation-based studies.

\section{Signal systematic uncertainties}
\label{sec:systematic}

\begin{table*}[tb]
\topcaption{
Systematic uncertainties in the yield of signal events,
averaged over all search bins.
The variations correspond to different signal
models and choices for the SUSY particle masses.
Results reported as 0.0 correspond to values less than~0.05\%.
}
\centering
\begin{tabular}{lc}
Item & Relative uncertainty (\%) \\
\hline
Renormalization and factorization scales \rscale \& \fscale        & 0.0--5.7 \\
Initial-state radiation                         & 0.0--14 \\
Jet energy scale                                & 0.0--14 \\
Jet energy resolution                           & 0.0--10 \\
Pileup modeling                                 & 0.0--2.4 \\
Isolated-lepton \& isolated-track vetoes        & 2.0 \\
\hspace*{5mm} (T1tttt, T5qqqqVV, and T2tt models) & \\
Integrated luminosity                           & 2.3--2.5 \\
Trigger efficiency (statistical)                & 0.2--2.6 \\
Trigger efficiency (systematic)                 & 2.0 \\
Statistical uncertainty in simulated samples    & 1.2--31 \\
\HT and \mht modeling                           & 0.0--11 \\
Jet quality requirements                        & 1.0 \\[\cmsTabSkip]
Total                                           & 4.0--33 \\
\end{tabular}
\label{tab:sig-syst}
\end{table*}

Systematic uncertainties in the
signal event yield are listed in Table~\ref{tab:sig-syst}.
To evaluate the uncertainty associated with the
renormalization (\rscale) and factorization (\fscale) scales,
each scale is varied independently
by a factor of 2.0 and 0.5~\cite{Kalogeropoulos:2018cke,Catani:2003zt,Cacciari:2003fi}.
The uncertainties associated with \rscale, \fscale, and ISR,
integrated over all search bins, typically lie below 0.1\%.
Nonetheless,
they can be as large as the maximum values noted in Table~\ref{tab:sig-syst}
if $\dmass\approx 0$,
where \dmass is the difference between the gluino or squark mass
and the sum of the masses of the particles into which
the gluino or squark decays.
For example,
for the T1tttt model,
$\dmass=\mgluino - (\mlsp+2\mtop)$,
with \mtop the top quark mass.
The uncertainties associated with the jet energy scale and jet energy resolution
are evaluated as functions of jet \pt and~$\eta$.
To evaluate the uncertainty associated with the pileup reweighting,
the value of the total inelastic cross section is varied by 5\%~\cite{Sirunyan:2018nqx}.
The isolated-lepton and isolated-track vetoes have a
minimal impact on the T1bbbb, T1qqqq, T2bb, and T2qq models because events in
these models rarely contain an isolated lepton.
Thus, the associated uncertainty is negligible ($\lesssim$0.1\%).
The systematic uncertainty in the determination of the
integrated luminosity varies between
2.3 and 2.5\%~\cite{CMS-PAS-LUM-17-001,CMS-PAS-LUM-17-004,CMS-PAS-LUM-18-002},
depending on the year of data collection.

Systematic uncertainties in the signal predictions
associated with the {\cPqb} jet tagging and misidentification efficiencies
are also evaluated.
These uncertainties do not affect the signal yield
but can potentially alter the shape of signal distributions.
The systematic uncertainties associated with
\rscale, \fscale, ISR, jet energy scale, jet energy resolution, the trigger,
statistical precision in the event samples, and \mht modeling
can also affect the shapes of the signal distributions.
We account for these potential changes in shape,
\ie, migration of events among search bins,
in the limit setting procedure
described in Section~\ref{sec:results}.

\section{Results}
\label{sec:results}

\begin{figure}[tbh]
\centering
\includegraphics[width=0.95\textwidth]{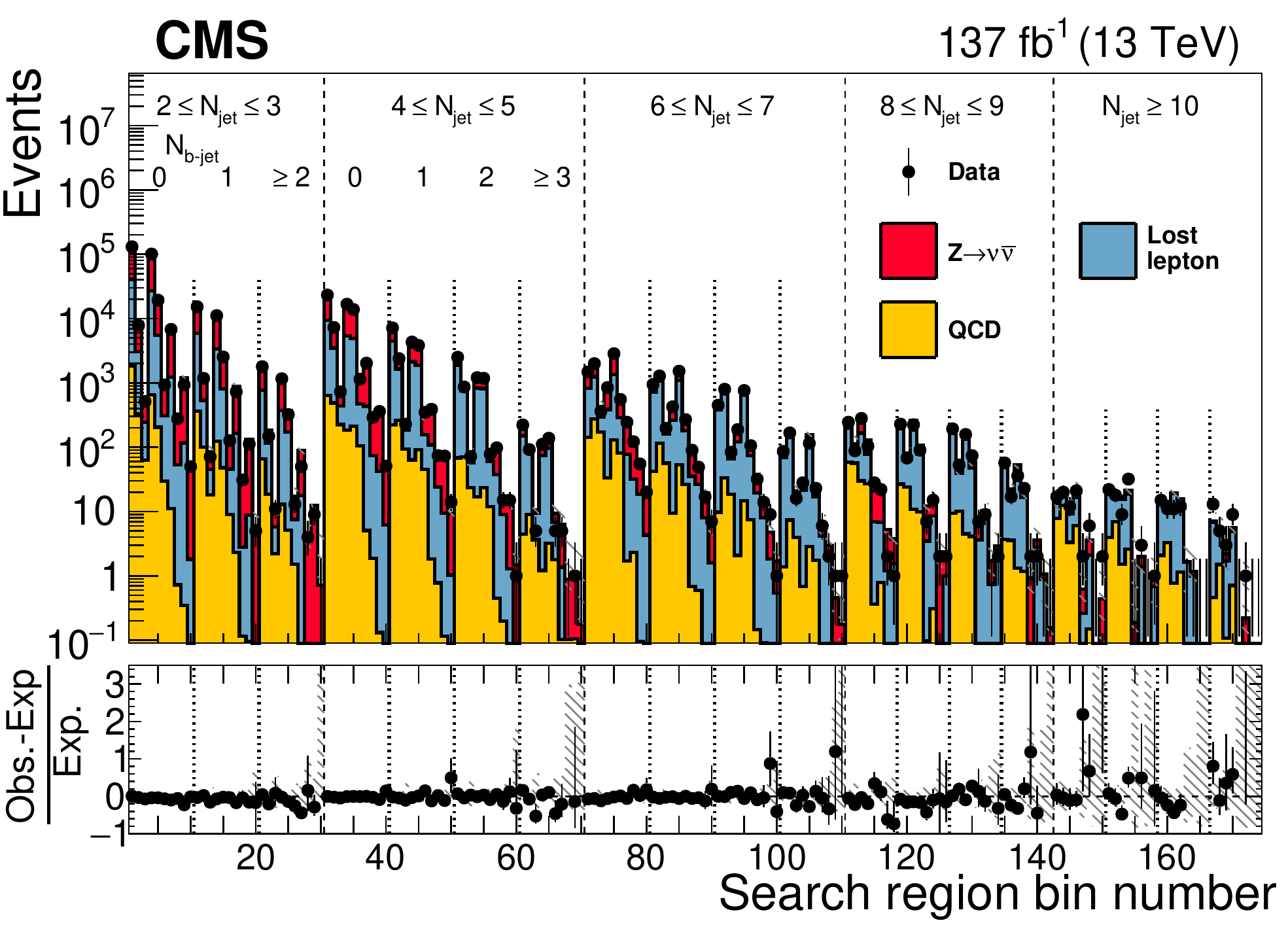}
\caption{
  The observed numbers of events and pre-fit SM background predictions
  in the 174 search bins of the analysis,
  where ``pre-fit'' means there is no constraint from the likelihood fit.
  The labeling of the bin numbers is the same as in
  Fig.~\ref{fig:extrapMCtest}.
  Numerical values are given in \cmsAppendix\ref{sec:prefit}.
  The hatching indicates the total uncertainty in the background
  predictions.
  The lower panel displays the fractional differences
  between the data and SM predictions.
}
\label{fig:fit-results}
\end{figure}

Figure~\ref{fig:fit-results} presents
the observed numbers of events in the 174 search bins.
The data are shown in comparison to the stacked
pre-fit predictions for the SM backgrounds,
where ``pre-fit'' refers to the predictions determined as described
in Section~\ref{sec:background},
before constraints from the fitting procedure have been applied.
Numerical values are given in \cmsAppendix\ref{sec:prefit}.
The uncertainties in the background predictions are mainly from
systematic uncertainties in the transfer factors,
statistical uncertainties in control sample yields,
and systematic uncertainties in the modeling of the search variables.
\cmsAppendix\ref{sec:prefit} lists the overall statistical
and systematic uncertainties for the individual background components
and for their sum.
In addition to the finely segmented search bins of Fig.~\ref{fig:fit-results},
we determine the results for 12 aggregate search bins,
each representing a potentially interesting signal topology.
These results are presented in \cmsAppendix\ref{app:aggbins}.

The observed event counts are consistent with the predicted backgrounds.
Thus we do not obtain evidence for supersymmetry.

\begin{figure*}[htb]
\centering
\includegraphics[width=0.45\textwidth]{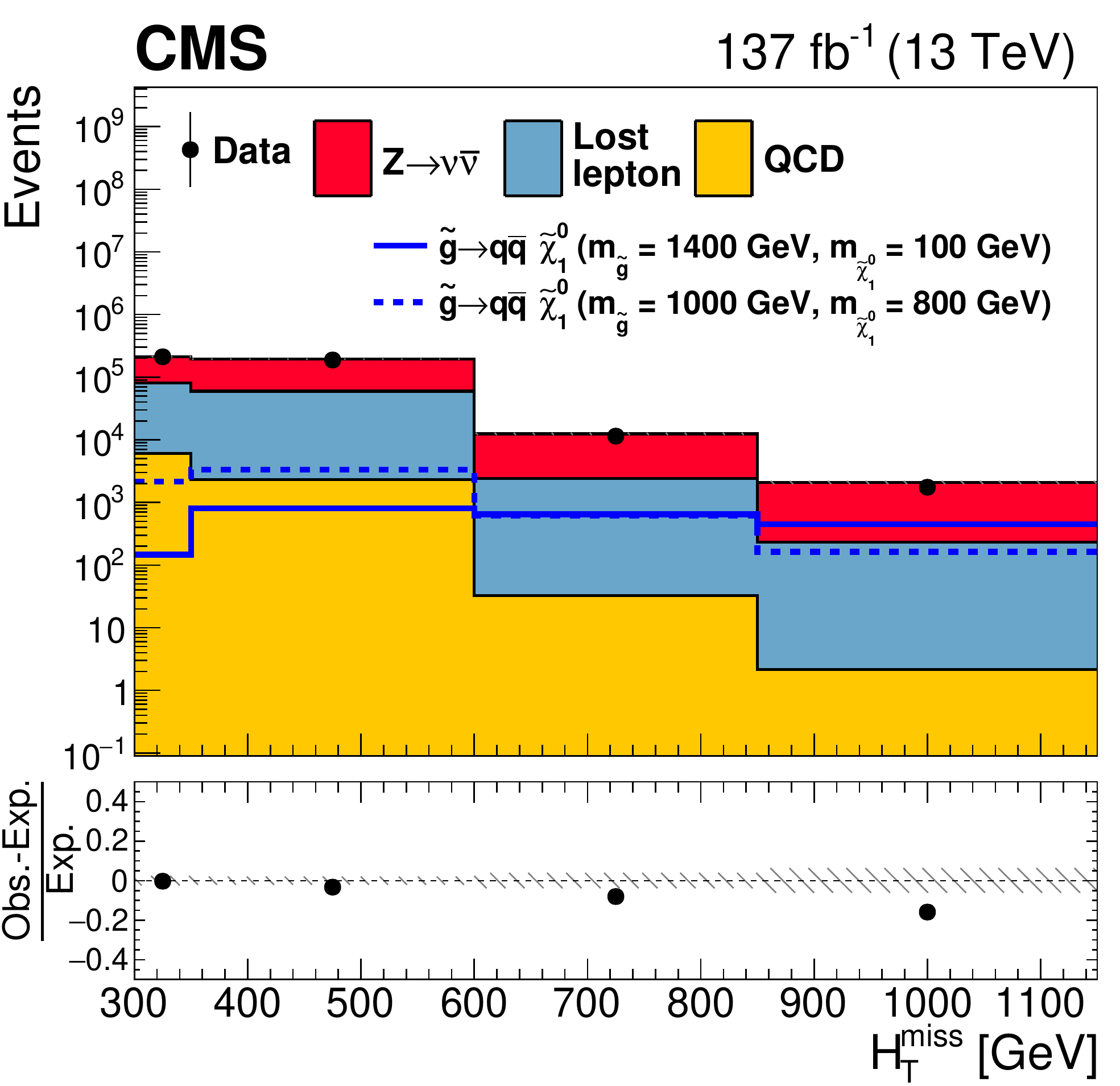}
\includegraphics[width=0.45\textwidth]{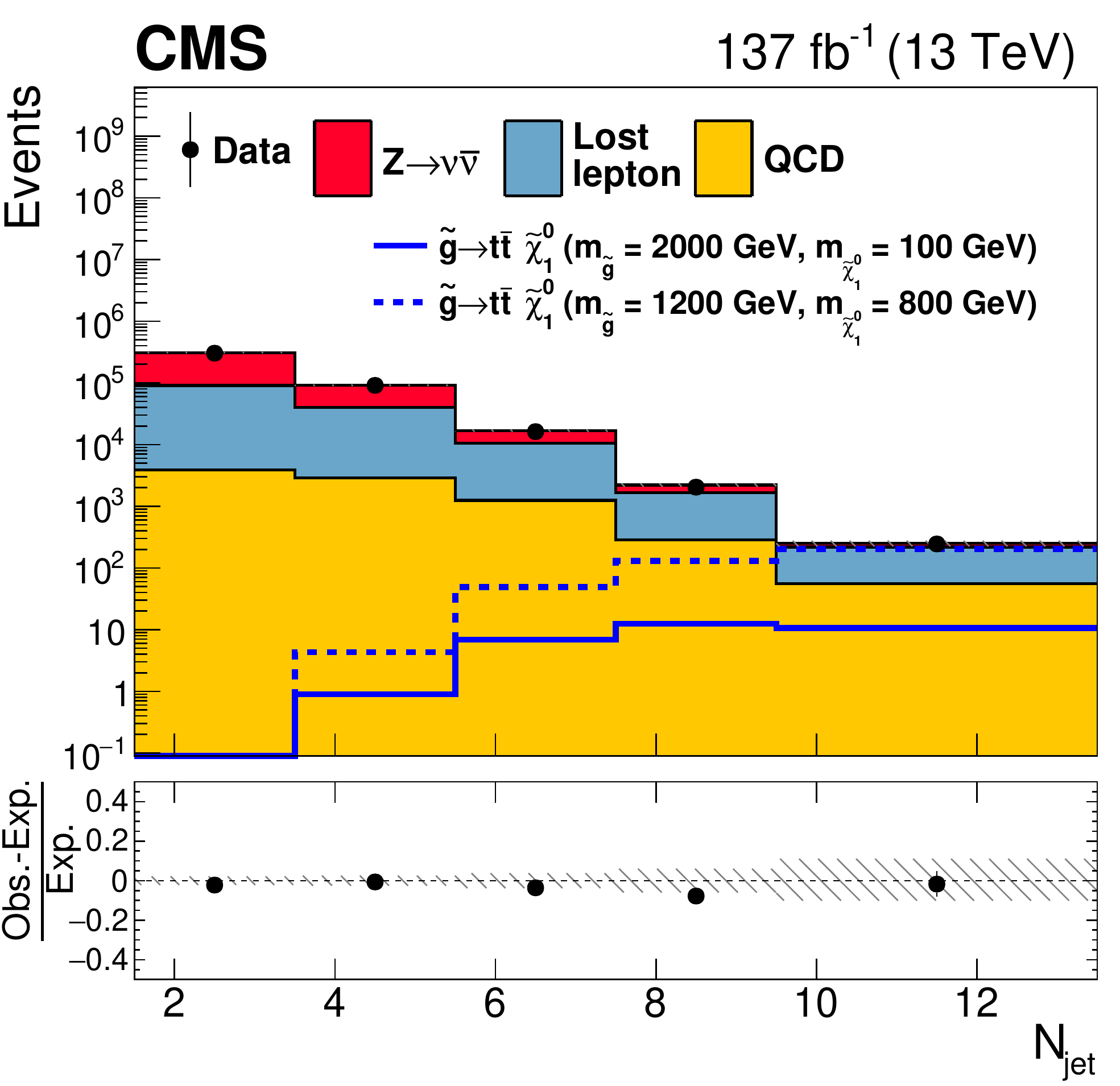}\\
\includegraphics[width=0.45\textwidth]{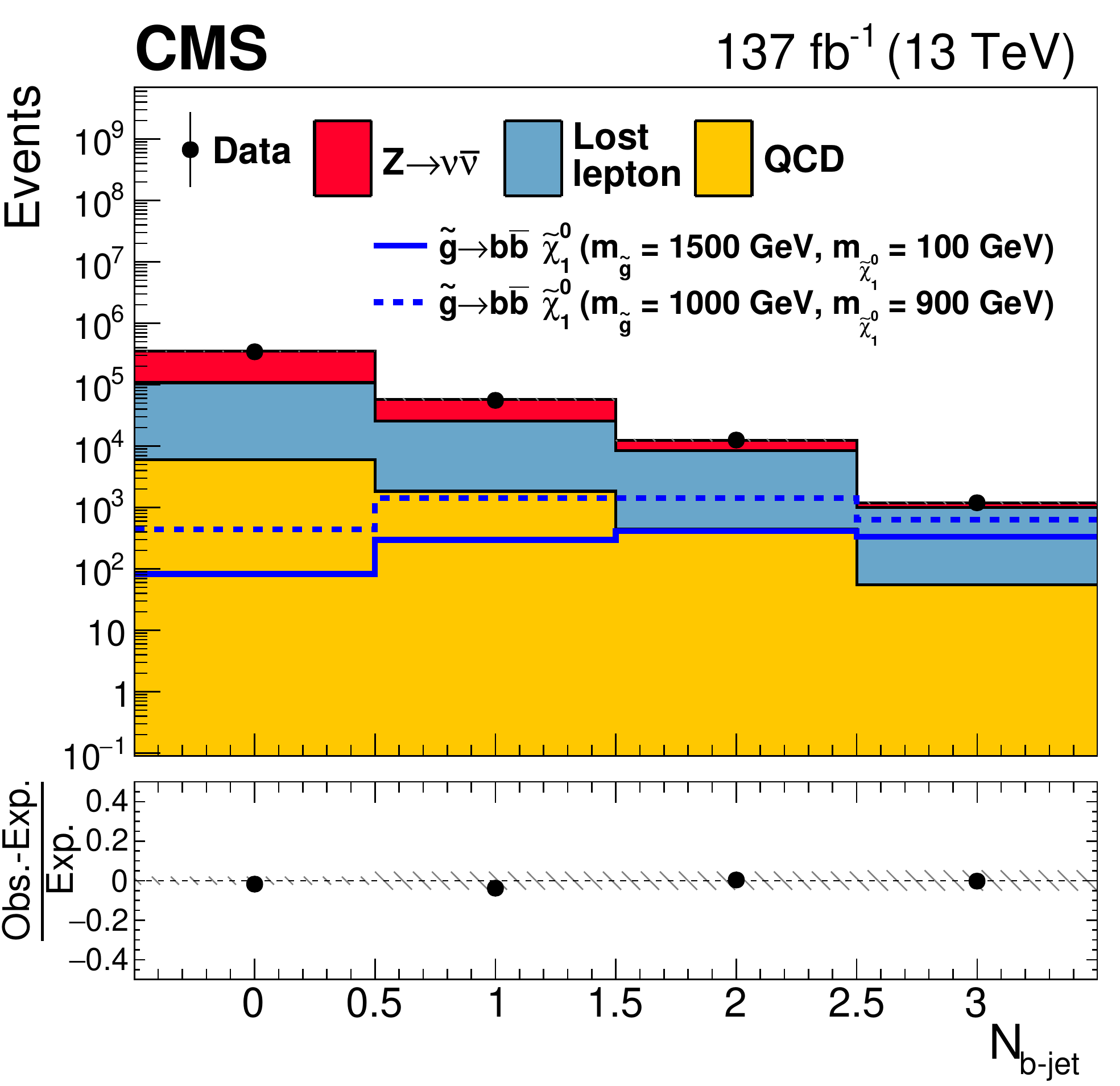}
\caption{
  One-dimensional projections of
  the data and pre-fit SM predictions in \mht, \njets, and \nbjets.
  The hatched regions indicate the total uncertainties in the background predictions.
  The (unstacked) results for two example signal scenarios are shown in each instance,
  one with $\dmass\gg 0$ and the other with $\dmass\approx 0$,
  where \dmass is the difference between the gluino or squark mass
  and the sum of the masses of the particles into which it decays.
 }
\label{fig:projections}
\end{figure*}
\begin{figure*}[htbp]
\centering
\includegraphics[width=0.45\textwidth]{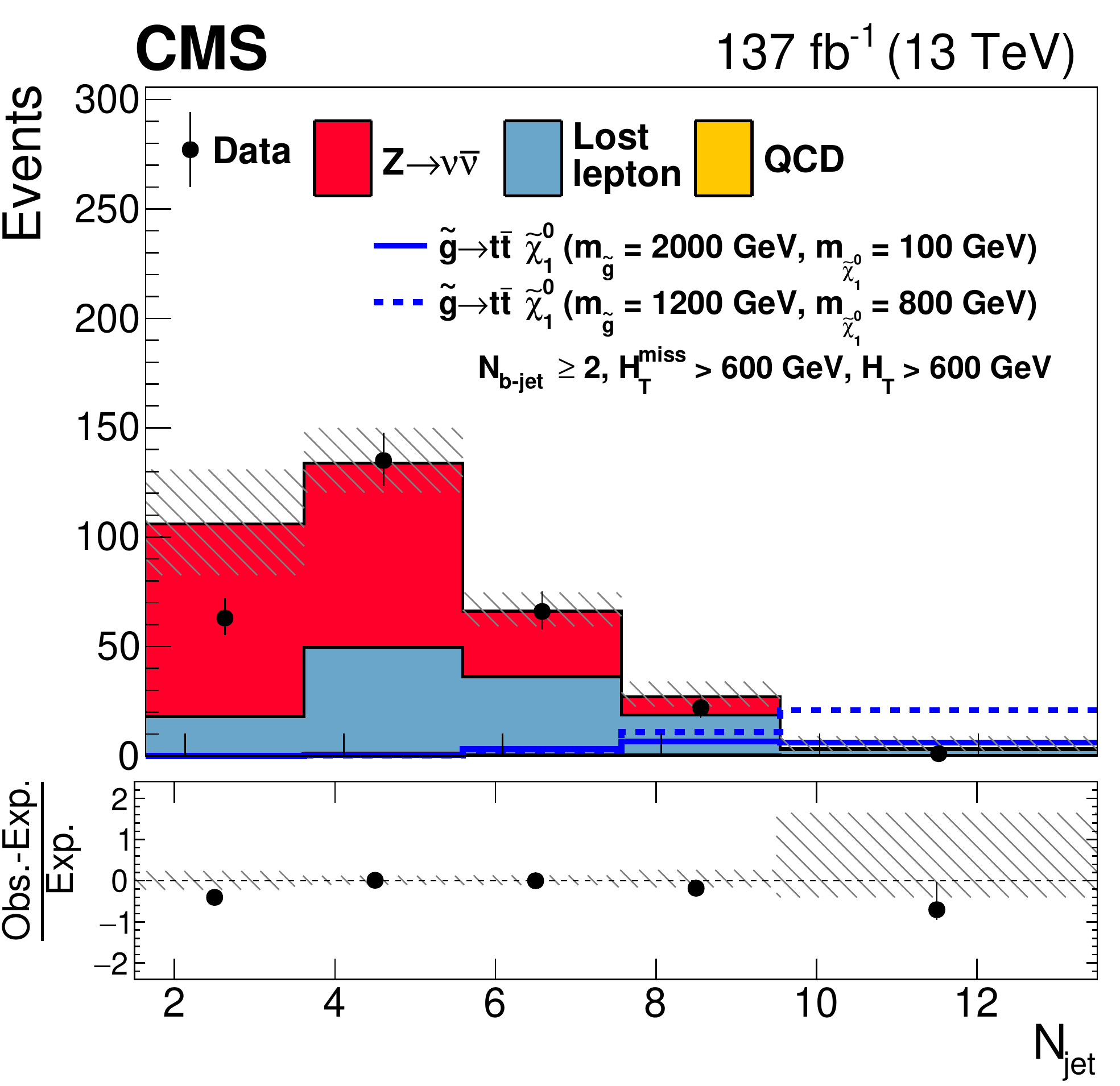}
\includegraphics[width=0.45\textwidth]{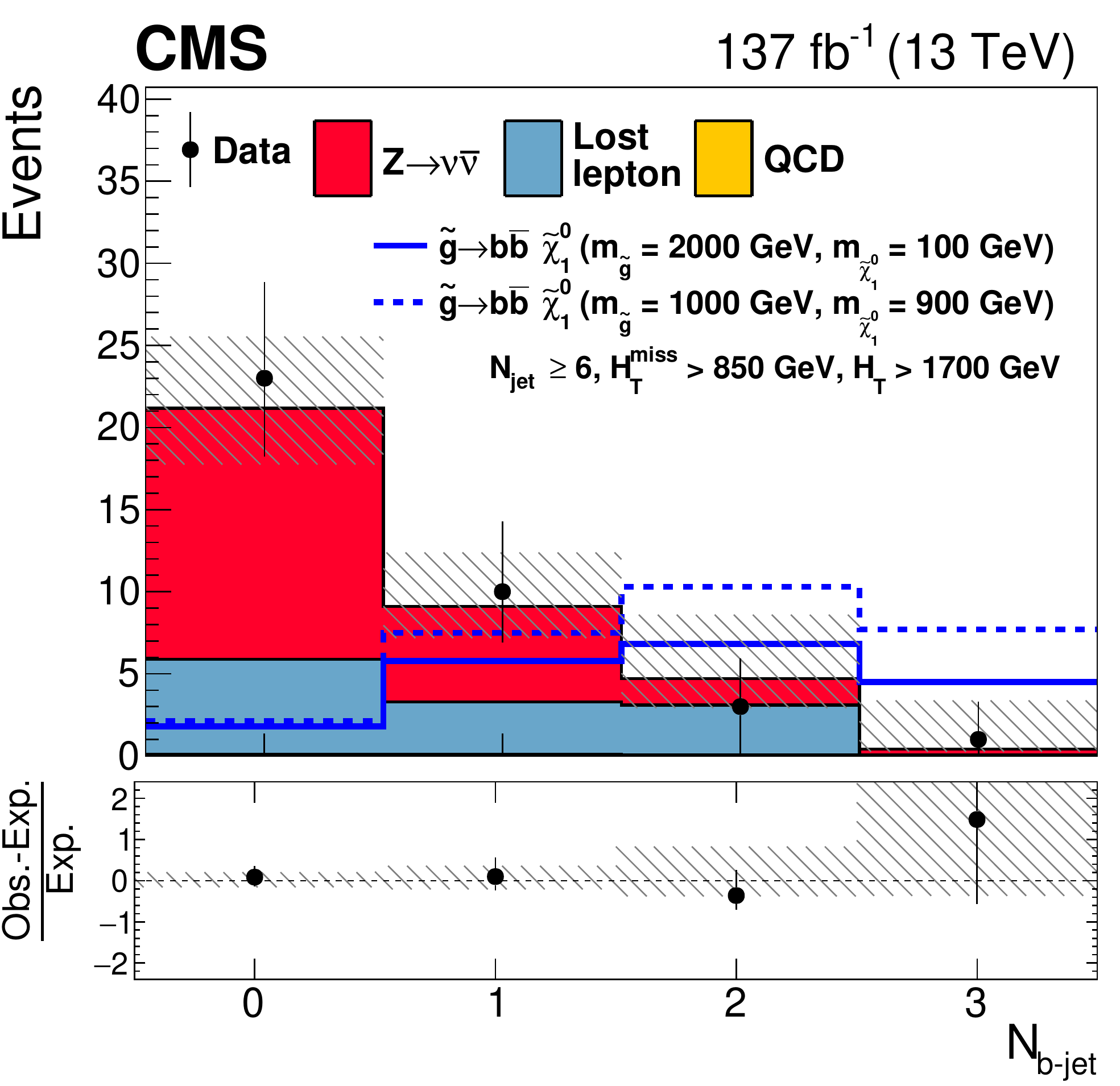}\\
\includegraphics[width=0.45\textwidth]{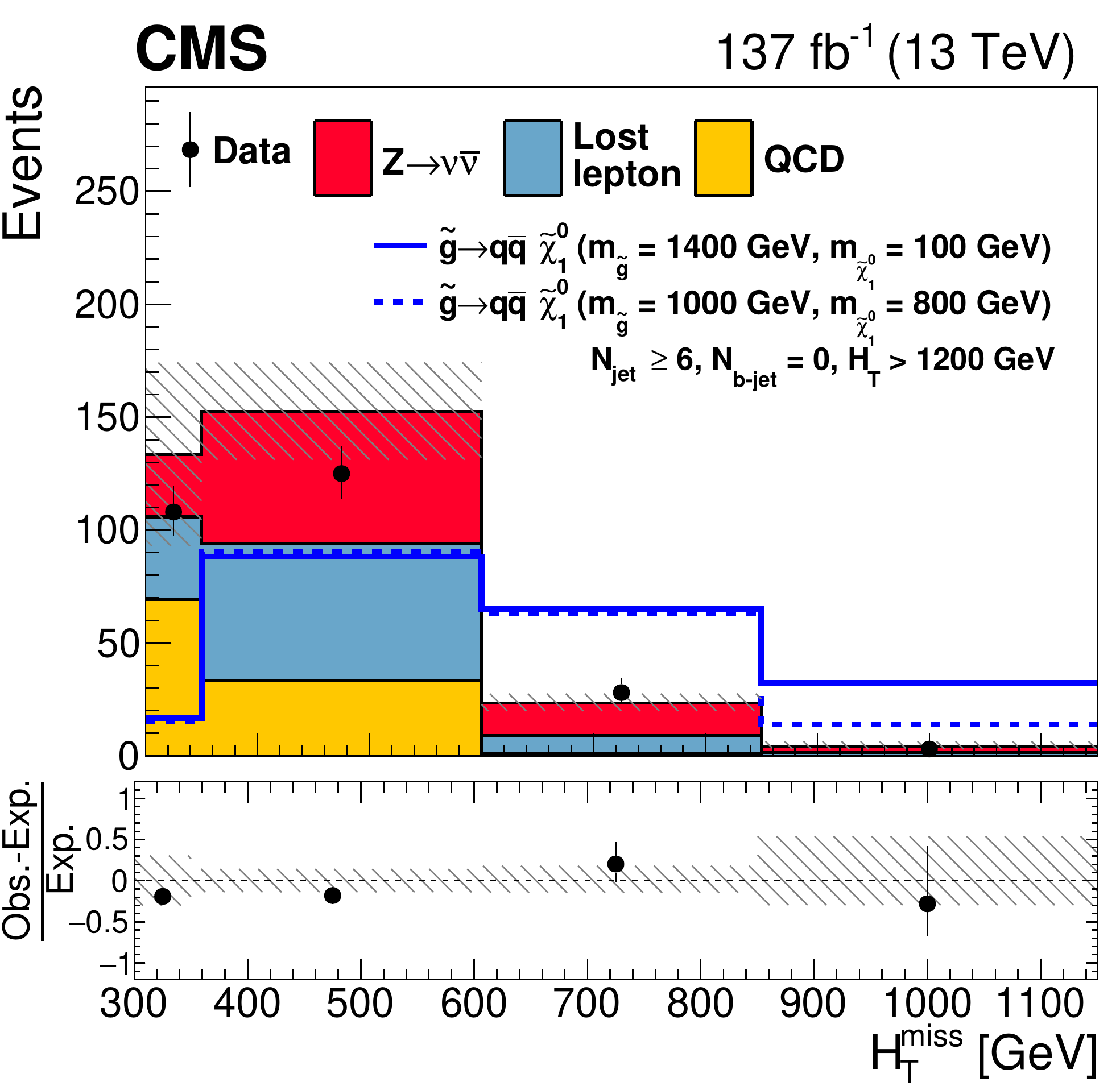}
\includegraphics[width=0.45\textwidth]{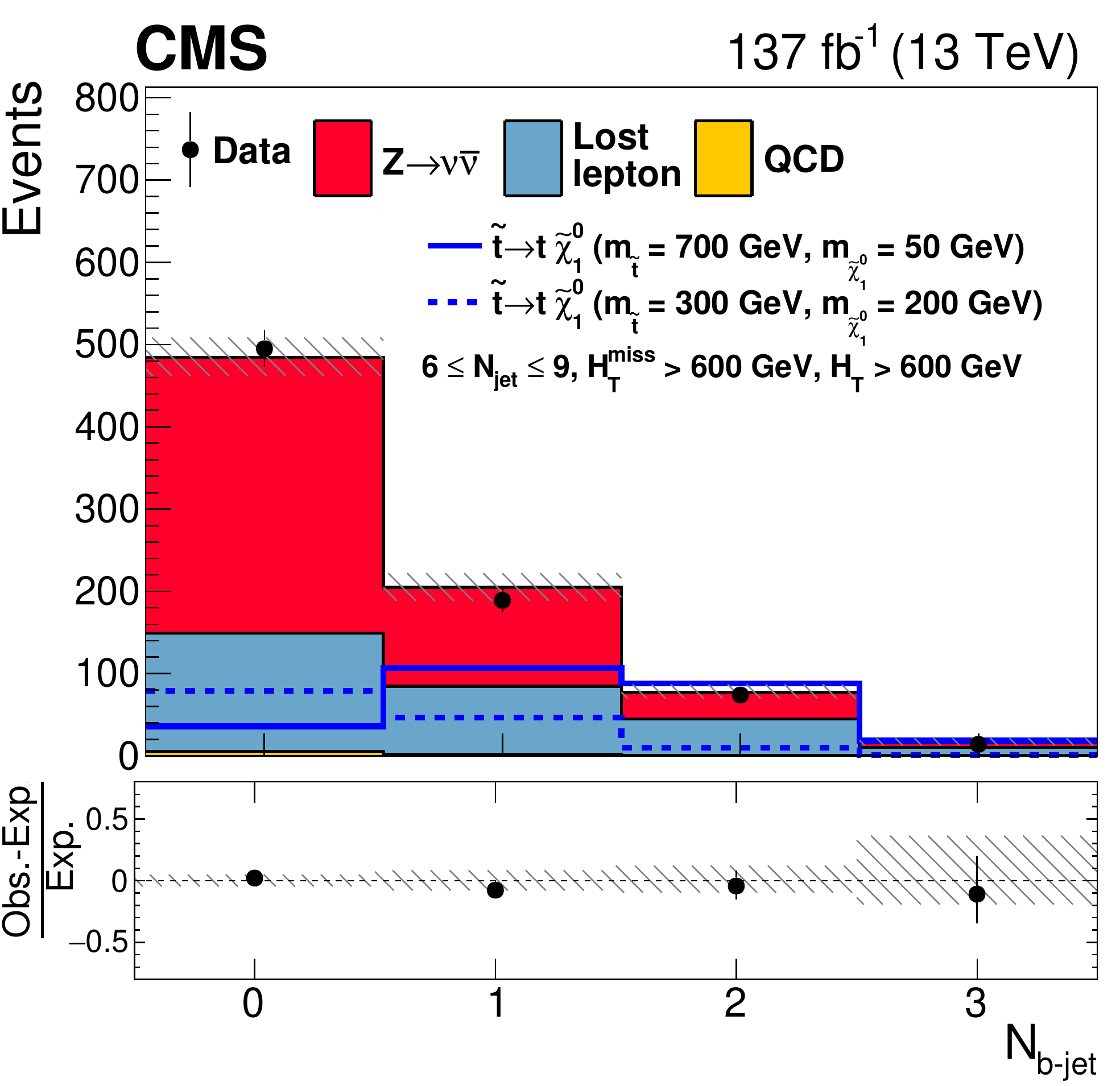}\\
\includegraphics[width=0.45\textwidth]{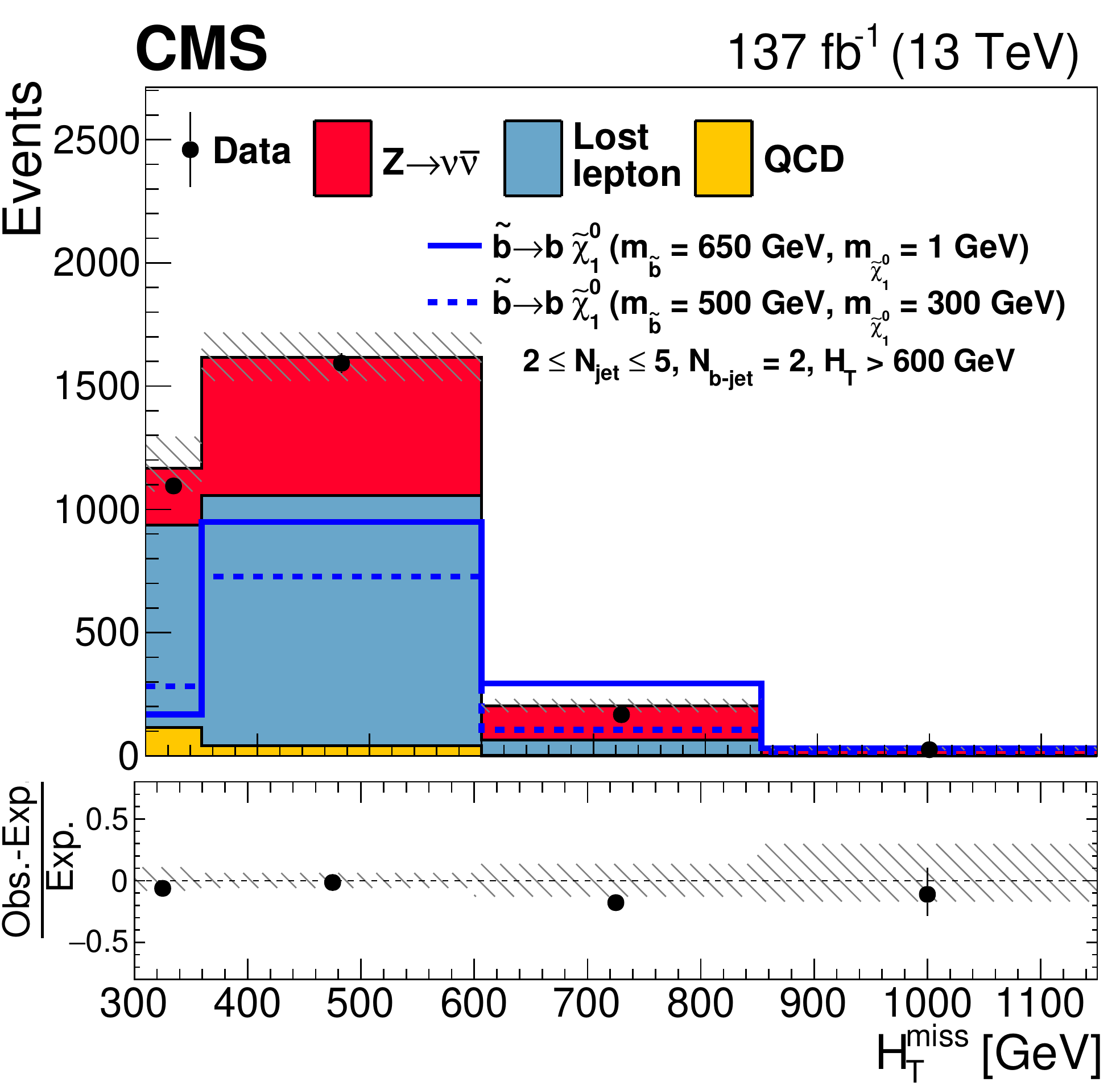}
\includegraphics[width=0.45\textwidth]{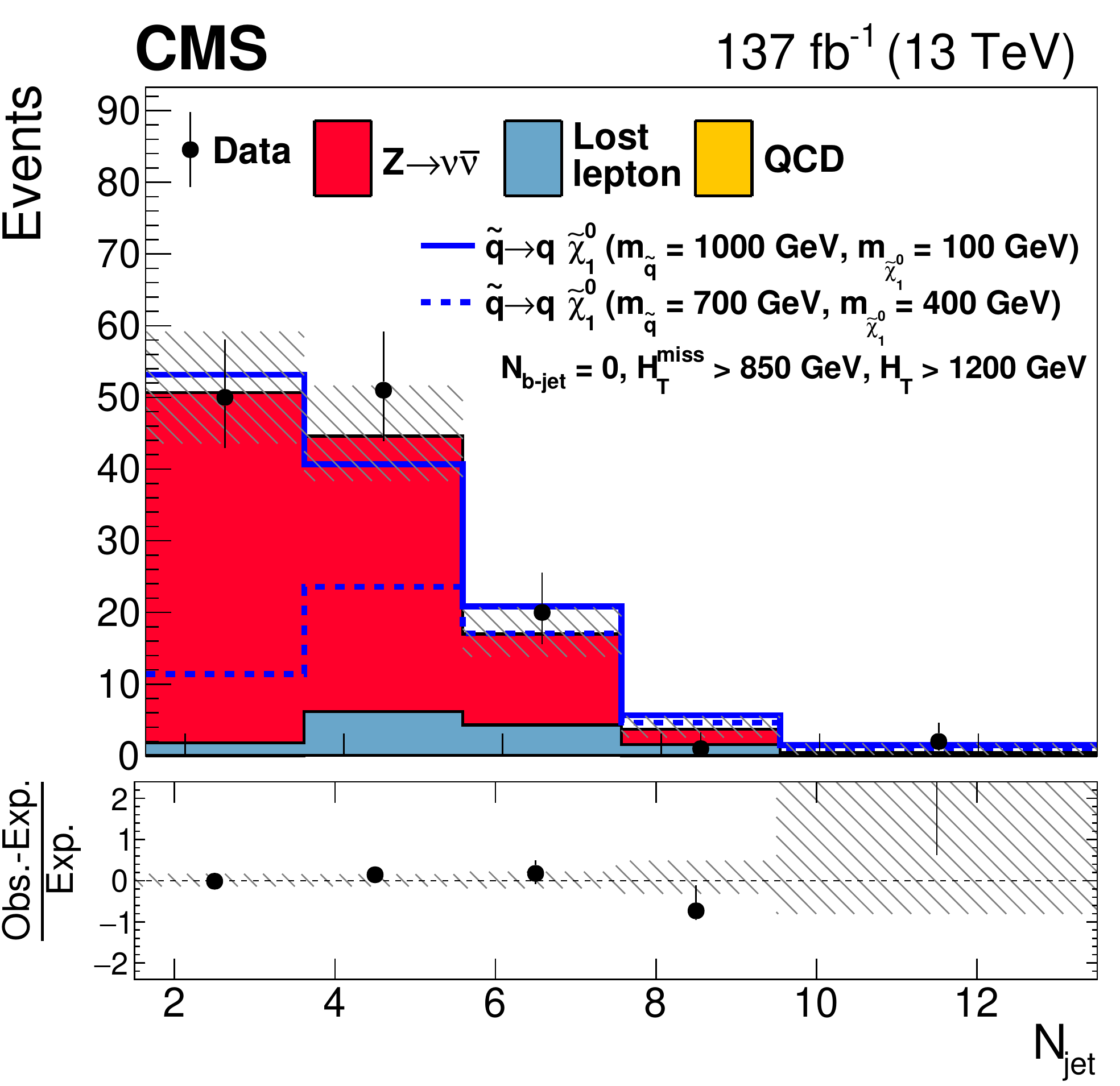}
\caption{
  One-dimensional projections of the data and pre-fit SM predictions in either
  \mht, \njets, or \nbjets
  after applying additional selection criteria,
  given in the figure legends,
  to enhance the sensitivity to the
  (upper left) T1tttt,
  (upper right) T1bbbb,
  (middle left) T1qqqq,
  (middle right) T2tt,
  (lower left) T2bb, and
  (lower right) T2qq signal processes.
  The (unstacked) results for two example signal scenarios are shown in each instance,
  one with $\dmass\gg 0$ and the other with $\dmass\approx 0$,
  where \dmass is the difference between the gluino or squark mass
  and the sum of the masses of the particles into which it decays.
 }
\label{fig:Sigprojections}
\end{figure*}

Figure~\ref{fig:projections} presents one-dimensional projections
of the data and SM predictions in \mht, \njets, and \nbjets.
Additional projections are shown in Fig.~\ref{fig:Sigprojections}.
For these latter results,
criteria have been imposed,
as indicated in the legends,
to enhance the sensitivity for a particular signal process.
For both Figs.~\ref{fig:projections} and~\ref{fig:Sigprojections},
two example signal distributions are shown:
one with $\dmass\gg 0$ and one with $\dmass\approx 0$,
where both example scenarios lie well within the
parameter space excluded by the present study.
The notation $\dmass\gg 0$ means that the mass difference \dmass
is large compared to the sum of the masses of the particles into which
the gluino or squark decays.

Upper limits are evaluated for the production cross sections
of the signal scenarios using a likelihood fit.
The SUSY signal strength $\mu$,
defined by the ratio of cross sections $\mu\equiv\sigma_{\text{SUSY}}/\sigma_{\text{SM}}$,
the signal uncertainties described in Section~\ref{sec:systematic},
the predicted SM background contributions shown in Fig.~\ref{fig:fit-results},
the uncertainties in these backgrounds listed in \cmsAppendix\ref{sec:prefit},
and the control sample yields
are all inputs to the fit.
The background uncertainties,
uncertainties in the signal shape and normalization,
and control sample statistical uncertainties are assigned as nuisance parameters,
which are constrained in the fit.

For the models of gluino (squark) pair production,
the limits are derived as a function of \mgluino (\msquark) and \mlsp.
All 174 search bins are used for each
choice of the SUSY particle masses.
The likelihood function is given by a product of
probability density
functions, one for each search bin.  Each of these is a product of
Poisson functions for the CR yields and log-normal constraint functions
for the nuisance parameters.
Correlations among bins are taken into account.
The signal yield uncertainties associated with
the renormalization and factorization scales, ISR,
jet energy scale,
\cPqb jet tagging,
pileup,
and statistical fluctuations
are evaluated as a function of \mgluino and~\mlsp,
or \msquark and \mlsp.
The test statistic is
$q_\mu =  - 2 \ln ( \mathcal{L}_\mu/\mathcal{L}_\text{max})$,
where $\mathcal{L}_\text{max}$ is the maximum likelihood
determined by allowing all parameters including the
SUSY signal strength $\mu$ to vary,
and $\mathcal{L}_\mu$ is the maximum likelihood for a fixed signal strength.
Limits are set under the asymptotic approximation~\cite{Cowan:2010js},
with $q_\mu$ approximated with an Asimov data set and used
in conjunction with the \CLs criterion described in Refs.~\cite{Junk1999,bib-cls}.

\begin{figure*}[htb]
\centering
\includegraphics[width=0.45\textwidth]{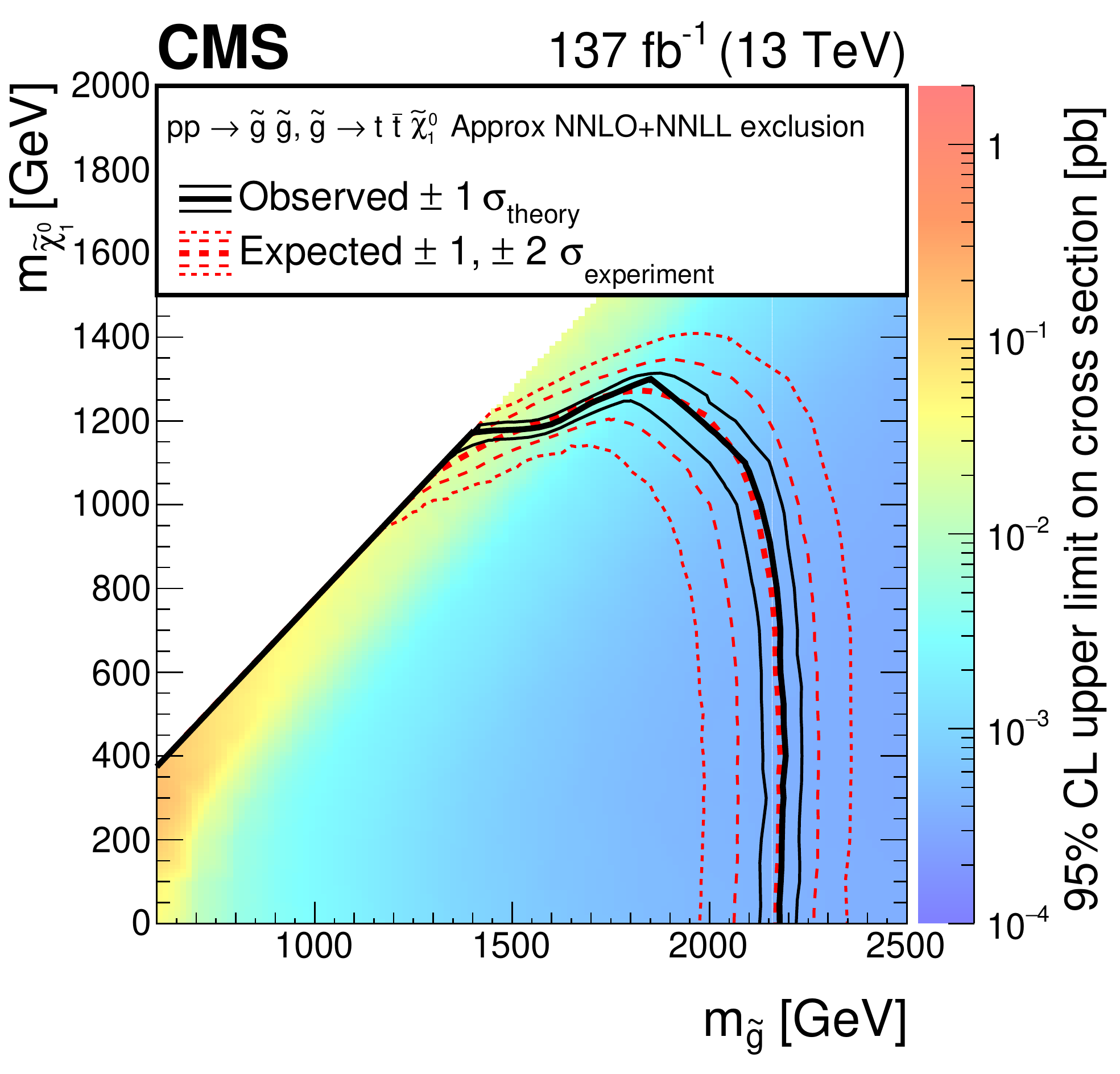}
\includegraphics[width=0.45\textwidth]{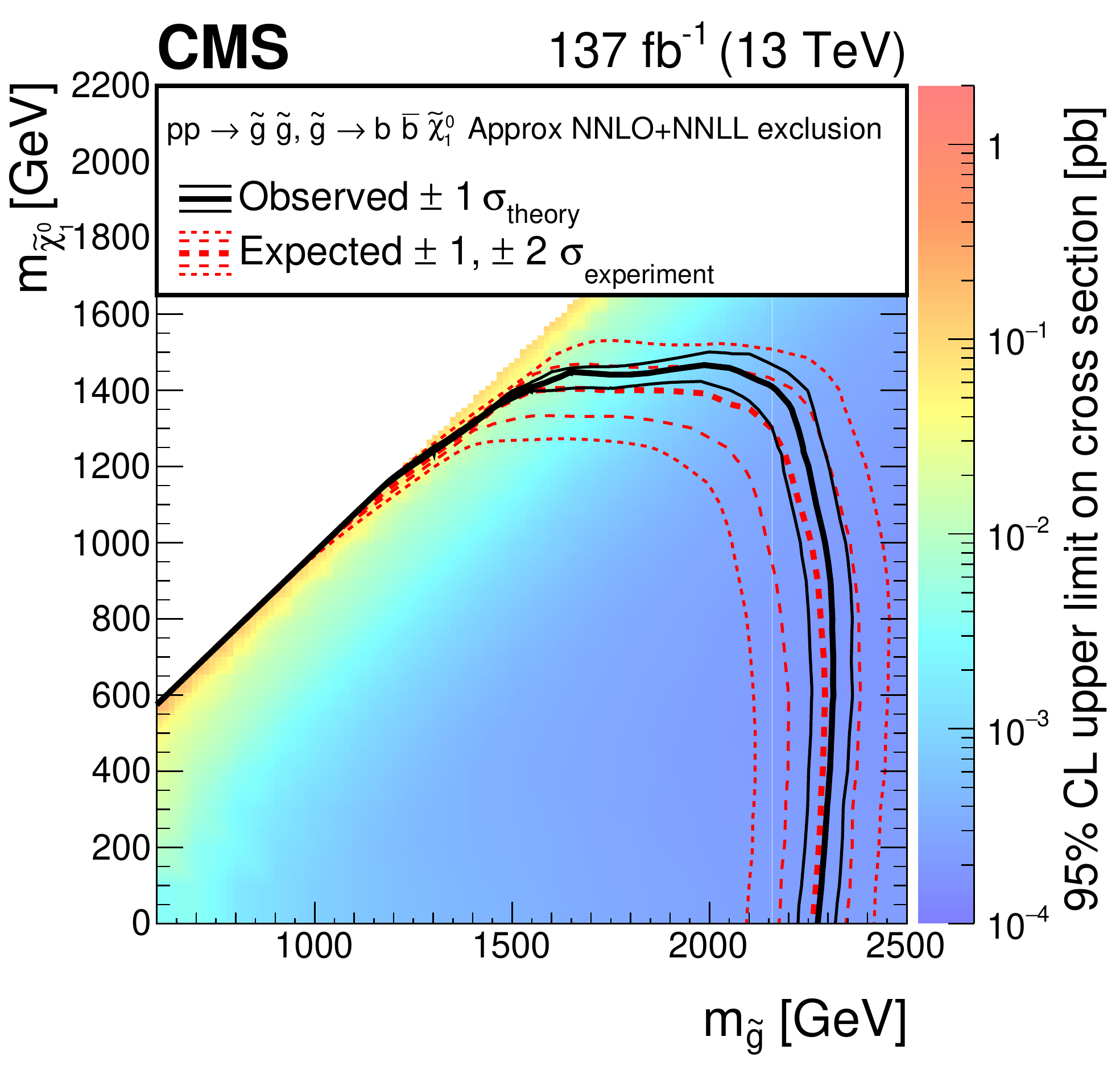}\\
\includegraphics[width=0.45\textwidth]{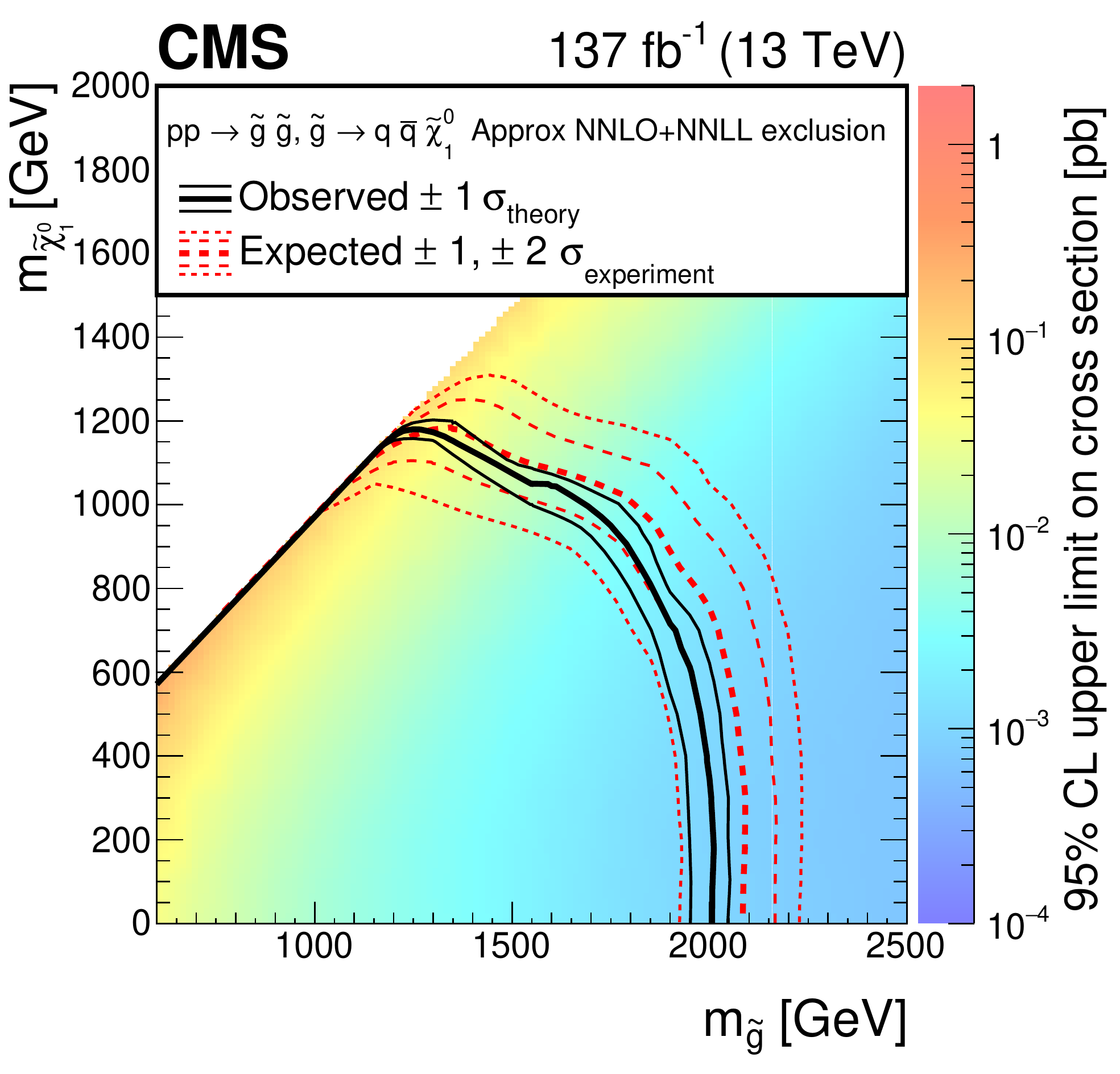}
\includegraphics[width=0.45\textwidth]{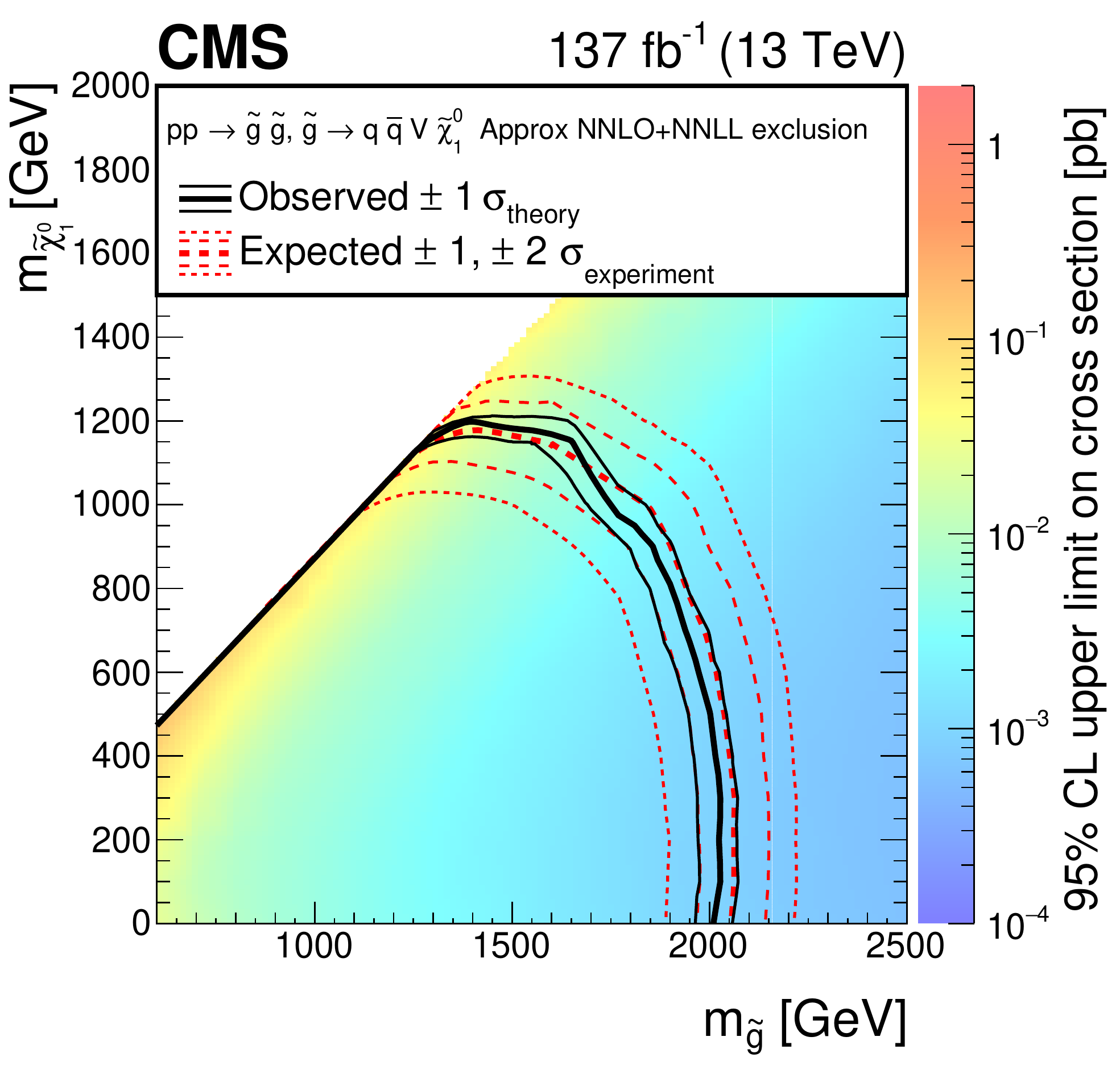}
    \caption{
      The 95\% \CL upper limits on the production
      cross sections of the (upper left) T1tttt,
      (upper right) T1bbbb,
      (lower left) T1qqqq,
      and (lower right) T5qqqqVV signal models
      as a function of the gluino and LSP masses \mgluino and~\mlsp.
      The thick solid (black) curves show the observed exclusion limits
      assuming the approximate-NNLO+NNLL cross
      sections~\cite{bib-nlo-nll-01,bib-nlo-nll-02,bib-nlo-nll-03,
      bib-nlo-nll-04,bib-nlo-nll-05,
      Beenakker:2016lwe,Beenakker:2011sf,Beenakker:2013mva,Beenakker:2014sma,
      Beenakker:1997ut,Beenakker:2010nq,Beenakker:2016gmf}.
      The thin solid (black) curves show the changes in these limits
      as the signal cross sections are varied
      by their theoretical uncertainties~\cite{Borschensky:2014cia}.
      The thick dashed (red) curves present the expected limits
      under the background-only hypothesis,
      while the two sets of thin dotted (red) curves
      indicate the region containing 68 and 95\%
      of the distribution of limits expected under this hypothesis.
    }
    \label{fig:limits-gluinos}
\end{figure*}

We evaluate 95\% confidence level (\CL) upper limits
on the signal cross sections.
The approximate NNLO+NNLL cross section is used
to determine corresponding exclusion curves.
Before computing these limits,
the signal yields are corrected to account for
the predicted signal contamination in the CRs
from the signal model under consideration.
Beyond the observed exclusion limits,
we derive expected exclusion limits
by evaluating the test statistic using the predicted
numbers of background events with their
expected Poisson fluctuations.

The results for the T1tttt, T1bbbb, T1qqqq, and T5qqqqVV
models are shown in Fig.~\ref{fig:limits-gluinos}.
Depending on the value of \mlsp,
gluinos with masses as large as
2180, 2310, 2000, and 2030\GeV, respectively,
are excluded.
These results significantly extend those
of our previous study~\cite{Sirunyan:2017cwe},
for which the corresponding limits
are 1960, 1950, 1825, and 1800\GeV.

\begin{figure*}[htb]
\centering
    \includegraphics[width=0.45\textwidth]{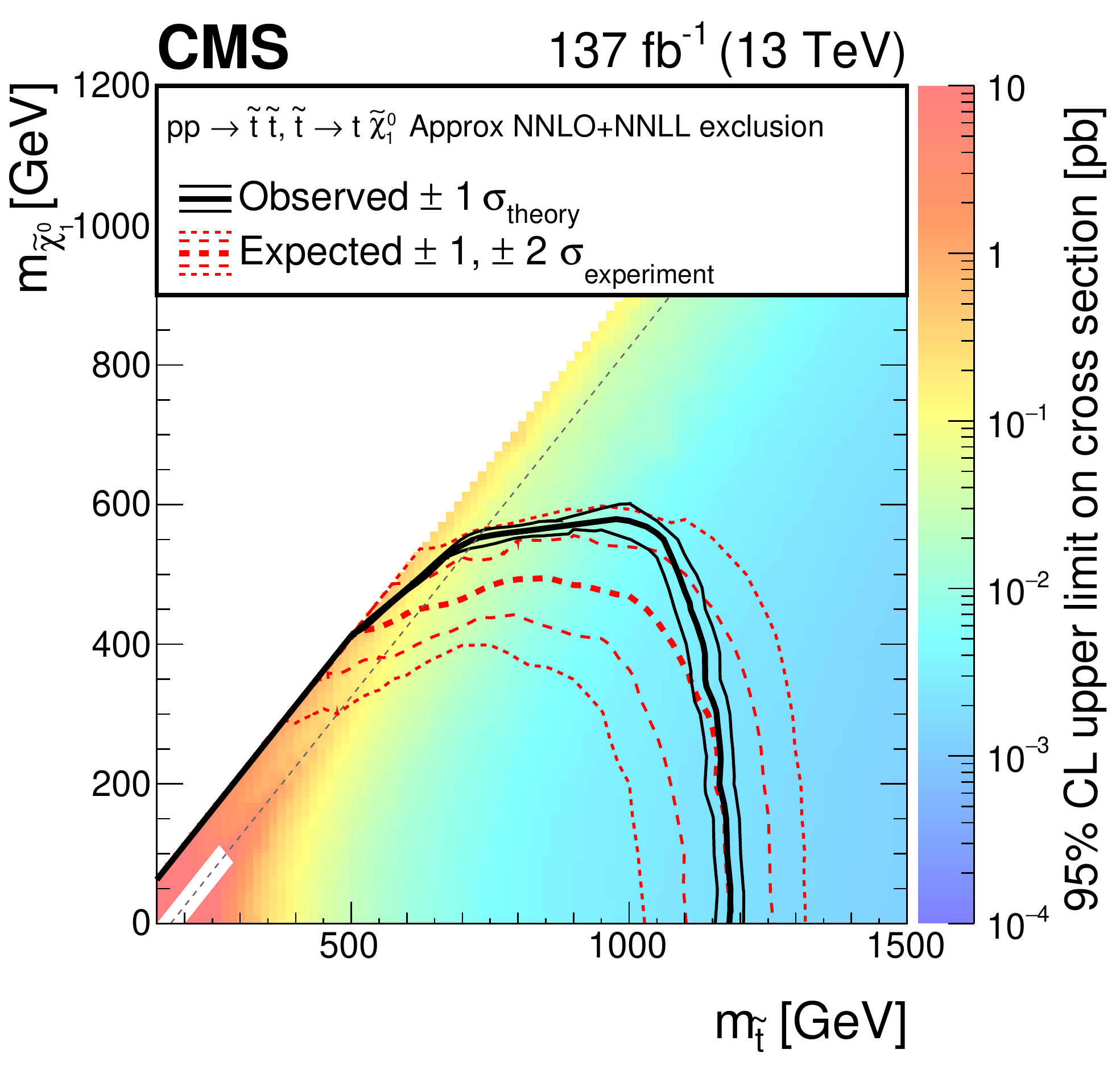}
    \includegraphics[width=0.45\textwidth]{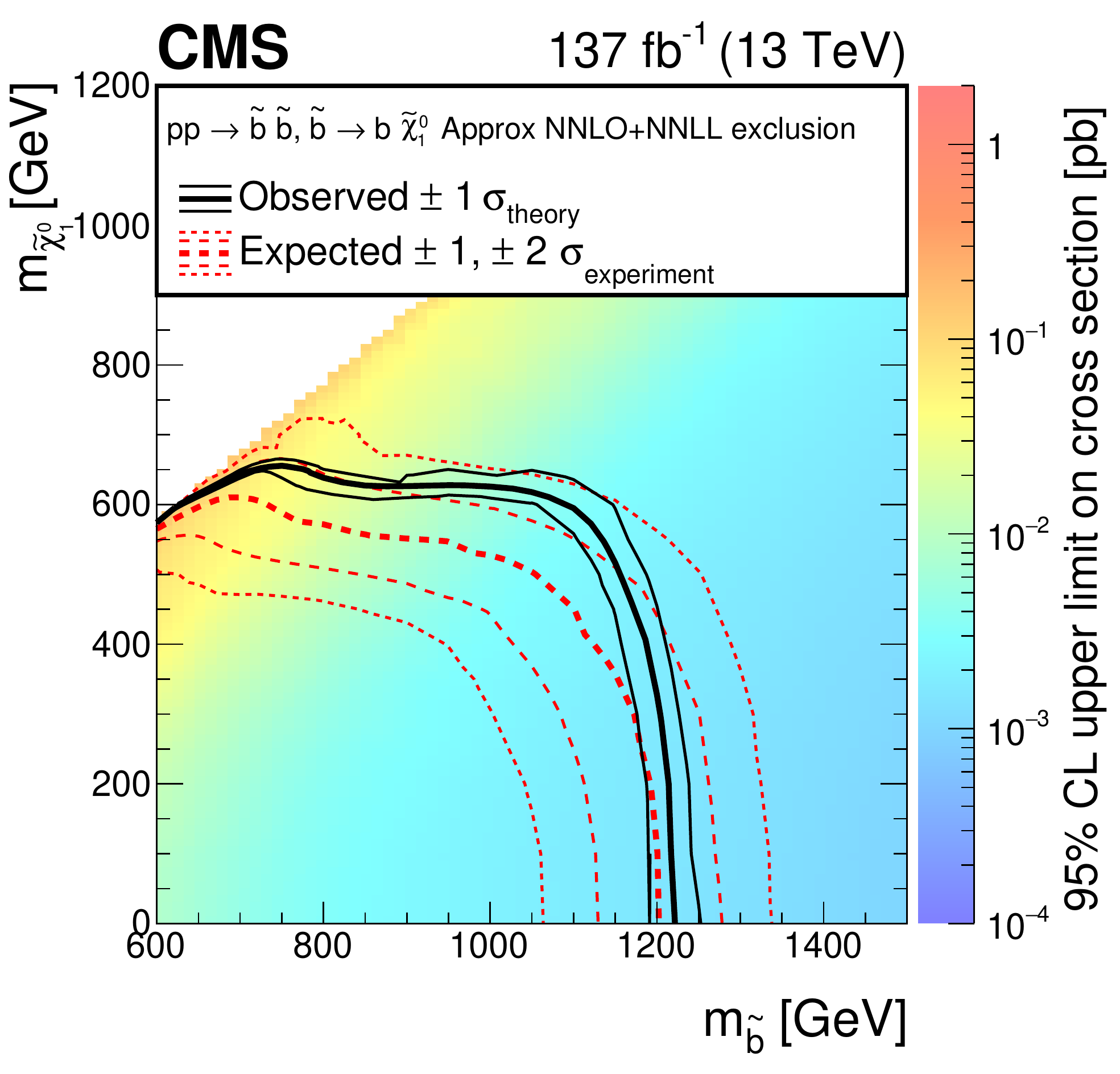}\\
    \includegraphics[width=0.45\textwidth]{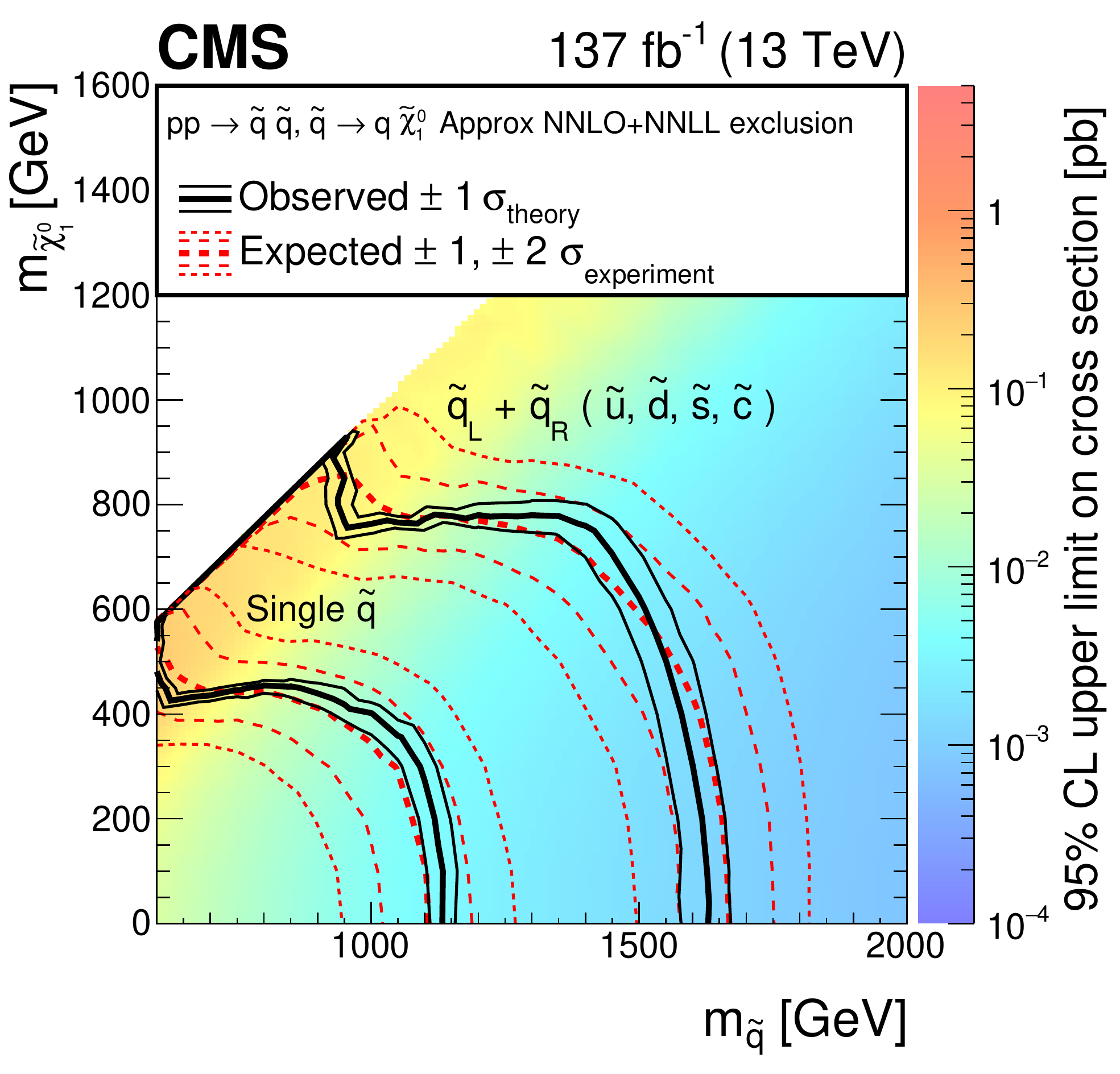}
    \caption{
      The 95\% \CL upper limits on the production
      cross sections of the (upper left) T2tt,
      (upper right) T2bb, and (lower) T2qq signal models
      as a function of the squark and LSP masses \msquark and~\mlsp.
      The meaning of the curves is described in the
      Fig.~\ref{fig:limits-gluinos} caption.
      For the T2tt model,
      we do not present cross section upper limits
      in the unshaded diagonal region at low \mlsp
      for the reason discussed in the text.
      The diagonal dotted line shown for this model
      corresponds to $\mstop-\mlsp=\mtop$.
    }
    \label{fig:limits-squarks}
\end{figure*}

Figure~\ref{fig:limits-squarks} shows
the corresponding results for the T2tt, T2bb, and T2qq models.
Squarks with masses up to 1190, 1220, and 1630\GeV,
respectively,
are excluded,
compared to 960, 990, and 1390\GeV
in our previous study~\cite{Sirunyan:2017cwe}.
Note that for the T2tt model
we do not present cross section upper limits at
small values of \mlsp if $\mstop-\mlsp\approx \mtop$,
corresponding to the unshaded diagonal region at low \mlsp
in Fig.~\ref{fig:limits-squarks} (upper left),
because
signal events are essentially
indistinguishable from SM \ttbar events in this region,
resulting in large signal contamination of the CRs and
rendering the signal event acceptance difficult to model.

In addition to the main T2qq model,
with four mass-degenerate squark flavors
(up, down, strange, and charm),
each arising from two different quark spin states,
Fig.~\ref{fig:limits-squarks} (lower) shows the results
should only one of these eight states
(``Single~\PSQ'')
be accessible at the LHC.
In this case,
the upper limit on the squark mass
is reduced to 1130\GeV.

\section{Summary}
\label{sec:summary}

Using essentially the full CMS Run 2 data sample
of proton-proton collisions at $\sqrt{s} = 13\TeV$,
corresponding to an integrated luminosity of 137\fbinv collected in 2016--2018,
a search for supersymmetry has been performed based on
events containing multiple jets and large missing transverse momentum.
The event yields are measured in 174 nonoverlapping
search bins defined in a four-dimensional space of
missing transverse momentum (\mht),
the scalar sum of jet transverse momenta (\HT),
the number of jets,
and the number of tagged bottom quark jets.
The events are required to satisfy $\mht>300\gev$, $\HT>300\gev$,
and to have at least two jets with transverse momentum $\pt>30\gev$.
Events with isolated high \pt leptons or photons are vetoed.

The results are compared to the expected number of background
events from standard model (SM) processes.
The principal backgrounds arise from events with neutrino production
or jet mismeasurement.
The SM background is evaluated using control regions in data
supplemented by information from Monte Carlo event simulation.
The observed event yields are found to be consistent with the SM background
and no evidence for supersymmetry is obtained.

The results are interpreted in the context of simplified models
for gluino and squark pair production.
For the gluino models,
each of the produced gluinos decays either to a \ttbar pair and an undetected,
stable, lightest supersymmetric particle,
assumed to be the {\PSGczDo} neutralino (T1tttt model);
to a $\bbbar$ pair and the {\PSGczDo} (T1bbbb model);
to a light-flavored (\cPqu, \cPqd, \cPqs, \cPqc)
\qqbar pair and the {\PSGczDo} (T1qqqq model);
or to a light-flavored quark and antiquark and either the
second-lightest neutralino \PSGczDt
or the lightest chargino $\PSGc_1^{\pm}$,
followed by decay of the \PSGczDt ($\PSGc_1^{\pm}$)
to the \PSGczDo and an on- or off-mass-shell
{\cPZ} ({$\PW^\pm$}) boson (T5qqqqVV model).
For the squark models,
each of the produced squarks decays
either to a top quark and the {\PSGczDo} (T2tt model),
to a bottom quark and the {\PSGczDo} (T2bb model),
or to a light-flavored quark and the {\PSGczDo} (T2qq model).

Using the predicted cross sections with next-to-leading order plus
approximate next-to-leading logarithm accuracy as a reference,
gluinos with masses as large as from 2000 to 2310\GeV
are excluded at 95\% confidence level,
depending on the signal model.
The corresponding limits on the masses of directly produced
squarks range from 1190 for top squarks to 1630\GeV for light-flavored squarks.
The results presented here supersede those of Ref.~\cite{Sirunyan:2017cwe},
extending the mass limits of this previous study by,
typically, 200\GeV or more.

\begin{acknowledgments}
We congratulate our colleagues in the CERN accelerator departments for the excellent performance of the LHC and thank the technical and administrative staffs at CERN and at other CMS institutes for their contributions to the success of the CMS effort. In addition, we gratefully acknowledge the computing centers and personnel of the Worldwide LHC Computing Grid for delivering so effectively the computing infrastructure essential to our analyses. Finally, we acknowledge the enduring support for the construction and operation of the LHC and the CMS detector provided by the following funding agencies: BMBWF and FWF (Austria); FNRS and FWO (Belgium); CNPq, CAPES, FAPERJ, FAPERGS, and FAPESP (Brazil); MES (Bulgaria); CERN; CAS, MoST, and NSFC (China); COLCIENCIAS (Colombia); MSES and CSF (Croatia); RPF (Cyprus); SENESCYT (Ecuador); MoER, ERC IUT, PUT and ERDF (Estonia); Academy of Finland, MEC, and HIP (Finland); CEA and CNRS/IN2P3 (France); BMBF, DFG, and HGF (Germany); GSRT (Greece); NKFIA (Hungary); DAE and DST (India); IPM (Iran); SFI (Ireland); INFN (Italy); MSIP and NRF (Republic of Korea); MES (Latvia); LAS (Lithuania); MOE and UM (Malaysia); BUAP, CINVESTAV, CONACYT, LNS, SEP, and UASLP-FAI (Mexico); MOS (Montenegro); MBIE (New Zealand); PAEC (Pakistan); MSHE and NSC (Poland); FCT (Portugal); JINR (Dubna); MON, RosAtom, RAS, RFBR, and NRC KI (Russia); MESTD (Serbia); SEIDI, CPAN, PCTI, and FEDER (Spain); MOSTR (Sri Lanka); Swiss Funding Agencies (Switzerland); MST (Taipei); ThEPCenter, IPST, STAR, and NSTDA (Thailand); TUBITAK and TAEK (Turkey); NASU and SFFR (Ukraine); STFC (United Kingdom); DOE and NSF (USA).

\hyphenation{Rachada-pisek} Individuals have received support from the Marie-Curie program and the European Research Council and Horizon 2020 Grant, contract Nos.\ 675440, 752730, and 765710 (European Union); the Leventis Foundation; the A.P.\ Sloan Foundation; the Alexander von Humboldt Foundation; the Belgian Federal Science Policy Office; the Fonds pour la Formation \`a la Recherche dans l'Industrie et dans l'Agriculture (FRIA-Belgium); the Agentschap voor Innovatie door Wetenschap en Technologie (IWT-Belgium); the F.R.S.-FNRS and FWO (Belgium) under the ``Excellence of Science -- EOS" -- be.h project n.\ 30820817; the Beijing Municipal Science \& Technology Commission, No. Z181100004218003; the Ministry of Education, Youth and Sports (MEYS) of the Czech Republic; the Lend\"ulet (``Momentum") Program and the J\'anos Bolyai Research Scholarship of the Hungarian Academy of Sciences, the New National Excellence Program \'UNKP, the NKFIA research grants 123842, 123959, 124845, 124850, 125105, 128713, 128786, and 129058 (Hungary); the Council of Science and Industrial Research, India; the HOMING PLUS program of the Foundation for Polish Science, cofinanced from European Union, Regional Development Fund, the Mobility Plus program of the Ministry of Science and Higher Education, the National Science Center (Poland), contracts Harmonia 2014/14/M/ST2/00428, Opus 2014/13/B/ST2/02543, 2014/15/B/ST2/03998, and 2015/19/B/ST2/02861, Sonata-bis 2012/07/E/ST2/01406; the National Priorities Research Program by Qatar National Research Fund; the Ministry of Science and Education, grant no. 3.2989.2017 (Russia); the Programa Estatal de Fomento de la Investigaci{\'o}n Cient{\'i}fica y T{\'e}cnica de Excelencia Mar\'{\i}a de Maeztu, grant MDM-2015-0509 and the Programa Severo Ochoa del Principado de Asturias; the Thalis and Aristeia programs cofinanced by EU-ESF and the Greek NSRF; the Rachadapisek Sompot Fund for Postdoctoral Fellowship, Chulalongkorn University and the Chulalongkorn Academic into Its 2nd Century Project Advancement Project (Thailand); the Welch Foundation, contract C-1845; and the Weston Havens Foundation (USA).
\end{acknowledgments}

\bibliography{auto_generated}

\clearpage
\appendix

\section{Numerical results for the full set of search bins}
\label{sec:prefit}

In this appendix,
we present numerical values for the results in the 174 search bins
shown in Fig.~\ref{fig:fit-results}.
\begin{table}[h]
\renewcommand{\arraystretch}{1.25}
\centering
\caption{Observed number of events and pre-fit background predictions in the $2\leq\njets\leq3$ search bins. For the background predictions,
the first uncertainty is statistical and the second systematic.}
\cmsTable{
\label{tab:pre-fit-results-nj1}
\begin{tabular}{ cccccccccc }
\multirow{2}{*}{Bin} & \mht & \HT & \multirow{2}{*}{$\njets$} & \multirow{2}{*}{$\nbjets$} & Lost-lepton & \znn & QCD & Total & \multirow{2}{*}{Observed} \\[-1.5mm]
    & [{\GeVns}] & [{\GeVns}] & & & background & background & background & background & \\
\hline
1 & 300--350 & 300--600 & 2--3 & 0 & $38\,870\pm320\pm580$ & $89\,100\pm200\pm2600$ & $1800\pm1000^{+1200}_{-800}$ & $129\,800\pm1100\pm2800$ & 130\,718 \\
2 & 300--350 & 600--1200 & 2--3 & 0 & $2760\pm61\pm39$ & $4970\pm50\pm150$ & $330\pm180\pm160$ & $8060\pm200\pm220$ & 7820 \\
3 & 300--350 & $\geq$1200 & 2--3 & 0 & $181\pm17\pm3$ & $308\pm12\pm18$ & $62\pm34\pm27$ & $552\pm40\pm32$ & 514 \\
4 & 350--600 & 350--600 & 2--3 & 0 & $26\,230\pm240\pm540$ & $78\,000\pm200\pm2200$ & $660\pm360\pm300$ & $104\,900\pm500\pm2300$ & 100\,828 \\
5 & 350--600 & 600--1200 & 2--3 & 0 & $5319\pm81\pm78$ & $14\,570\pm80\pm430$ & $210\pm110\pm100$ & $20\,100\pm160\pm450$ & 19\,319 \\
6 & 350--600 & $\geq$1200 & 2--3 & 0 & $279\pm21\pm6$ & $689\pm17^{+41}_{-36}$ & $29\pm16\pm13$ & $997\pm32\pm40$ & 933 \\
7 & 600--850 & 600--1200 & 2--3 & 0 & $1220\pm43\pm25$ & $6290\pm50\pm370$ & $11.1\pm6.0^{+5.4}_{-5.1}$ & $7520\pm70\pm360$ & 6786 \\
8 & 600--850 & $\geq$1200 & 2--3 & 0 & $52\pm9\pm2$ & $240\pm11\pm15$ & $0.73\pm0.65^{+0.31}_{-0.07}$ & $293\pm14\pm16$ & 277 \\
9 & $\geq$850 & 850--1700 & 2--3 & 0 & $116\pm14\pm3$ & $1088\pm23\pm98$ & $0.35\pm0.21\pm0.15$ & $1205\pm28\pm98$ & 933 \\
10 & $\geq$850 & $\geq$1700 & 2--3 & 0 & $1.8^{+4.1}_{-1.5}\pm0.1$ & $48.9^{+5.3}_{-4.8}\pm0.5$ & $0.02\pm0.02^{+0.01}_{-0.00}$ & $50.7^{+6.7}_{-5.0}\pm5.1$ & 50 \\
11 & 300--350 & 300--600 & 2--3 & 1 & $5590\pm100\pm100$ & $9800\pm20\pm1500$ & $360\pm200^{+330}_{-160}$ & $15\,800\pm200\pm1500$ & 15\,272 \\
12 & 300--350 & 600--1200 & 2--3 & 1 & $436\pm25\pm6$ & $616\pm6\pm95$ & $99\pm54^{+79}_{-45}$ & $1150\pm60\pm110$ & 1177 \\
13 & 300--350 & $\geq$1200 & 2--3 & 1 & $27.4^{+7.9}_{-6.3}\pm0.4$ & $38.4\pm1.5\pm6.1$ & $18\pm10^{+14}_{-8}$ & $84\pm13^{+15}_{-10}$ & 71 \\
14 & 350--600 & 350--600 & 2--3 & 1 & $3237\pm75\pm99$ & $8600\pm20\pm1300$ & $124\pm67^{+96}_{-57}$ & $11\,900\pm100\pm1300$ & 11\,121 \\
15 & 350--600 & 600--1200 & 2--3 & 1 & $757\pm32\pm14$ & $1780\pm10\pm270$ & $48\pm27^{+38}_{-21}$ & $2590\pm40\pm270$ & 2530 \\
16 & 350--600 & $\geq$1200 & 2--3 & 1 & $36.7^{+8.9}_{-7.3}\pm0.5$ & $86\pm2\pm14$ & $9.1\pm5.0^{+6.9}_{-4.1}$ & $132\pm10\pm15$ & 127 \\
17 & 600--850 & 600--1200 & 2--3 & 1 & $162\pm17\pm4$ & $710\pm10\pm120$ & $2.3\pm1.3^{+1.8}_{-1.0}$ & $880\pm20\pm110$ & 728 \\
18 & 600--850 & $\geq$1200 & 2--3 & 1 & $2.7^{+3.5}_{-1.7}\pm0.1$ & $29.5\pm1.3\pm4.8$ & $0.12\pm0.10^{+0.09}_{-0.02}$ & $32.3^{+3.8}_{-2.1}\pm4.8$ & 31 \\
19 & $\geq$850 & 850--1700 & 2--3 & 1 & $8.7^{+5.2}_{-3.5}\pm0.2$ & $124\pm3\pm22$ & $0.10\pm0.07^{+0.07}_{-0.02}$ & $133\pm5\pm22$ & 112 \\
20 & $\geq$850 & $\geq$1700 & 2--3 & 1 & $0.0^{+3.6}_{-0.0}\pm0.0$ & $6.0\pm0.7\pm1.1$ & $0.03^{+0.04+0.02}_{-0.03-0.00}$ & $6.0^{+3.6}_{-0.6}\pm1.1$ & 5 \\
21 & 300--350 & 300--600 & 2--3 & $\geq$2 & $706\pm37\pm13$ & $940\pm2\pm290$ & $66^{+68+72}_{-66- 0}$ & $1710\pm80\pm290$ & 1787 \\
22 & 300--350 & 600--1200 & 2--3 & $\geq$2 & $96\pm13\pm1$ & $71\pm1\pm22$ & $19\pm11^{+19}_{-8}$ & $186\pm18^{+29}_{-23}$ & 148 \\
23 & 300--350 & $\geq$1200 & 2--3 & $\geq$2 & $3.5^{+4.7}_{-2.3}\pm0.1$ & $4.4\pm0.2\pm1.4$ & $2.2\pm1.3^{+2.1}_{-0.9}$ & $10.2^{+4.8+2.5}_{-2.6-1.7}$ & 11 \\
24 & 350--600 & 350--600 & 2--3 & $\geq$2 & $362\pm27\pm14$ & $810\pm2\pm250$ & $13\pm8^{+13}_{-5}$ & $1190\pm30\pm250$ & 1159 \\
25 & 350--600 & 600--1200 & 2--3 & $\geq$2 & $166\pm18\pm5$ & $201\pm1\pm61$ & $5.1\pm3.3^{+5.1}_{-1.8}$ & $373\pm18\pm62$ & 322 \\
26 & 350--600 & $\geq$1200 & 2--3 & $\geq$2 & $6.0^{+4.8}_{-2.9}\pm0.1$ & $9.9\pm0.2\pm3.1$ & $1.5\pm0.9^{+1.5}_{-0.6}$ & $17.5^{+4.9+3.4}_{-3.1-3.1}$ & 13 \\
27 & 600--850 & 600--1200 & 2--3 & $\geq$2 & $17.5^{+7.6}_{-5.6}\pm0.3$ & $72\pm1\pm22$ & $0.09\pm0.09^{+0.09}_{-0.00}$ & $89\pm7\pm22$ & 50 \\
28 & 600--850 & $\geq$1200 & 2--3 & $\geq$2 & $0.0^{+2.9}_{-0.0}\pm0.0$ & $3.4\pm0.1\pm1.0$ & $0.08\pm0.08^{+0.07}_{-0.00}$ & $3.4^{+2.9}_{-0.2}\pm1.0$ & 4 \\
29 & $\geq$850 & 850--1700 & 2--3 & $\geq$2 & $0.0^{+4.4}_{-0.0}\pm0.0$ & $12.5\pm0.3\pm4.0$ & $0.09\pm0.07^{+0.09}_{-0.02}$ & $12.6^{+4.5}_{-0.3}\pm4.0$ & 9 \\
30 & $\geq$850 & $\geq$1700 & 2--3 & $\geq$2 & $0.0^{+3.7}_{-0.0}\pm0.0$ & $0.68\pm0.07\pm0.22$ & $0.04\pm0.04^{+0.03}_{-0.00}$ & $0.7^{+3.7}_{-0.1}\pm0.2$ & 0 \\
\end{tabular}}
\end{table}

\begin{table}[h]
\renewcommand{\arraystretch}{1.25}
\centering
\caption{Observed number of events and pre-fit background predictions in the $4\leq\njets\leq5$ search bins. For the background predictions,
the first uncertainty is statistical and the second systematic.}
\cmsTable{
\label{tab:pre-fit-results-nj2}
\begin{tabular}{ cccccccccc }
\multirow{2}{*}{Bin} & \mht & \HT & \multirow{2}{*}{$\njets$} & \multirow{2}{*}{$\nbjets$} & Lost-lepton & \znn & QCD & Total & \multirow{2}{*}{Observed} \\[-1.5mm]
    & [{\GeVns}] & [{\GeVns}] & & & background & background & background & background & \\
\hline
31 & 300--350 & 300--600 & 4--5 & 0 & $8720\pm110\pm120$ & $13\,930\pm70\pm590$ & $630\pm350^{+410}_{-290}$ & $23\,280\pm370^{+740}_{-660}$ & 23\,241 \\
32 & 300--350 & 600--1200 & 4--5 & 0 & $2990\pm48\pm54$ & $3960\pm40\pm150$ & $490\pm260\pm230$ & $7440\pm270\pm280$ & 7277 \\
33 & 300--350 & $\geq$1200 & 4--5 & 0 & $216\pm14\pm5$ & $317\pm12\pm18$ & $230\pm120\pm100$ & $760\pm120\pm100$ & 726 \\
34 & 350--600 & 350--600 & 4--5 & 0 & $5230\pm90\pm160$ & $11\,410\pm70\pm450$ & $180\pm100\pm80$ & $16\,820\pm150\pm490$ & 16\,720 \\
35 & 350--600 & 600--1200 & 4--5 & 0 & $4654\pm59\pm68$ & $9000\pm60\pm350$ & $210\pm110\pm100$ & $13\,870\pm140\pm370$ & 13\,837 \\
36 & 350--600 & $\geq$1200 & 4--5 & 0 & $364\pm17\pm6$ & $680\pm17\pm37$ & $104\pm56\pm45$ & $1148\pm61\pm59$ & 1141 \\
37 & 600--850 & 600--1200 & 4--5 & 0 & $428\pm19\pm9$ & $1592\pm25\pm94$ & $5.1\pm2.8\pm2.3$ & $2025\pm32\pm94$ & 2028 \\
38 & 600--850 & $\geq$1200 & 4--5 & 0 & $72.2^{+8.1}_{-7.3}\pm1.1$ & $225\pm10\pm14$ & $1.9\pm1.1\pm0.8$ & $299\pm13\pm14$ & 291 \\
39 & $\geq$850 & 850--1700 & 4--5 & 0 & $42.4\pm6.9\pm0.8$ & $351\pm13\pm32$ & $0.13\pm0.09\pm0.5$ & $393\pm15\pm32$ & 360 \\
40 & $\geq$850 & $\geq$1700 & 4--5 & 0 & $6.1^{+3.3}_{-2.3}\pm0.1$ & $38.4\pm4.2\pm4.4$ & $0.06\pm0.05^{+0.02}_{-0.01}$ & $44.6^{+5.5}_{-4.6}\pm4.4$ & 51 \\
41 & 300--350 & 300--600 & 4--5 & 1 & $4217\pm69\pm77$ & $2850\pm15\pm450$ & $220\pm120^{+200}_{-100}$ & $7290\pm140\pm480$ & 7157 \\
42 & 300--350 & 600--1200 & 4--5 & 1 & $1389\pm35\pm23$ & $850\pm10\pm130$ & $260\pm140^{+210}_{-120}$ & $2500\pm150^{+250}_{-180}$ & 2387 \\
43 & 300--350 & $\geq$1200 & 4--5 & 1 & $93\pm10\pm3$ & $69\pm3\pm11$ & $93\pm50^{+71}_{-43}$ & $255\pm51^{+72}_{-44}$ & 229 \\
44 & 350--600 & 350--600 & 4--5 & 1 & $2068\pm50\pm41$ & $2330\pm10\pm370$ & $64\pm35^{+49}_{-29}$ & $4460\pm60\pm370$ & 4317 \\
45 & 350--600 & 600--1200 & 4--5 & 1 & $1777\pm40\pm29$ & $1910\pm10\pm300$ & $92\pm50^{+73}_{-42}$ & $3780\pm70\pm300$ & 3822 \\
46 & 350--600 & $\geq$1200 & 4--5 & 1 & $112\pm11\pm3$ & $148\pm4\pm24$ & $45\pm24^{+34}_{-21}$ & $305\pm27^{+42}_{-32}$ & 350 \\
47 & 600--850 & 600--1200 & 4--5 & 1 & $107\pm11\pm3$ & $332\pm5\pm54$ & $1.8\pm1.1^{+1.5}_{-0.8}$ & $441\pm12\pm54$ & 388 \\
48 & 600--850 & $\geq$1200 & 4--5 & 1 & $23.1^{+5.5}_{-4.6}\pm0.4$ & $48.6\pm2.2\pm8.0$ & $0.78\pm0.51^{+0.59}_{-0.27}$ & $72.5\pm5.5\pm8.1$ & 74 \\
49 & $\geq$850 & 850--1700 & 4--5 & 1 & $9.4^{+4.0}_{-3.0}\pm0.3$ & $73\pm3\pm13$ & $0.12\pm0.09^{+0.09}_{-0.03}$ & $82\pm5\pm13$ & 73 \\
50 & $\geq$850 & $\geq$1700 & 4--5 & 1 & $1.0^{+2.3}_{-0.8}\pm0.0$ & $8.3\pm1.0\pm1.6$ & $0.03^{+0.04+0.02}_{-0.03-0.00}$ & $9.4^{+2.5}_{-1.2}\pm1.6$ & 14 \\
51 & 300--350 & 300--600 & 4--5 & 2 & $1806\pm49\pm30$ & $468\pm2\pm79$ & $68\pm45^{+74}_{-24}$ & $2340\pm70^{+110}_{-90}$ & 2505 \\
52 & 300--350 & 600--1200 & 4--5 & 2 & $687\pm26\pm10$ & $144\pm1\pm24$ & $71\pm39^{+70}_{-32}$ & $902\pm47^{+75}_{-41}$ & 864 \\
53 & 300--350 & $\geq$1200 & 4--5 & 2 & $34.0^{+7.4}_{-6.2}\pm0.7$ & $12.0\pm0.4\pm2.1$ & $24\pm13^{+23}_{-11}$ & $70\pm14^{+23}_{-11}$ & 72 \\
54 & 350--600 & 350--600 & 4--5 & 2 & $820\pm35\pm20$ & $381\pm2\pm64$ & $17\pm10^{+17}_{-7}$ & $1218\pm36\pm68$ & 1208 \\
55 & 350--600 & 600--1200 & 4--5 & 2 & $794\pm29\pm12$ & $324\pm2\pm54$ & $23\pm13^{+23}_{-10}$ & $1141\pm32\pm58$ & 1180 \\
56 & 350--600 & $\geq$1200 & 4--5 & 2 & $47.8^{+8.2+1.1}_{-7.2-1.1}$ & $25.6^{+0.6+4.4}_{-0.6-4.4}$ & $12^{+ 7+12}_{- 7- 5}$ & $85^{+11+12}_{-10- 7}$ & 78 \\
57 & 600--850 & 600--1200 & 4--5 & 2 & $37.1^{+8.0}_{-6.7}\pm0.7$ & $55.5\pm0.9\pm9.6$ & $0.45\pm0.30^{+0.45}_{-0.16}$ & $93.1^{+8.0}_{-6.8}\pm9.7$ & 98 \\
58 & 600--850 & $\geq$1200 & 4--5 & 2 & $8.8^{+5.3}_{-3.5}\pm0.1$ & $8.4\pm0.4\pm1.5$ & $0.20\pm0.18^{+0.19}_{-0.02}$ & $17.4^{+5.3}_{-3.6}\pm1.5$ & 15 \\
59 & $\geq$850 & 850--1700 & 4--5 & 2 & $1.2^{+2.8}_{-1.0}\pm0.0$ & $12.0\pm0.4\pm2.2$ & $0.09\pm0.07^{+0.09}_{-0.02}$ & $13.3^{+2.8}_{-1.1}\pm2.2$ & 15 \\
60 & $\geq$850 & $\geq$1700 & 4--5 & 2 & $0.0^{+2.6}_{-0.0}\pm0.0$ & $1.44\pm0.16\pm0.28$ & $0.04\pm0.04^{+0.03}_{-0.00}$ & $1.5^{+2.6}_{-0.1}\pm0.3$ & 1 \\
61 & 300--350 & 300--600 & 4--5 & $\geq$3 & $147\pm15\pm2$ & $40\pm0\pm14$ & $4.4\pm4.2^{+6.1}_{-0.2}$ & $192\pm15\pm15$ & 222 \\
62 & 300--350 & 600--1200 & 4--5 & $\geq$3 & $76.7\pm9.0\pm1.3$ & $13.5\pm0.1\pm4.8$ & $ 9\pm6^{+12}_{-3}$ & $99\pm10^{+13}_{-6}$ & 92 \\
63 & 300--350 & $\geq$1200 & 4--5 & $\geq$3 & $5.8^{+3.9}_{-2.5}\pm0.1$ & $1.14\pm0.04\pm0.41$ & $3.7\pm2.2^{+4.7}_{-1.5}$ & $10.6^{+4.5+4.7}_{-3.3-1.5}$ & 5 \\
64 & 350--600 & 350--600 & 4--5 & $\geq$3 & $73\pm11\pm1$ & $33\pm0\pm12$ & $1.2\pm1.1^{+1.6}_{-0.1}$ & $107\pm11\pm12$ & 111 \\
65 & 350--600 & 600--1200 & 4--5 & $\geq$3 & $92^{+11+ 2}_{-10- 2}$ & $30^{+ 0+11}_{- 0-11}$ & $3.2^{+2.0+4.2}_{-2.0-1.2}$ & $125^{+11+12}_{-10-11}$ & 138 \\
66 & 350--600 & $\geq$1200 & 4--5 & $\geq$3 & $5.0^{+3.4}_{-2.2}\pm0.1$ & $2.45\pm0.06\pm0.87$ & $1.8\pm1.2^{+2.3}_{-0.6}$ & $9.3^{+3.6+2.5}_{-2.5-1.1}$ & 5 \\
67 & 600--850 & 600--1200 & 4--5 & $\geq$3 & $1.3^{+2.9}_{-1.1}\pm0.0$ & $4.9\pm0.1\pm1.8$ & $0.10^{+0.12+0.13}_{-0.10-0.00}$ & $6.3^{+2.9}_{-1.1}\pm1.8$ & 5 \\
68 & 600--850 & $\geq$1200 & 4--5 & $\geq$3 & $0.0^{+2.6}_{-0.0}\pm0.0$ & $0.79\pm0.04\pm0.28$ & $0.10^{+0.12+0.13}_{-0.10-0.00}$ & $0.9^{+2.6}_{-0.1}\pm0.3$ & 0 \\
69 & $\geq$850 & 850--1700 & 4--5 & $\geq$3 & $0.0^{+3.2}_{-0.0}\pm0.0$ & $1.05\pm0.04\pm0.38$ & $0.10\pm0.09^{+0.13}_{-0.02}$ & $1.2^{+3.2}_{-0.1}\pm0.4$ & 1 \\
70 & $\geq$850 & $\geq$1700 & 4--5 & $\geq$3 & $0.0^{+2.3}_{-0.0}\pm0.0$ & $0.13\pm0.01\pm0.05$ & $0.04^{+0.05+0.05}_{-0.04-0.00}$ & $0.2^{+2.3}_{-0.0}\pm0.1$ & 0 \\
\end{tabular}}
\end{table}

\begin{table}[h]
\renewcommand{\arraystretch}{1.25}
\centering
\caption{Observed number of events and pre-fit background predictions in the $6\leq\njets\leq7$ search bins. For the background predictions,
the first uncertainty is statistical and the second systematic.}
\label{tab:pre-fit-results-nj3}
\cmsTable{
\begin{tabular}{ cccccccccc }
\multirow{2}{*}{Bin} & \mht & \HT & \multirow{2}{*}{$\njets$} & \multirow{2}{*}{$\nbjets$} & Lost-lepton & \znn & QCD & Total & \multirow{2}{*}{Observed} \\[-1.5mm]
    & [{\GeVns}] & [{\GeVns}] & & & background & background & background & background & \\
\hline
71 & 300--350 & 300--600 & 6--7 & 0 & $686\pm29\pm11$ & $761\pm17\pm63$ & $144\pm83^{+92}_{-61}$ & $1590\pm90^{+110}_{-90}$ & 1480 \\
72 & 300--350 & 600--1200 & 6--7 & 0 & $967\pm25\pm14$ & $873\pm18\pm65$ & $280\pm140\pm130$ & $2110\pm140\pm150$ & 1993 \\
73 & 300--350 & $\geq$1200 & 6--7 & 0 & $121.5\pm8.8\pm2.8$ & $116.8\pm7.3\pm9.2$ & $172\pm86\pm74$ & $410\pm87\pm75$ & 362 \\
74 & 350--600 & 350--600 & 6--7 & 0 & $353\pm21\pm8$ & $514\pm14\pm40$ & $33\pm20\pm14$ & $901\pm32\pm44$ & 847 \\
75 & 350--600 & 600--1200 & 6--7 & 0 & $1219\pm28\pm28$ & $1540\pm20\pm110$ & $130\pm65\pm63$ & $2890\pm80\pm130$ & 2842 \\
76 & 350--600 & $\geq$1200 & 6--7 & 0 & $208\pm11\pm4$ & $258\pm11\pm18$ & $81\pm40\pm35$ & $547\pm43\pm39$ & 553 \\
77 & 600--850 & 600--1200 & 6--7 & 0 & $76.1^{+1.0}_{-1.0}\pm1.0$ & $182\pm8\pm15$ & $1.70\pm0.88\pm0.81$ & $259\pm11\pm15$ & 245 \\
78 & 600--850 & $\geq$1200 & 6--7 & 0 & $29.7\pm4.2\pm0.5$ & $72.8\pm5.6\pm5.7$ & $2.3\pm1.2\pm1.0$ & $104.8^{+7.4}_{-6.7}\pm5.8$ & 122 \\
79 & $\geq$850 & 850--1700 & 6--7 & 0 & $18.5\pm3.5\pm0.3$ & $35.2\pm3.6\pm3.8$ & $0.10\pm0.07^{+0.04}_{-0.02}$ & $53.8^{+5.4}_{-4.7}\pm3.9$ & 55 \\
80 & $\geq$850 & $\geq$1700 & 6--7 & 0 & $4.3^{+2.0}_{-1.4}\pm0.2$ & $12.7\pm2.3\pm1.9$ & $0.05\pm0.04^{+0.02}_{-0.01}$ & $17.0^{+3.2}_{-2.6}\pm1.9$ & 20 \\
81 & 300--350 & 300--600 & 6--7 & 1 & $675\pm25\pm12$ & $248\pm6\pm45$ & $42\pm22^{+27}_{-20}$ & $965\pm34\pm53$ & 946 \\
82 & 300--350 & 600--1200 & 6--7 & 1 & $950\pm26\pm15$ & $289\pm6\pm52$ & $115\pm58\pm55$ & $1355\pm63\pm77$ & 1282 \\
83 & 300--350 & $\geq$1200 & 6--7 & 1 & $105.6^{+9.1}_{-8.4}\pm2.7$ & $39.3\pm2.5\pm7.1$ & $57\pm28\pm24$ & $201\pm30\pm26$ & 197 \\
84 & 350--600 & 350--600 & 6--7 & 1 & $252\pm16\pm5$ & $168\pm5\pm30$ & $9.5\pm5.0\pm4.3$ & $429\pm18\pm31$ & 425 \\
85 & 350--600 & 600--1200 & 6--7 & 1 & $1050\pm28\pm19$ & $510\pm8\pm91$ & $53\pm27\pm26$ & $1614\pm39\pm96$ & 1521 \\
86 & 350--600 & $\geq$1200 & 6--7 & 1 & $155\pm11\pm4$ & $86\pm4\pm15$ & $26\pm13\pm11$ & $268\pm17\pm20$ & 269 \\
87 & 600--850 & 600--1200 & 6--7 & 1 & $34.7^{+5.4}_{-4.8}\pm0.6$ & $60\pm3\pm11$ & $0.69\pm0.41^{+0.33}_{-0.28}$ & $95\pm6\pm11$ & 90 \\
88 & 600--850 & $\geq$1200 & 6--7 & 1 & $25.9\pm4.3\pm0.4$ & $24.4\pm1.9\pm4.4$ & $0.59\pm0.34\pm0.25$ & $50.9^{+5.1}_{-4.4}\pm4.4$ & 49 \\
89 & $\geq$850 & 850--1700 & 6--7 & 1 & $7.9^{+2.9}_{-2.2}\pm0.1$ & $11.5\pm1.1\pm2.3$ & $0.05\pm0.04^{+0.02}_{-0.00}$ & $19.4^{+3.2}_{-2.5}\pm2.3$ & 17 \\
90 & $\geq$850 & $\geq$1700 & 6--7 & 1 & $1.5^{+2.0}_{-1.0}\pm0.0$ & $4.29^{+0.85}_{-0.72}\pm0.95$ & $0.04^{+0.05+0.02}_{-0.04-0.00}$ & $5.9^{+2.2}_{-1.2}\pm0.9$ & 7 \\
91 & 300--350 & 300--600 & 6--7 & 2 & $376\pm19\pm8$ & $64\pm2\pm13$ & $9.8\pm5.5^{+6.3}_{-4.2}$ & $450\pm20\pm16$ & 450 \\
92 & 300--350 & 600--1200 & 6--7 & 2 & $693\pm23\pm10$ & $76\pm2\pm15$ & $34\pm17\pm16$ & $803\pm28\pm25$ & 797 \\
93 & 300--350 & $\geq$1200 & 6--7 & 2 & $46.7^{+6.4}_{-5.7}\pm0.7$ & $10.5\pm0.7\pm2.1$ & $18.7\pm9.4\pm8.1$ & $76\pm11\pm8$ & 84 \\
94 & 350--600 & 350--600 & 6--7 & 2 & $120\pm12\pm2$ & $43.6\pm1.2\pm8.9$ & $2.1\pm1.2\pm0.9$ & $165\pm12\pm9$ & 188 \\
95 & 350--600 & 600--1200 & 6--7 & 2 & $661\pm23\pm11$ & $134\pm2\pm27$ & $14.6\pm7.5\pm7.0$ & $809\pm24\pm30$ & 762 \\
96 & 350--600 & $\geq$1200 & 6--7 & 2 & $66.6\pm7.3\pm2.2$ & $22.8\pm0.9\pm4.6$ & $7.5\pm3.8\pm3.2$ & $96.9\pm8.3\pm6.0$ & 106 \\
97 & 600--850 & 600--1200 & 6--7 & 2 & $19.3^{+4.7}_{-3.9}\pm0.3$ & $15.7\pm0.7\pm3.2$ & $0.15\pm0.10\pm0.06$ & $35.2\pm4.3\pm3.2$ & 32 \\
98 & 600--850 & $\geq$1200 & 6--7 & 2 & $8.0^{+3.2}_{-2.4}\pm0.2$ & $6.5\pm0.5\pm1.3$ & $0.09\pm0.07^{+0.04}_{-0.01}$ & $14.5^{+3.3}_{-2.4}\pm1.3$ & 14 \\
99 & $\geq$850 & 850--1700 & 6--7 & 2 & $1.8^{+1.7}_{-1.0}\pm0.0$ & $2.98\pm0.30\pm0.65$ & $0.05\pm0.04^{+0.02}_{-0.01}$ & $4.8^{+1.8}_{-1.0}\pm0.7$ & 9 \\
100 & $\geq$850 & $\geq$1700 & 6--7 & 2 & $0.5^{+1.2}_{-0.4}\pm0.0$ & $1.15^{+0.23}_{-0.19}\pm0.28$ & $0.02\pm0.02^{+0.01}_{-0.00}$ & $1.7^{+1.2}_{-0.5}\pm0.3$ & 1 \\
101 & 300--350 & 300--600 & 6--7 & $\geq$3 & $67.8^{+8.8}_{-7.9}\pm1.6$ & $8.8\pm0.2\pm3.7$ & $1.4\pm1.0^{+0.9}_{-0.4}$ & $78.0\pm8.5\pm4.0$ & 86 \\
102 & 300--350 & 600--1200 & 6--7 & $\geq$3 & $136\pm11\pm2$ & $10.5\pm0.2\pm4.3$ & $7.4\pm4.2^{+3.6}_{-3.2}$ & $154\pm11\pm6$ & 167 \\
103 & 300--350 & $\geq$1200 & 6--7 & $\geq$3 & $15.7^{+4.1}_{-3.4}\pm0.2$ & $1.44\pm0.09\pm0.59$ & $3.9\pm2.2\pm1.7$ & $21.1\pm4.3\pm1.8$ & 16 \\
104 & 350--600 & 350--600 & 6--7 & $\geq$3 & $20.6^{+5.3}_{-4.3}\pm0.5$ & $6.0\pm0.2\pm2.5$ & $0.68\pm0.62^{+0.31}_{-0.07}$ & $27.2^{+5.4}_{-4.4}\pm2.5$ & 28 \\
105 & 350--600 & 600--1200 & 6--7 & $\geq$3 & $137\pm11\pm4$ & $18.5\pm0.3\pm7.6$ & $2.8\pm1.6\pm1.3$ & $158\pm11\pm9$ & 115 \\
106 & 350--600 & $\geq$1200 & 6--7 & $\geq$3 & $15.4^{+4.4}_{-3.5}\pm0.6$ & $3.1\pm0.1\pm1.3$ & $1.7\pm1.0^{+0.8}_{-0.7}$ & $20.2^{+4.5}_{-3.7}\pm1.6$ & 23 \\
107 & 600--850 & 600--1200 & 6--7 & $\geq$3 & $4.1^{+2.5}_{-1.7}\pm0.0$ & $2.16\pm0.10\pm0.89$ & $0.05^{+0.06+0.02}_{-0.05-0.00}$ & $6.3^{+2.5}_{-1.7}\pm0.9$ & 6 \\
108 & 600--850 & $\geq$1200 & 6--7 & $\geq$3 & $2.1^{+2.0}_{-1.1}\pm0.0$ & $0.89\pm0.07\pm0.37$ & $0.07\pm0.06^{+0.03}_{-0.01}$ & $3.0^{+2.0}_{-1.1}\pm0.4$ & 2 \\
109 & $\geq$850 & 850--1700 & 6--7 & $\geq$3 & $0.0^{+1.2}_{-0.0}\pm0.0$ & $0.41\pm0.04\pm0.17$ & $0.05\pm0.04^{+0.02}_{-0.01}$ & $0.5^{+1.2}_{-0.1}\pm0.2$ & 1 \\
110 & $\geq$850 & $\geq$1700 & 6--7 & $\geq$3 & $0.0^{+1.9}_{-0.0}\pm0.0$ & $0.16\pm0.03\pm0.07$ & $0.02\pm0.02^{+0.01}_{-0.00}$ & $0.2^{+1.9}_{-0.0}\pm0.1$ & 1 \\
\end{tabular}}
\end{table}

\begin{table}[h]
\renewcommand{\arraystretch}{1.25}
\centering
\caption{Observed number of events and pre-fit background predictions in the $8\leq\njets\leq9$ search bins. For the background predictions,
the first uncertainty is statistical and the second systematic.}
\cmsTable{
\label{tab:pre-fit-results-nj4}
\begin{tabular}{ cccccccccc }
\multirow{2}{*}{Bin} & \mht & \HT & \multirow{2}{*}{$\njets$} & \multirow{2}{*}{$\nbjets$} & Lost-lepton & \znn & QCD & Total & \multirow{2}{*}{Observed} \\[-1.5mm]
    & [{\GeVns}] & [{\GeVns}] & & & background & background & background & background & \\
\hline
111 & 300--350 & 600--1200 & 8--9 & 0 & $139.5\pm9.5\pm1.9$ & $60.0\pm4.6\pm9.7$ & $58\pm29\pm28$ & $258\pm31\pm30$ & 245 \\
112 & 300--350 & $\geq$1200 & 8--9 & 0 & $31.0\pm4.3\pm1.1$ & $25.1\pm3.5\pm2.7$ & $57\pm28\pm24$ & $113\pm29\pm25$ & 88 \\
113 & 350--600 & 600--1200 & 8--9 & 0 & $136.1\pm9.3\pm1.7$ & $123\pm7\pm14$ & $30\pm15\pm14$ & $289\pm19\pm20$ & 280 \\
114 & 350--600 & $\geq$1200 & 8--9 & 0 & $49.9\pm5.3\pm0.9$ & $52.2\pm4.8\pm5.4$ & $27\pm14\pm12$ & $129\pm16\pm13$ & 104 \\
115 & 600--850 & 600--1200 & 8--9 & 0 & $6.6^{+2.3}_{-1.8}\pm0.2$ & $13.9\pm2.4\pm1.5$ & $0.37\pm0.21\pm0.17$ & $20.9^{+3.5}_{-2.9}\pm1.5$ & 28 \\
116 & 600--850 & $\geq$1200 & 8--9 & 0 & $6.1^{+2.1}_{-1.6}\pm0.1$ & $12.9\pm2.4\pm1.6$ & $0.79\pm0.44\pm0.34$ & $19.7\pm3.0\pm1.6$ & 22 \\
117 & $\geq$850 & 850--1700 & 8--9 & 0 & $1.1^{+1.1}_{-0.6}\pm0.0$ & $4.1^{+1.5}_{-1.2}\pm0.6$ & $0.06\pm0.04^{+0.03}_{-0.02}$ & $5.3^{+1.9}_{-1.3}\pm0.6$ & 2 \\
118 & $\geq$850 & $\geq$1700 & 8--9 & 0 & $1.5^{+1.2}_{-0.7}\pm0.1$ & $2.2^{+1.3}_{-0.9}\pm0.3$ & $0.02\pm0.02^{+0.01}_{-0.00}$ & $3.7^{+1.8}_{-1.1}\pm0.3$ & 1 \\
119 & 300--350 & 600--1200 & 8--9 & 1 & $183\pm11\pm3$ & $37\pm3\pm11$ & $27\pm13\pm13$ & $247\pm18\pm17$ & 229 \\
120 & 300--350 & $\geq$1200 & 8--9 & 1 & $43.8\pm5.3\pm0.7$ & $13.8\pm1.9\pm3.8$ & $24\pm12\pm10$ & $82\pm13\pm11$ & 68 \\
121 & 350--600 & 600--1200 & 8--9 & 1 & $176\pm11\pm3$ & $75\pm4\pm21$ & $10.9\pm5.5\pm5.3$ & $262\pm13\pm22$ & 224 \\
122 & 350--600 & $\geq$1200 & 8--9 & 1 & $68.4\pm6.5\pm1.2$ & $29.5\pm 2.7\pm8.1$ & $9.8\pm5.0\pm4.2$ & $107.8\pm8.5\pm9.3$ & 90 \\
123 & 600--850 & 600--1200 & 8--9 & 1 & $3.4^{+2.0}_{-1.4}\pm0.2$ & $8.7\pm1.5\pm2.4$ & $0.10\pm0.08^{+0.05}_{-0.02}$ & $12.2\pm2.3\pm2.4$ & 7 \\
124 & 600--850 & $\geq$1200 & 8--9 & 1 & $8.3^{+2.8}_{-2.1}\pm0.1$ & $8.1\pm1.5\pm2.3$ & $0.31\pm0.18\pm0.12$ & $16.7^{+3.2}_{-2.6}\pm2.3$ & 15 \\
125 & $\geq$850 & 850--1700 & 8--9 & 1 & $0.0^{+1.2}_{-0.0}\pm0.0$ & $2.08^{+0.79}_{-0.59}\pm0.61$ & $0.05\pm0.04^{+0.02}_{-0.01}$ & $2.1^{+1.5}_{-0.6}\pm0.6$ & 2 \\
126 & $\geq$850 & $\geq$1700 & 8--9 & 1 & $1.0^{+1.3}_{-0.7}\pm0.0$ & $1.35^{+0.81}_{-0.54}\pm0.40$ & $0.02\pm0.02^{+0.01}_{-0.00}$ & $2.4^{+1.5}_{-0.8}\pm0.4$ & 2 \\
127 & 300--350 & 600--1200 & 8--9 & 2 & $169\pm11\pm4$ & $11.0\pm0.9\pm4.1$ & $9.5\pm4.9\pm4.6$ & $190\pm12\pm7$ & 193 \\
128 & 300--350 & $\geq$1200 & 8--9 & 2 & $28.9\pm4.4\pm0.5$ & $5.5\pm0.8\pm1.9$ & $10.1\pm5.1\pm4.4$ & $44.6\pm6.8\pm4.8$ & 53 \\
129 & 350--600 & 600--1200 & 8--9 & 2 & $146\pm10\pm2$ & $23.1\pm1.3\pm8.1$ & $4.5\pm2.4\pm2.1$ & $174\pm11\pm9$ & 158 \\
130 & 350--600 & $\geq$1200 & 8--9 & 2 & $42.9\pm5.3\pm0.9$ & $11.0\pm1.1\pm3.9$ & $4.1\pm2.1\pm1.8$ & $58.0^{+6.1}_{-5.5}\pm4.4$ & 74 \\
131 & 600--850 & 600--1200 & 8--9 & 2 & $3.6^{+2.4}_{-1.6}\pm0.2$ & $2.52\pm0.44\pm0.89$ & $0.09\pm0.08^{+0.04}_{-0.01}$ & $6.2^{+2.5}_{-1.6}\pm0.9$ & 7 \\
132 & 600--850 & $\geq$1200 & 8--9 & 2 & $8.0^{+2.9}_{-2.2}\pm0.3$ & $2.30\pm0.42\pm0.82$ & $0.08^{+0.09+0.04}_{-0.08-0.00}$ & $10.4^{+3.0}_{-2.3}\pm0.9$ & 9 \\
133 & $\geq$850 & 850--1700 & 8--9 & 2 & $0.7^{+1.6}_{-0.6}\pm0.0$ & $0.96^{+0.37}_{-0.27}\pm0.35$ & $0.05\pm0.04^{+0.02}_{-0.01}$ & $1.7^{+1.6}_{-0.7}\pm0.3$ & 0 \\
134 & $\geq$850 & $\geq$1700 & 8--9 & 2 & $2.5^{+3.3}_{-1.7}\pm0.1$ & $0.40^{+0.24}_{-0.16}\pm0.15$ & $0.02\pm0.02^{+0.01}_{-0.00}$ & $2.9^{+3.4}_{-1.7}\pm0.2$ & 2 \\
135 & 300--350 & 600--1200 & 8--9 & $\geq$3 & $46.8^{+6.1}_{-5.5}\pm0.7$ & $3.8\pm0.3\pm2.3$ & $3.7\pm2.6^{+1.8}_{-1.2}$ & $54.3\pm6.3\pm2.9$ & 57 \\
136 & 300--350 & $\geq$1200 & 8--9 & $\geq$3 & $17.3^{+4.0}_{-3.3}\pm0.5$ & $1.26\pm0.17\pm0.76$ & $3.6\pm2.0\pm1.5$ & $22.2^{+4.4}_{-3.8}\pm1.8$ & 17 \\
137 & 350--600 & 600--1200 & 8--9 & $\geq$3 & $44.4\pm5.6\pm1.0$ & $7.5\pm0.4\pm4.6$ & $1.31\pm0.81^{+0.63}_{-0.51}$ & $53.2\pm5.7\pm4.7$ & 36 \\
138 & 350--600 & $\geq$1200 & 8--9 & $\geq$3 & $15.2^{+3.6}_{-2.9}\pm0.3$ & $2.8\pm0.3\pm1.7$ & $1.17\pm0.68\pm0.50$ & $19.2\pm3.3\pm1.8$ & 23 \\
139 & 600--850 & 600--1200 & 8--9 & $\geq$3 & $0.0^{+1.7+0.0}_{-0.0-0.0}$ & $0.88^{+0.16+0.54}_{-0.14-0.53}$ & $0.04^{+0.04+0.02}_{-0.04-0.00}$ & $0.9^{+1.7+0.5}_{-0.1-0.5}$ & 2 \\
140 & 600--850 & $\geq$1200 & 8--9 & $\geq$3 & $2.7^{+2.2}_{-1.3}\pm0.1$ & $0.83\pm0.15\pm0.51$ & $0.05\pm0.05^{+0.02}_{-0.00}$ & $3.6^{+2.2}_{-1.3}\pm0.5$ & 2 \\
141 & $\geq$850 & 850--1700 & 8--9 & $\geq$3 & $0.8^{+2.0}_{-0.7}\pm0.0$ & $0.18^{+0.07}_{-0.05}\pm0.11$ & $0.05\pm0.04^{+0.02}_{-0.01}$ & $1.1^{+2.0}_{-0.7}\pm0.1$ & 0 \\
142 & $\geq$850 & $\geq$1700 & 8--9 & $\geq$3 & $0.0^{+1.8}_{-0.0}\pm0.0$ & $0.14^{+0.08}_{-0.05}\pm0.08$ & $0.02\pm0.02^{+0.01}_{-0.00}$ & $0.2^{+1.8}_{-0.1}\pm0.1$ & 0 \\
\end{tabular}}
\end{table}

\begin{table}[h]
\renewcommand{\arraystretch}{1.25}
\centering
\caption{Observed number of events and pre-fit background predictions in the $\njets\geq10$ search bins. For the background predictions,
the first uncertainty is statistical and the second systematic.}
\cmsTable{
\label{tab:pre-fit-results-nj5}
\begin{tabular}{ cccccccccc }
\multirow{2}{*}{Bin} & \mht & \HT & \multirow{2}{*}{$\njets$} & \multirow{2}{*}{$\nbjets$} & Lost-lepton & \znn & QCD & Total & \multirow{2}{*}{Observed} \\[-1.5mm]
    & [{\GeVns}] & [{\GeVns}] & & & background & background & background & background & \\
\hline
143 & 300--350 & 600--1200 & $\geq$10 & 0 & $5.7^{+2.2}_{-1.7}\pm0.3$ & $2.9^{+1.3+0.6}_{-1.0-0.5}$ & $7.8\pm4.5^{+3.7}_{-3.3}$ & $16.4\pm5.0^{+3.8}_{-3.3}$ & 17 \\
144 & 300--350 & $\geq$1200 & $\geq$10 & 0 & $5.7^{+2.5}_{-1.8}\pm0.2$ & $2.5^{+1.5}_{-1.0}\pm0.3$ & $12.6\pm6.3\pm5.4$ & $20.8^{+7.0}_{-6.7}\pm5.4$ & 20 \\
145 & 350--600 & 600--1200 & $\geq$10 & 0 & $6.0^{+2.4}_{-1.8}\pm0.1$ & $4.2^{+1.6}_{-1.2}\pm0.6$ & $3.3\pm1.8\pm1.5$ & $13.6^{+3.4}_{-2.8}\pm1.6$ & 12 \\
146 & 350--600 & $\geq$1200 & $\geq$10 & 0 & $10.7^{+2.9}_{-2.3}\pm0.2$ & $6.5^{+2.1}_{-1.6}\pm0.9$ & $6.0\pm3.1\pm2.6$ & $23.2^{+4.7}_{-4.2}\pm2.8$ & 21 \\
147 & 600--850 & 600--1200 & $\geq$10 & 0 & $0.19^{+0.44}_{-0.17}\pm0.00$ & $0.36^{+0.84}_{-0.30}\pm0.05$ & $0.07\pm0.07^{+0.03}_{-0.00}$ & $0.63^{+0.95}_{-0.35}\pm0.05$ & 2 \\
148 & 600--850 & $\geq$1200 & $\geq$10 & 0 & $2.0^{+1.6}_{-1.0}\pm0.0$ & $1.5^{+1.2}_{-0.7}\pm0.2$ & $0.15\pm0.13^{+0.06}_{-0.02}$ & $3.6^{+2.0}_{-1.2}\pm0.2$ & 6 \\
149 & $\geq$850 & 850--1700 & $\geq$10 & 0 & $0.0^{+2.3}_{-0.0}\pm0.0$ & $0.00^{+0.64}_{-0.00}\pm0.00$ & $0.05\pm0.04^{+0.02}_{-0.01}$ & $0.0^{+2.4}_{-0.0}\pm0.0$ & 0 \\
150 & $\geq$850 & $\geq$1700 & $\geq$10 & 0 & $0.00^{+0.91}_{-0.00}\pm0.00$ & $0.42^{+0.96}_{-0.35}\pm0.07$ & $0.02\pm0.02^{+0.01}_{-0.00}$ & $0.4^{+1.3}_{-0.3}\pm0.1$ & 2 \\
151 & 300--350 & 600--1200 & $\geq$10 & 1 & $15.2^{+3.3}_{-2.8}\pm0.2$ & $1.24^{+0.56}_{-0.40}\pm0.90$ & $4.0\pm2.1\pm1.9$ & $20.4^{+4.0}_{-3.5}\pm2.1$ & 22 \\
152 & 300--350 & $\geq$1200 & $\geq$10 & 1 & $11.2^{+3.2}_{-2.6}\pm0.4$ & $1.05^{+0.63}_{-0.42}\pm0.76$ & $6.9\pm3.5\pm3.0$ & $19.2^{+4.8}_{-4.4}\pm3.1$ & 18 \\
153 & 350--600 & 600--1200 & $\geq$10 & 1 & $13.8^{+3.3}_{-2.7}\pm0.3$ & $1.8^{+0.7}_{-0.5}\pm1.3$ & $1.53\pm0.85^{+0.74}_{-0.68}$ & $17.1^{+3.5}_{-2.9}\pm1.5$ & 9 \\
154 & 350--600 & $\geq$1200 & $\geq$10 & 1 & $16.2^{+3.4}_{-2.9}\pm0.4$ & $2.7^{+0.9}_{-0.7}\pm2.0$ & $2.6\pm1.3\pm1.1$ & $21.5^{+3.8}_{-3.2}\pm2.3$ & 32 \\
155 & 600--850 & 600--1200 & $\geq$10 & 1 & $0.0^{+3.6}_{-0.0}\pm0.0$ & $0.15^{+0.35+0.11}_{-0.13-0.09}$ & $0.04\pm0.04^{+0.02}_{-0.00}$ & $0.2^{+3.6}_{-0.1}\pm0.1$ & 0 \\
156 & 600--850 & $\geq$1200 & $\geq$10 & 1 & $1.3^{+1.3}_{-0.7}\pm0.0$ & $0.61^{+0.49}_{-0.29}\pm0.44$ & $0.06\pm0.05^{+0.03}_{-0.01}$ & $2.0^{+1.4+0.5}_{-0.8-0.4}$ & 3 \\
157 & $\geq$850 & 850--1700 & $\geq$10 & 1 & $0.0^{+3.2}_{-0.0}\pm0.0$ & $0.00^{+0.27}_{-0.00}\pm0.00$ & $0.05\pm0.04^{+0.02}_{-0.01}$ & $0.0^{+3.2}_{-0.0}\pm0.0$ & 0 \\
158 & $\geq$850 & $\geq$1700 & $\geq$10 & 1 & $0.7^{+1.5}_{-0.6}\pm0.0$ & $0.18^{+0.41+0.13}_{-0.15-0.10}$ & $0.03^{+0.04+0.01}_{-0.03-0.00}$ & $0.9^{+1.6}_{-0.6}\pm0.1$ & 1 \\
159 & 300--350 & 600--1200 & $\geq$10 & 2 & $13.1^{+3.2}_{-2.6}\pm0.3$ & $0.38^{+0.18+0.42}_{-0.13-0.36}$ & $2.1\pm1.5^{+1.0}_{-0.6}$ & $15.5^{+3.5+1.1}_{-3.0-0.8}$ & 15 \\
160 & 300--350 & $\geq$1200 & $\geq$10 & 2 & $10.8^{+3.0}_{-2.4}\pm0.4$ & $0.33^{+0.19+0.36}_{-0.13-0.30}$ & $3.3\pm1.7\pm1.4$ & $14.4^{+3.5}_{-3.0}\pm1.5$ & 11 \\
161 & 350--600 & 600--1200 & $\geq$10 & 2 & $18.2^{+3.8}_{-3.2}\pm0.3$ & $0.55^{+0.21+0.60}_{-0.16-0.53}$ & $0.77\pm0.52^{+0.37}_{-0.26}$ & $19.5\pm3.5\pm0.7$ & 11 \\
162 & 350--600 & $\geq$1200 & $\geq$10 & 2 & $13.7^{+3.2}_{-2.6}\pm0.3$ & $0.85^{+0.27+0.92}_{-0.21-0.82}$ & $1.15\pm0.66\pm0.50$ & $15.7^{+3.3}_{-2.7}\pm1.0$ & 12 \\
163 & 600--850 & 600--1200 & $\geq$10 & 2 & $1.6^{+2.2}_{-1.2}\pm0.0$ & $0.05^{+0.11+0.05}_{-0.04-0.03}$ & $0.04\pm0.04^{+0.02}_{-0.00}$ & $1.7^{+2.2+0.1}_{-1.2-0.0}$ & 0 \\
164 & 600--850 & $\geq$1200 & $\geq$10 & 2 & $0.9^{+1.2}_{-0.6}\pm0.0$ & $0.19^{+0.15+0.21}_{-0.09-0.17}$ & $0.06\pm0.05^{+0.03}_{-0.01}$ & $1.2^{+1.2}_{-0.6}\pm0.2$ & 0 \\
165 & $\geq$850 & 850--1700 & $\geq$10 & 2 & $0.0^{+2.4}_{-0.0}\pm0.0$ & $0.00^{+0.08}_{-0.00}\pm0.00$ & $0.05\pm0.04^{+0.02}_{-0.01}$ & $0.0^{+2.4}_{-0.0}\pm0.0$ & 0 \\
166 & $\geq$850 & $\geq$1700 & $\geq$10 & 2 & $0.0^{+1.5}_{-0.0}\pm0.0$ & $0.05^{+0.13+0.06}_{-0.04-0.03}$ & $0.02\pm0.02^{+0.01}_{-0.00}$ & $0.1^{+1.5+0.1}_{-0.0-0.0}$ & 0 \\
167 & 300--350 & 600--1200 & $\geq$10 & $\geq$3 & $6.4^{+2.4}_{-1.8}\pm0.1$ & $0.36^{+0.17+0.41}_{-0.12-0.34}$ & $0.46\pm0.32^{+0.22}_{-0.14}$ & $7.2^{+2.4}_{-1.8}\pm0.4$ & 13 \\
168 & 300--350 & $\geq$1200 & $\geq$10 & $\geq$3 & $3.8^{+2.1}_{-1.4}\pm0.1$ & $0.31^{+0.19+0.35}_{-0.12-0.28}$ & $1.50\pm0.87\pm0.64$ & $5.6^{+2.3}_{-1.7}\pm0.7$ & 5 \\
169 & 350--600 & 600--1200 & $\geq$10 & $\geq$3 & $1.6^{+1.5}_{-0.9}\pm0.0$ & $0.52^{+0.20+0.59}_{-0.15-0.50}$ & $0.11^{+0.12+0.05}_{-0.11-0.00}$ & $2.2^{+1.6+0.6}_{-0.9-0.5}$ & 3 \\
170 & 350--600 & $\geq$1200 & $\geq$10 & $\geq$3 & $4.2^{+2.1}_{-1.4}\pm0.1$ & $0.81^{+0.26+0.90}_{-0.20-0.78}$ & $0.71\pm0.44^{+0.31}_{-0.27}$ & $5.7^{+2.1+0.9}_{-1.5-0.8}$ & 9 \\
171 & 600--850 & 600--1200 & $\geq$10 & $\geq$3 & $0.0^{+3.0}_{-0.0}\pm0.0$ & $0.05^{+0.10+0.05}_{-0.04-0.03}$ & $0.04\pm0.04^{+0.02}_{-0.00}$ & $0.1^{+3.0+0.1}_{-0.1-0.0}$ & 0 \\
172 & 600--850 & $\geq$1200 & $\geq$10 & $\geq$3 & $0.0^{+1.4}_{-0.0}\pm0.0$ & $0.18^{+0.14+0.20}_{-0.09-0.16}$ & $0.04\pm0.04^{+0.02}_{-0.00}$ & $0.2^{+1.4}_{-0.1}\pm0.2$ & 1 \\
173 & $\geq$850 & 850--1700 & $\geq$10 & $\geq$3 & $0.0^{+2.0}_{-0.0}\pm0.0$ & $0.00^{+0.08}_{-0.00}\pm0.00$ & $0.05\pm0.04^{+0.02}_{-0.01}$ & $0.0^{+2.0}_{-0.0}\pm0.0$ & 0 \\
174 & $\geq$850 & $\geq$1700 & $\geq$10 & $\geq$3 & $0.0^{+1.3}_{-0.0}\pm0.0$ & $0.05^{+0.12+0.06}_{-0.04-0.03}$ & $0.02\pm0.02^{+0.01}_{-0.00}$ & $0.1^{+1.3+0.1}_{-0.0-0.0}$ & 0 \\
\end{tabular}}
\end{table}

\clearpage
\section{Aggregate search bins}
\label{app:aggbins}

\begin{figure}[tbh]
\centering
\includegraphics[width=0.95\textwidth]{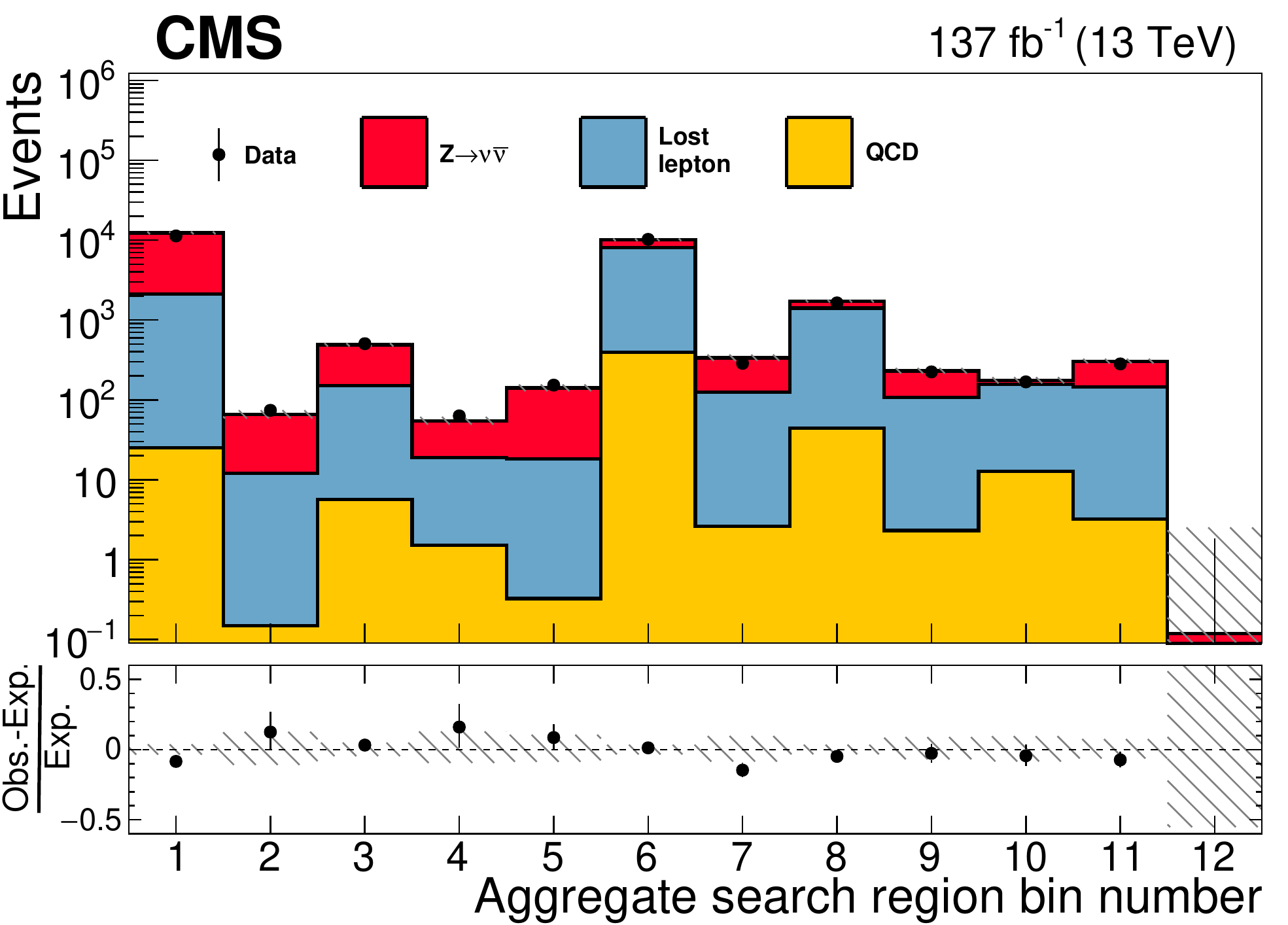}
\caption{
The observed numbers of events and pre-fit SM background predictions
in the aggregate search bins.
The total background uncertainty is shown by the hatched regions.
The lower panel displays the fractional differences between the data and the SM predictions.
}
\label{fig:results-asr}
\end{figure}

To simplify the results from the full set of search bins,
we present in this appendix the observed number of events
and corresponding SM background prediction in 12 aggregate search bins,
obtained by summing the results from the nominal search bins while
taking correlations into account.
The aggregate bins are intended to represent 12 general topologies of interest,
as indicated in Table~\ref{tab:aggbinDef}.
The intervals used to define the aggregate bins are optimized
using the signal models described in this paper.
The definitions of the aggregate bins,
along with the corresponding background predictions and observed event counts,
are given in Table~\ref{tab:pre-fit-results-asrs}.
The corresponding data are presented in Fig.~\ref{fig:results-asr}.

\begin{table}[h]
\centering
\caption{Targeted event topologies for the 12 aggregate search bins.
The variable
\dmass states the difference between the gluino or squark mass
and the sum of the masses of the particles into which
the gluino or squark decays.
}
\begin{tabular}{ c c c c}
Bin & Parton multiplicity & Heavy flavor & \dmass \\
\hline
1 & Low    & No  & Small \\
2 & Low    & No  & Large \\
3 & Medium & No  & Small \\
4 & Medium & No  & Large \\
5 & High   & No  & All \\
6 & Low    & Yes & Small \\
7 & Low    & Yes & Large \\
8 & Medium & Yes & Small \\
9 & Medium & Yes & Large \\
10 & High   & Yes & Small \\
11 & High   & Yes & Large \\
12 & High   & Yes & All \\
\end{tabular}
\label{tab:aggbinDef}
\end{table}

\begin{table}[h]
\renewcommand{\arraystretch}{1.25}
\centering
\caption{
Selection criteria,
pre-fit background predictions,
and observed number of events for the 12 aggregate search bins.
For the background predictions,
the first uncertainty is statistical and the second systematic.
}
\label{tab:pre-fit-results-asrs}
\cmsTable{
\begin{tabular}{ cccccccccc }
\multirow{2}{*}{Bin} & \mht & \HT & \multirow{2}{*}{$\njets$} & \multirow{2}{*}{$\nbjets$} & Lost-lepton & \znn & QCD & Total & \multirow{2}{*}{Observed} \\[-1.5mm]
    & [{\GeVns}] & [{\GeVns}] & & & background & background & background & background & \\
\hline
1  & $\geq$600 & $\geq$600  & $\geq$2 & $0$     & $2087\pm51\pm28$ & $10\,210\pm70\pm440$ & $25.0\pm7.0\pm9.8$ & $12\,320\pm80\pm450$ & 11\,281 \\
2  & $\geq$850 & $\geq$1700 & $\geq$4 & $0$     & $11.9^{+4.1}_{-2.8}\pm0.2$ & $53.7\pm5.0\pm4.8$ & $0.15\pm0.07\pm0.04$ & $65.8^{+6.7}_{-5.4}\pm4.9$ & 74 \\
3  & $\geq$600 & $\geq$600  & $\geq$6 & $0$     & $146\pm10\pm2$ & $338\pm12\pm18$ & $5.7\pm1.6\pm2.1$ & $489\pm15\pm18$ & 505 \\
4  & $\geq$600 & $\geq$600  & $\geq$8 & $0$--$1$  & $17.6^{+4.6}_{-2.8}\pm0.2$ & $35.2^{+4.6}_{-3.5}\pm2.5$ & $1.51\pm0.51\pm0.56$ & $54.3^{+6.5}_{-4.5}\pm2.5$ & 63 \\
5  & $\geq$850 & $\geq$1700 & $\geq$10 & $0$--$1$ & $17.9^{+7.8}_{-3.5}\pm0.2$ & $122.7^{+9.1}_{-7.9}\pm8.8$ & $0.33\pm0.11\pm0.10$ & $141^{+12}_{-9}\pm9$ & 153 \\
6  & $\geq$300 & $\geq$300 & $\geq$4 & $\geq$2 & $7630\pm90\pm99$ & $2070\pm10\pm160$ & $390\pm70\pm270$ & $10\,090\pm120\pm330$ & 10\,216 \\
7  & $\geq$600 & $\geq$600 & $\geq$2 & $\geq$2 & $122^{+19}_{-12}\pm2$ & $211\pm2\pm26$ & $2.6\pm0.5\pm1.6$ & $336^{+19}_{-12}\pm26$ & 287 \\
8  & $\geq$350 & $\geq$350 & $\geq$6 & $\geq$2 & $1362\pm33\pm17$ & $314\pm4\pm41$ & $45\pm9\pm16$ & $1720\pm35\pm47$ & 1637 \\
9  & $\geq$600 & $\geq$600 & $\geq$4 & $\geq$2 & $105^{+16}_{-10}\pm1$ & $123\pm2\pm12$ & $2.3\pm0.5\pm1.4$ & $230^{+16}_{-10}\pm12$ & 224 \\
10 & $\geq$300 & $\geq$300 & $\geq$8 & $\geq$3 & $143^{+12}_{- 9}\pm2$ & $19.6\pm0.7\pm9.8$ & $12.8\pm3.5\pm4.7$ & $176^{+13}_{-10}\pm11$ & 168 \\
11 & $\geq$600 & $\geq$600 & $\geq$6 & $\geq$1 & $141^{+15}_{-10}\pm2$ & $160\pm6\pm16$ & $3.2\pm0.6\pm1.1$ & $304^{+16}_{-11}\pm16$ & 282 \\
12 & $\geq$850 & $\geq$850 & $\geq$10 & $\geq$3 & $0.0^{+2.4}_{-0.0}\pm0.0$ & $0.05^{+0.14+0.06}_{-0.04-0.01}$ & $0.07\pm0.04^{+0.03}_{-0.02}$ & $0.1^{+2.4+0.1}_{-0.1-0.0}$ & 0 \\

\end{tabular}
}
\end{table}

\cleardoublepage \section{The CMS Collaboration \label{app:collab}}\begin{sloppypar}\hyphenpenalty=5000\widowpenalty=500\clubpenalty=5000\vskip\cmsinstskip
\textbf{Yerevan Physics Institute, Yerevan, Armenia}\\*[0pt]
A.M.~Sirunyan$^{\textrm{\dag}}$, A.~Tumasyan
\vskip\cmsinstskip
\textbf{Institut f\"{u}r Hochenergiephysik, Wien, Austria}\\*[0pt]
W.~Adam, F.~Ambrogi, T.~Bergauer, J.~Brandstetter, M.~Dragicevic, J.~Er\"{o}, A.~Escalante~Del~Valle, M.~Flechl, R.~Fr\"{u}hwirth\cmsAuthorMark{1}, M.~Jeitler\cmsAuthorMark{1}, N.~Krammer, I.~Kr\"{a}tschmer, D.~Liko, T.~Madlener, I.~Mikulec, N.~Rad, J.~Schieck\cmsAuthorMark{1}, R.~Sch\"{o}fbeck, M.~Spanring, D.~Spitzbart, W.~Waltenberger, C.-E.~Wulz\cmsAuthorMark{1}, M.~Zarucki
% ...
\vskip\cmsinstskip
\dag: Deceased\\
1:  Also at Vienna University of Technology, Vienna, Austria\\
% ...
\end{sloppypar}
\end{document}